%% file: main.tex
\documentclass[twocolumn,appendixfloats]{aastex6}
\usepackage{morefloats}
\usepackage{color}
\usepackage{mathtools}
\usepackage{xcolor}
\usepackage{pgf}
\usepackage{import}
\usepackage[version=4]{mhchem}

\usepackage[T1]{fontenc}
\usepackage[utf8]{inputenc}
\usepackage{lmodern}
\usepackage{textcomp}
\usepackage{textgreek}

\usepackage{savesym}
\savesymbol{tablenum}
\usepackage[separate-uncertainty=true]{siunitx}
\restoresymbol{SIX}{tablenum}

\DeclareSIUnit\solarmass{\ensuremath{M_\sun}}
\DeclareSIUnit\solarradius{\ensuremath{R_\sun}}
\DeclareSIUnit\solarluminosity{\ensuremath{L_\sun}}
\DeclareSIUnit\Rstar{\ensuremath{R_\text{star}}}
\DeclareSIUnit\year{yr}
\DeclareSIUnit\astronomicalunit{AU}
\DeclareSIUnit\erg{erg}
\DeclareSIUnit\pixel{pixel}
\DeclareSIUnit\jansky{Jy}
\DeclareSIUnit\magnitude{mag}
\DeclareSIUnit\parsec{pc}

\setcitestyle{comma}

\hypersetup{
    colorlinks,
    linkcolor={red!60!black},
    citecolor={blue!40!black},
    urlcolor={blue!40!black}
}

\let\pgfimageWithoutPath\pgfimage
\renewcommand{\pgfimage}[2][]{\pgfimageWithoutPath[#1]{media/#2}}

\input{autoref.tex}

\begin{document}

\title{An \textit{HST} Survey of Protostellar Outflow Cavities: Does Feedback Clear Envelopes?}
\author{Nolan M. Habel\altaffilmark{1}, S. Thomas Megeath\altaffilmark{1}, Joseph Jon Booker\altaffilmark{1}, William J. Fischer\altaffilmark{2}, Marina Kounkel\altaffilmark{3}, Charles Poteet\altaffilmark{1,2}, Elise Furlan\altaffilmark{4}, Amelia Stutz\altaffilmark{5,6}, P. Manoj\altaffilmark{7}, John J. Tobin\altaffilmark{8}, Zsofia Nagy\altaffilmark{1,9}, Riwaj Pokhrel\altaffilmark{1}, Dan Watson\altaffilmark{10}}
\altaffiltext{1}{Ritter Astrophysical Research Center, Department of Physics and Astronomy, University of Toledo, 2801 W. Bancroft Street, Toledo, OH 43606, USA}
\altaffiltext{2}{Space Telescope Science Institute, 3700 San Martin Drive, Baltimore, MD 21218, USA}
\altaffiltext{3}{Department of Physics and Astronomy, Western Washington University, 516 High st., Bellingham, WA, 98225}
\altaffiltext{4}{NASA Exoplanet Science Institute, Caltech/IPAC, Mail Code 100-22, 1200 E. California Blvd., 
Pasadena, CA 91125, USA}
\altaffiltext{5}{Max-Planck-Institut f\"{u}r Astronomie, K\"{o}nigstuhl 17, D-69117 Heidelberg, Germany}
\altaffiltext{6}{Departmento de Astronom\'{\i}a, Facultad Ciencias F\'{\i}sicas y Matem\'{a}ticas, Universidad de Concepci\'{o}n, Av. Esteban Iturra s/n Barro Universitario, Casilla 160-C,  Concepci\'{o}n, Chile}
\altaffiltext{7}{Department of Astronomy and Astrophysics, Tata Institute of Fundamental Research, Homi Bhabha Road, Colaba, Mumbai 400005, India}
\altaffiltext{8}{National Radio Astronomy Observatory, 520
Edgemont Rd., Charlottesville,VA 22903, USA}
\altaffiltext{9}{Konkoly Observatory, Research Centre for Astronomy and Earth Sciences, H-1121 Budapest, Konkoly Thege \'ut 15--17, Hungary}
\altaffiltext{10}{Department of Physics and Astronomy, University of Rochester, Rochester, NY 14627, USA}

\begin{abstract}
We study protostellar envelope and outflow evolution using Hubble Space Telescope NICMOS or WFC3  images of 304 protostars in the Orion Molecular clouds. These near-IR images resolve structures in the envelopes delineated by the scattered light of the central protostars with 80~AU resolution and they complement the \SIrange{1.2}{870}{\micro\meter} spectral energy distributions obtained with the Herschel Orion Protostar Survey program (HOPS). Based on their \SI{1.60}{\micro\meter} morphologies, we classify the protostars into five categories: non-detections, point sources without nebulosity, bipolar cavity sources, unipolar cavity sources, and irregulars. We find point sources without associated nebulosity are the most numerous, and show through monochromatic Monte Carlo radiative transfer modeling that this morphology occurs when protostars are observed at low inclinations or have low envelope densities. We also find that the morphology is correlated with the SED-determined evolutionary class with Class 0 protostars more likely to be non-detections, Class I protostars to show cavities and flat-spectrum protostars to be point sources. Using an edge detection algorithm to trace the projected edges of the cavities, we fit power-laws to the resulting cavity shapes, thereby measuring the cavity half-opening angles and power-law exponents. We find no evidence for the growth of outflow cavities as protostars evolve through the Class I protostar phase, in contradiction with previous studies of smaller samples. We conclude that  the decline of mass infall with time cannot be explained by  the progressive  clearing of envelopes by growing outflow cavities. Furthermore, the low star formation efficiency inferred for molecular cores  cannot be explained by envelope clearing alone. 
 
\end{abstract}

\section{Introduction}
Low mass protostars are characterized by a rapid evolution, with the accretion of the stellar mass, the formation of disks and potentially the initiation of planet formation occurring within \SI{0.5}{\mega\year} \citep{arce_evolution_2006,cassen_formation_1981, dunham_evolution_2014, alma_partnership_2014_2015,dipierro_planet_2015}. The defining characteristic of the protostellar phase is the presence of a dusty, infalling envelope which absorbs and reprocesses most of the luminosity from the central protostar. In the initial phases of protostellar evolution, the envelope dominates the mass, while in the later phases, most of the mass is already accreted onto the star. Even in these later phases, the mass of the envelope typically exceeds that of the circumstellar disks surrounding the central protostar \citep[e.g.][]{fischer_hops_2014}; hence, infall in these phases shapes the properties of circumstellar disks and sets the stage for planet formation. Understanding the factors that govern the evolution of the envelopes, and thereby influence mass accretion and the properties of nascent disks, is a key problem in star and planet formation studies.

This evolution is accompanied by a rapid change in the shape of the SEDs produced by the reprocessing and scattering of radiative energy in the evolving disks and envelopes \citep{furlan_herschel_2016}. Since the central protostar is deeply embedded in its  envelope, the effective temperatures and photospheric luminosities of protostars typically cannot be measured directly. In most cases, unlike pre-main sequence stars, they cannot be reliably placed on HR diagrams and compared to evolutionary tracks to estimate masses and ages.  Instead, the evolution of protostars is largely inferred from the shape of their SEDs. This evolution is typically measured by sorting protostars into bulk evolutionary classes based on the percentage of luminosity radiated in the sub-millimeter, their near- to mid-infrared spectral index or $T_\text{bol}$, the bolometeric temperature \citep[e.g.,][]{adams_infrared_1985,myers_bolometric_1993,andre_submillimeter_1993,stutz_herschel_2013,dunham_evolution_2014,furlan_herschel_2016}. The observed sequence of evolutionary classes, Class 0, Class I and flat-spectrum, shows the peak of the SED shifting from the far-infrared to the mid-infrared and the SED flattening as the protostars evolve and the envelopes dissipate \citep[e.g.,][]{furlan_herschel_2016}. Class II objects can be identified by their decreasing near- to mid-IR SED slopes and are primarily pre-main sequence stars with disks that have exited the protostellar phase.

Due to the flattening of envelopes by rotation and the clearing of cavities in the envelopes by outflows, the luminosity of the protostars is not radiated isotropically, but is preferentially beamed along the rotation axis of the protostars. The resulting SEDs depend on the inclination of the protostars, and the effects of inclination on the SEDs are difficult to disentangle from those due to evolution \citep{kenyon_embedded_1993,whitney_two-dimensional_2003}. To circumvent this degeneracy, \citet{whitney_model_1993} and \citet{robitaille_interpreting_2007} proposed a set of evolutionary stages which are dependent on the physical properties of the envelopes and not inclination; however, it is often difficult to reliably infer the stage  of a protostar from the obseved SED alone. Nevertheless, taking into account the uncertainties due to inclination, \citet{furlan_herschel_2016} demonstrate that the observed SEDs of the distinct evolutionary classes require the dissipation of the envelope, with the density of the \replaced{infalling}{envelope} gas (as inferred by model fits to the SEDs) dropping by a factor of 50 between the Class 0 and flat-spectrum phases. This shows that the envelopes decrease in density dramatically during the Class I phase.

Although SEDs are currently the primary information we have on large samples of protostars, imaging at millimeter, submillimeter and near-infrared wavelengths can be used to study protostellar evolution by resolving structures in the envelope that may change as protostars evolve \citep[e.g.,][]{arce_evolution_2006}. \textit{HST} near-infrared images of protostars resolve structures seen directly in light scattered by dust grains in an envelope or in silhouette against \added{the} scattered light, placing constraints on the envelopes and disks that are complementary to those inferred from SEDs. \textit{HST} imaging of protostars by \citet{padgett_hst/wfpc2_1999}, \citet{allen_hubble_2002}, \citet{terebey_circumstellar_2006} and \citet{fischer_hops_2014} show outflow cavities illuminated in scattered light, edge-on disks seen in absorption and shadows cast into the envelopes by flared disks.

Of particular interest is the role of feedback from outflows in driving the evolution of protostars by clearing the envelope and halting infall. SED-based measurements cannot reliably constrain outflow cavity sizes \citep{furlan_herschel_2016}; hence, studies of the growth of outflow cavities must rely on observations that spatially resolve structures in envelopes. The \ce{CO} observations of nine Class 0, I and II sources by \citet{arce_evolution_2006} showed a widening in outflow size with evolutionary class. Bolstering their sample by nine sources in the literature, they found evidence that outflow cavity sizes increase progressively as protostars evolve. \citet{tobin_imaging_2007}\added{ and }\citet{seale_morphological_2008} used \textit{Spitzer} IRAC images of protostar outflow cavities illuminated in scattered light to study the growth of cavities, and the latter authors found some evidence of outflow cavity growth with evolution, \added{although} with significant scatter.

These studies suggest that feedback from outflows play a significant role in the decrease or halting of infall and accretion. Although accretion from the disk can continue after infall stops, the resulting increase in mass is small compared to the stellar mass. By reducing or halting infall, feedback can also play an important role in the star formation efficiency inferred from the core mass function. In particular, the mass function of cores identified in sub-mm measurements can reproduce the initial mass function if each core forms a star with a star formation efficiency (defined by the stellar to initial core mass) of \SIrange{30}{40}{\percent} \citep{alves_mass_2007,konyves_census_2015}. Furthermore, simulations of protostars including feedback can produce star formation efficiencies of \SI{50}{\percent} or lower \citep{machida_evolution_2013,machida_impact_2012,hansen_feedback_2012,offner_investigations_2014, offner_outflows_turbulence_2017}.

There are difficulties, however, in explaining the low star formation efficiency with feedback alone. Single dish radio observations suggest that outflows may carry too little mass to clear out the envelope in \SI{0.5}{\mega\year} \citep{hatchell_star_2007,curtis_submillimetre_2010}. Furthermore, even large cavities  clear less than half of the envelope mass \citep{frank_jets_2014}.
These studies, however, likely underestimate the amount of entrained, low-velocity gas in the outflow \citep{dunham_molecular_outflows_2014}.  ALMA observations can now map these lower-velocity flows and show whether they transport a significant fraction of the envelope gas over the lifetime of the protostar \citep{zhang_alma_2016}.

To further investigate the impact of outflows on protostellar envelopes, we use in this work the largest \textit{HST} survey of protostars to date. This survey focuses on the sample of protostars targeted by the Herschel Orion Protostar Survey, or HOPS.\ The protostars were identified using combined 2MASS and \textit{Spitzer} photometry from the \textit{Spitzer} Orion Survey \citep{megeath_spitzer_2012,megeath_spitzer_2016} and observed with Herschel and APEX to obtain well sampled \SIrange{1.2}{870}{\micro\meter} SEDs.  This sample was supplemented by very red protostars discovered with Herschel \citep{stutz_herschel_2013,tobin_herschel_2016}.  \citet{furlan_herschel_2016} published the  SEDs of the entire sample and then presented model fits to 319 of the protostars and eleven pre-main sequence stars after rejecting likely extragalactic contamination and sources without Herschel detections. The \textit{HST} survey examined 304 of these sources, (enumerated in \autoref{tab:bigtable}), using initially NICMOS at 1.60 and \SI{2.05}{\micro\meter}, and then after the failure of NICMOS, WFC3 at \SI{1.60}{\micro\meter}.  A search for binary systems using these data was published by \citet{kounkel_hst_2016}.

The morphologies of outflow cavities carved by the outflows can be seen by mapping the location of the cavity wall in scattered light. The volumes of the cavities carved by outflows can then be directly measured. The mechanism for creating these cavities, whether by jet precession, wide-angled winds or jet entrainment \citep{raga_molecular_1993,lee_hydrodynamic_2001,matzner_bipolar_1999,ybarra_first_2006}, is still debated. Independent of the underlying mechanism, the scattered-light cavities provide a direct measurement of the cleared gas with the \SI{80}{\astronomicalunit} (\SI{0.18}{\arcsecond}) resolution of \textit{HST}. These are used in this work to estimate the fraction of the volume cleared, which provides an estimate of the fraction of mass cleared.

In \autoref{sec:observations}, we discuss the observations used in this paper. We make use of radiative transfer modeling, described in \autoref{sec:modeling}, to understand the morphologies of the observed cavities and to calibrate the relationship between the scattered-light distributions and the cavity properties. In \autoref{sec:results}, we present the morphologies of the observed protostar  and our analysis of the cavity sizes. Finally, we discuss the implications for protostellar evolution in \autoref{sec:discussion}. Images of the protostars in our sample are shown in \autoref{sec:images}.

\section{\textit{HST} Observations of The Sample}\label{sec:observations}
The \textit{Hubble Space Telescope} observations were assembled from two GO programs and a snapshot program. The bulk of the sample was observed in program GO 11548. The Near Infrared Camera and Multi-Object Spectrometer's (NICMOS) F205W and F160W filters were used for a total of 87 orbits in August and September of 2008 to image 92 objects in the HOPS catalog, before the failure of the cryocooler of NICMOS. After the June 2009 deployment of the Wide Field Camera 3 (WFC3), 126 orbits were used between August 2009 and December 2010 to observe 237 HOPS objects with the F160W filter. The observation and reduction of these data is described in \citet{kounkel_hst_2016}. A subsequent program using WFC3, SNAP 14181, was designed to target multiple star forming regions within 500 pc. It completed observations during 114 orbits between December 2015 and September 2017, 10 of which imaged 13 objects in the Orion Molecular Clouds. A final WFC3 study, program GO 14695, targeted four objects in Orion with weak 24~$\mu$m fluxes atypical of protostars. These observations were conducted in September 2016 with four orbits. For these final two programs we used the standard data products produced from the \texttt{calwf3} data reduction pipeline which were then combined with \texttt{AstroDrizzle} from the \texttt{DrizzlePac} package using a drop size of 1 onto a pixel scale of \SI{0.13}{\arcsecond}.  

The NICMOS observations used the NIC2 camera, which has a \SI{0.075}{\arcsecond} pixel size and resolution of \SI{0.15}{\arcsecond}. Integration times were \SI{1215.4}{\second} and \SI{767.6}{\second} for F160W and F205W filters, respectively. The WFC3 integration times were \SI{2496.2}{\second}  for GO 11548, \SI{1596.9}{\second} for SNAP 14181, and  \SI{2396.9}{\second} for GO 14695.  All have a \SI{0.18}{\arcsecond} angular resolution and a pixel size of \SI{0.13}{\arcsecond}. In this work we adopt a distance to Orion of 420 parsecs for consistency with \citet{furlan_herschel_2016}. This is within the range of distances found in \citet{kounkel_distances_2018} and \cite{grossschedl_distances_2018} through APOGEE and \textit{Gaia} measurements. At this distance, both NICMOS and WFC3 resolve structures down to  \SI{80}{\astronomicalunit} scales.

Nine images taken with NICMOS are excluded from this analysis due to the lack of guide star tracking; these contain HOPS 46, 47, 134, 139, 149, 227, 250, 271 and 276. Three WFC3 images, containing HOPS 293, 330 and 336, were also excluded due to what appear to be tracking failures. One additional  WFC3 observation was excluded due to an apparent pointing error with its target object, HOPS 100, only partially appearing on the edge of the frame. Three images where only one guide star was used, those containing HOPS 10, 177, 316 and 358, may suffer from a small amount of rotation during the exposure, although this is not apparent in the data. These are included in our program. Twenty-seven of the HOPS targets were imaged by both NICMOS and WFC3 due to their proximity to other protostars. Of these sources, only HOPS 250 showed a clear difference between the two observations due to the tracking failure. 

Some of the HOPS targets were classified as potential extragalactic contamination by \citet{furlan_herschel_2016} based on the presence of PAH features in their \textit{Spitzer} IRS spectrum, lack of silicate absorption at $\SI{10}{\micro\meter}$, or the shape of the mid-infrared continuum \citep[see appendix of][]{furlan_herschel_2016}. The \textit{HST} observations provide an independent means for separating galaxies from protostars. Only one source, HOPS 339, is conclusively determined by its morphology to be a galaxy and is omitted from the table in the appendix of this work (\autoref{tab:bigtable}). The WFC3 image of this source is shown in \autoref{sec:vermin}.  Conversely, we add back into our sample and assign a class to HOPS 48, 67, and 301. These have morphologies in WFC3 imaging indicative of protostellar cavities. The nature of the remaining potentially extragalactic sources could not be clarified through WFC3 imaging. In program GO 14695, two of the four targeted sources were found not to be protostars; one was a galaxy and one an outflow knot; neither of these has a HOPS number (see \autoref{sec:vermin}).
In total, we imaged 304 objects in our sample.  We note that 7 of these were determined to be Class II objects by their SEDs in \citet{furlan_herschel_2016}.  Since these sources are in the HOPS sample and may have residual envelopes, we keep them in the analysis.  We typically use “protostars” to refer to this entire sample. In addition, we serendipitously observed two Class II sources with nebulosity in our images \citep{kounkel_hst_2016}. We describe these objects in \autoref{sec:extCII}.

\begin{figure}\centering
	\includegraphics[scale = 1.2]{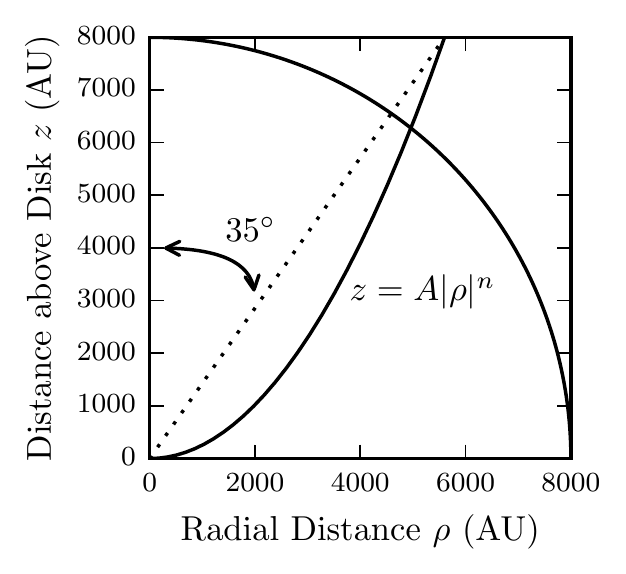}
	\caption{\label{fig:cavity-schematic}The definition of the cavity half-opening angle used in this paper and in the \texttt{HO-CHUNK} code. The circular region is the outer radius of the envelope of the protostar in these models, set to 8000 AU;  the parabolic ($n=2$) black line is the adopted boundary of the evacuated outflow cavity. The central protostar is located at the origin. The cavity half-opening angle is defined by the angle between the dotted line intersecting the cavity at \SI{8000}{\astronomicalunit} above the normal to the disk and the polar axis.}
\end{figure}

\begin{table*}\centering
	\caption{\label{tab:fixedpars}Parameters used in the model grid described in \autoref{sec:modeling}.}
	\begin{tabular}{lc}
		\firsthline
		\textbf{Parameter} & \textbf{Value(s)} \\
		\si{\Rstar}: Radius of star & \SI{2.09}{\solarradius} \\
		Temperature of central star & \SI{4.e3}{\kelvin} \\
		Mass of central star & \SI{0.5}{\solarmass} \\
		Minimum disk radius & \SI{7.00}{\Rstar} \\
		Disk Scale height at \si{\Rstar} \footnote{$h_0$ of \citet[eqn 5]{whitney_model_1992}} & \SI{0.018}{\astronomicalunit} \\
		Maximum envelope radius & \SI{8000}{\astronomicalunit} \\
		Minimum envelope radius & \SI{6.85}{\Rstar} \\
		Degree of polynomial shape of cavities & $2.0$ \\
		Height of cavity wall at $\rho$ = 0 & \SI{0}{\Rstar} \\
		Density of the cavity & \SI{0}{\gram\per\centi\meter\cubed} \\
		Ambient cloud density & \SI{0}{\gram\per\centi\meter\cubed} \\
		Minimum radius of outflow & \SI{.1}{\astronomicalunit} \\
		Maximum disk radius & \SIlist{5;50;100;200;300}{\astronomicalunit} \\
		Centrifugal radius & Always equal to maximum disk radius \\
		Mass of disk & \SIlist{0.001;0.005;0.01;0.05}{\solarmass} \\
		$\alpha_{disk}$: Radial exponent in disk density law & \numlist{2.125;2.25} \\
		$\beta_{disk}$: Vertical exponent in disk density law &  $\alpha_{disk} - 1$ \\
		Mass infall rate \footnote{See \citet[eqn 3]{whitney_model_1993}}  & \SIlist{1e-7;5e-7;1e-6;5e-6;1e-5;5e-5}{\solarmass\per\year} \\
		\replaced{Opening}{Half-opening} angle of inner cavity wall & \SIlist[list-units=repeat]{5;15;25;35;45}{\degree} \\
		Angle of inclination measured from polar axis & \SIlist[list-units=repeat]{1;10;20;30;40;50;60;70;80;90}{\degree} \\
		\lasthline
	\end{tabular}
\end{table*}

\section{Model Grid}\label{sec:modeling}
In order to quantify the shape of the observed cavities, we used a monochromatic Monte Carlo radiative transfer code, \texttt{HO-CHUNK.ttsscat.20090521} \citep[based on][]{whitney_model_1992,whitney_model_1993}. With this code, we simulated  \SI{1.60}{\micro\meter} images of a half solar mass star surrounded by a flared disk, with a power-law radial density and scale height and an envelope,  following the Terebey, Shu, and Cassen (TSC) model described in \citet{terebey_collapse_1984}, \citep[see also][]{ulrich_infall_1976,cassen_formation_1981}.
We examined six envelope densities (each corresponding to a different mass infall rate in Table 1), five cavity half-opening angles (see \autoref{fig:cavity-schematic} for the definition), five disk sizes, four disk masses, two variations on disk flaring and ten inclinations. \autoref{tab:fixedpars} shows the parameters used in our model grid. All models adopt an identical photon flux from the central star and assume fully-evacuated cavities containing no material. These model images were convolved with the \textit{HST} WFC3IR point spread function for the F160W filter. In this paper, we are primarily interested in variations in the observed near-IR morphology due to changes in envelope density, cavity half-opening angle, and inclination.

\begin{figure*}\centering
	\includegraphics{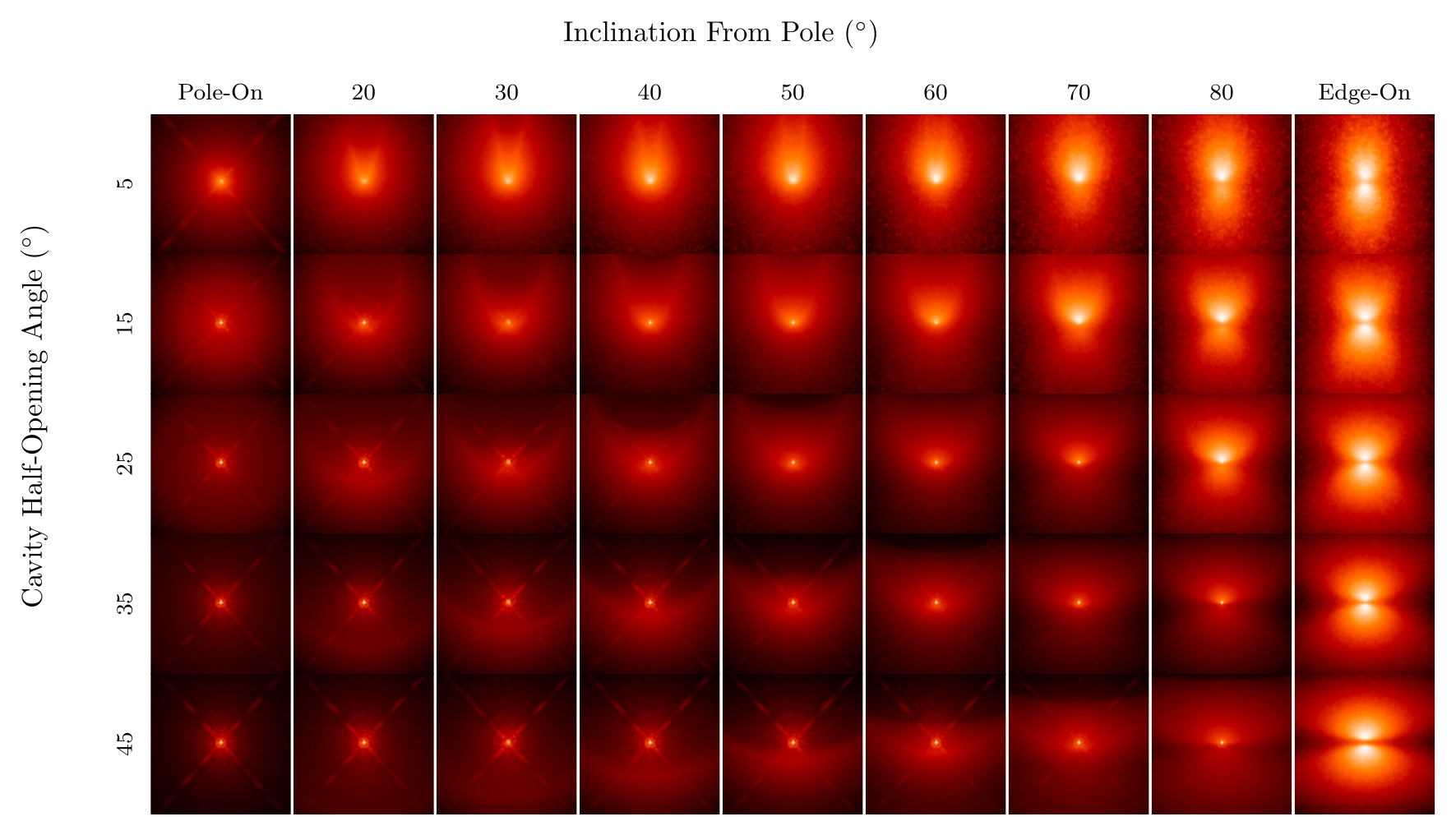}
	\caption{\label{fig:model-grid}A selection of models from the grid used in this work, showing variations in the observed scattered-light morphologies from models with a mass infall rate of $\SI{5e-6}{\solarmass\per\year}$. Note that the contrast between the cavity and the central point source is highest when observed at an inclination greater than the half-opening angle. Each model is shown with an approximately 8000 by 8000 AU field of view.}
\end{figure*}

In these models, the mass infall rate is used as a parameter to control the densities of the envelopes. The infall rate is combined with an adopted central stellar mass of \SI{0.5}{\solarmass}  to scale the envelope density using Eqn.\@ 3 from \citet{kenyon_embedded_1993}. See  \citet{furlan_herschel_2016} for further discussion on this scaling.

The disk and envelope dust \replaced{use a}{opacities are from a} model by \citet{ormel_dust_2011} that adopts a 2:1 mixture of ice-coated silicates and bare graphite grains, where the depth of the ice coating is \SI{10}{\percent} of the particle radius. The particles are subjected to time-dependent coagulation; we choose a coagulation time of \SI{0.3}{\mega\year}. This is identical to the dust model used in \citet{fischer_hops_2014} and \citet{furlan_herschel_2016}. In the near-infrared, the opacities predicted by this model are slightly smaller than those of the often-cited OH5 opacities \citep{ossenkopf_dust_1994}. The reasons for adopting this model are described in  \citet{furlan_herschel_2016}. 
In \autoref{sec:dust_law_comparison}, we assess the dependence of the cavity appearance on the dust law.

Motivated by the shape of the observed outflow cavities, we used a parabolic model ($n = 2$) shown in \autoref{fig:cavity-schematic} for outflow/envelope boundary in our models. In \autoref{sec:cavity-measure}, we relax this constraint and use the power-law fit 
\begin{equation}
z = A\left | \rho  \right |^{n}
\label{eqn:power_law}
\end{equation}
where the resulting power-law index, \textit{n}, may be 1 or greater. For a given power-law, the cavity half-opening angle depends on the adopted outer radius of the envelope; only for the case of a conical cavity ($n = 1$) is the half-opening angle independent of the adopted outer radius. 

Examples of models from the grid are shown in \autoref{fig:model-grid}, which displays the effect of differing inclinations and cavity half-opening angles. Several model parameters, such as the radius and temperature of the central protostar or the presence of hot spots, are not constrained by either the SEDs or the near-infrared images. The surface brightness found in an image depends on the  monochromatic luminosity of the protostar (which in turn depends on the temperature, radius and presence of hot spots), but the morphology of the image depends primarily on envelope density, outflow cavity shape and inclination. The rest of our model parameters are chosen to cover a range of physical parameters observed in the fitting done by \citet{furlan_herschel_2016}. This allows us to compare in \autoref{sec:sed-compare} the values for the parameters determined by the fits to the SEDs and those determined from the near infrared images.

As shown by the models, the observed morphologies of the cavities trace the light scattered at a discontinuity in the dust density; in this case, the discontinuity is the boundary of a cleared cavity. If the protostar is seen edge-on, both cavities carved by the bipolar outflow are apparent. For these edge-on cases, a dust disk obscures the scattered light creating a dust lane (\Autoref{fig:model-grid}). If the system is inclined such that the extinction toward the far cavity is significantly higher than that toward the nearer one, a bowl-shaped unipolar structure is seen due to the obscuration of the more distant cavity. The envelope itself can be directly illuminated if the density is low enough for near-infrared photons to penetrate past the cavity walls and scatter off grains deep in the envelope. In these cases, the disk can cast shadows in the envelope which are also apparent for edge-on inclinations. 

\begin{figure}
	\includegraphics[angle=0,scale=0.6]{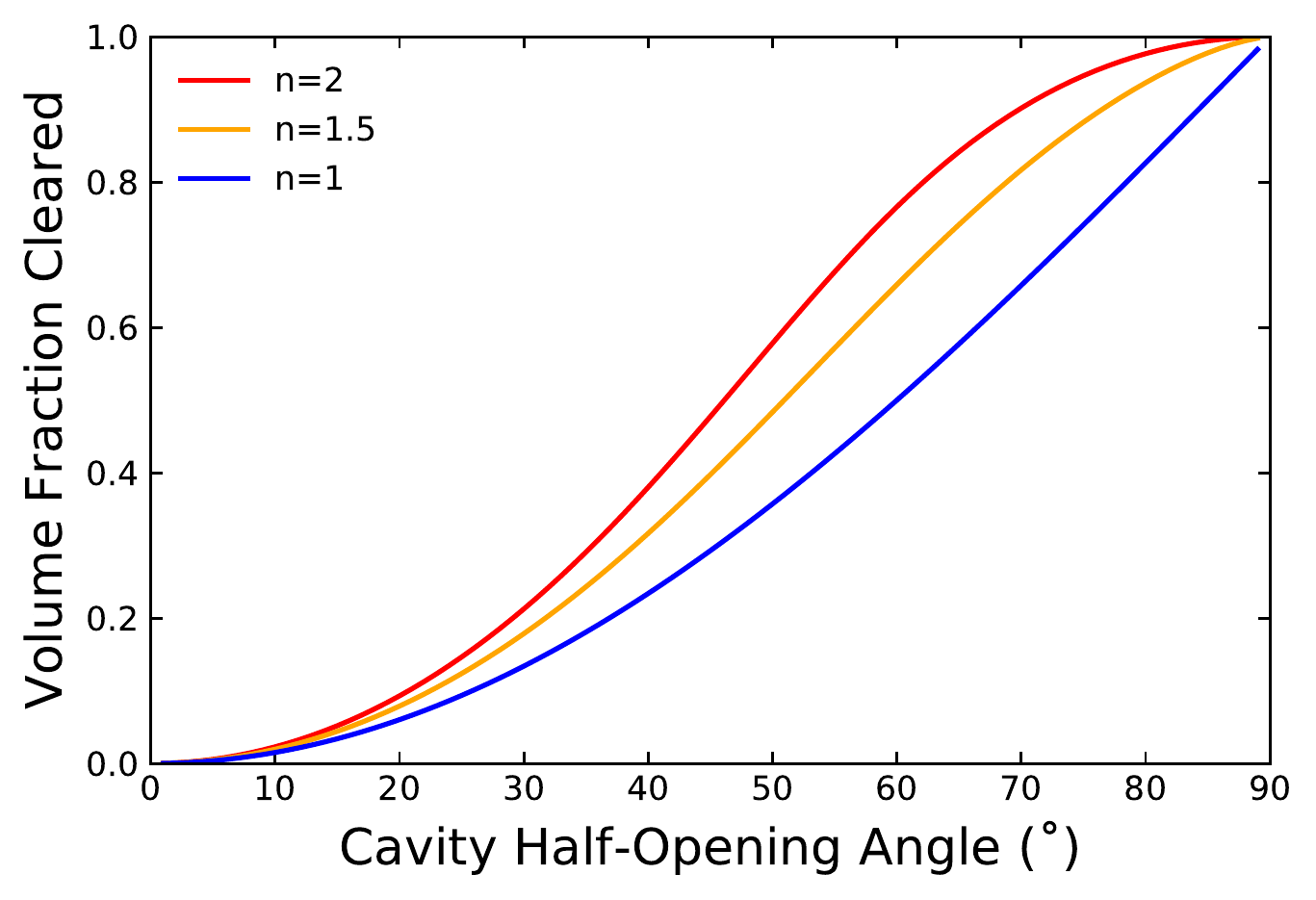}
	\caption{\label{fig:varying-n}The dependence of cleared cavity volume as a fraction of total envelope volume on cavity exponent and half-opening angle. The cavities are carved in a spherically symmetric envelope with an outer radius of 8000 AU. \explain{Added figure and associated paragraph to show how the physical meaning of cavity half-opening angle depends on choice of $n$.}}
\end{figure}

To compare our cavities to those measured in other analyses that adopt different models for their shapes \citep[e.g. this work, ][]{furlan_herschel_2016,arce_evolution_2006}, we will determine the fraction of the envelope volume within the cavities.  This is a measure of the amount of gas cleared by the outflow. The volume of the cavities in these models depends only on the  power-law exponent  $n$, the half-opening angle $\theta$ and the outer envelope radius $R_\text{max}$ (\autoref{fig:cavity-schematic}). 
In~\autoref{fig:varying-n}, we show the dependence of the fraction of the envelope volume cleared by the cavity on the cavity half-opening angle and the cavity exponent.

\begin{figure}
	\includegraphics[angle=0,scale=0.6]{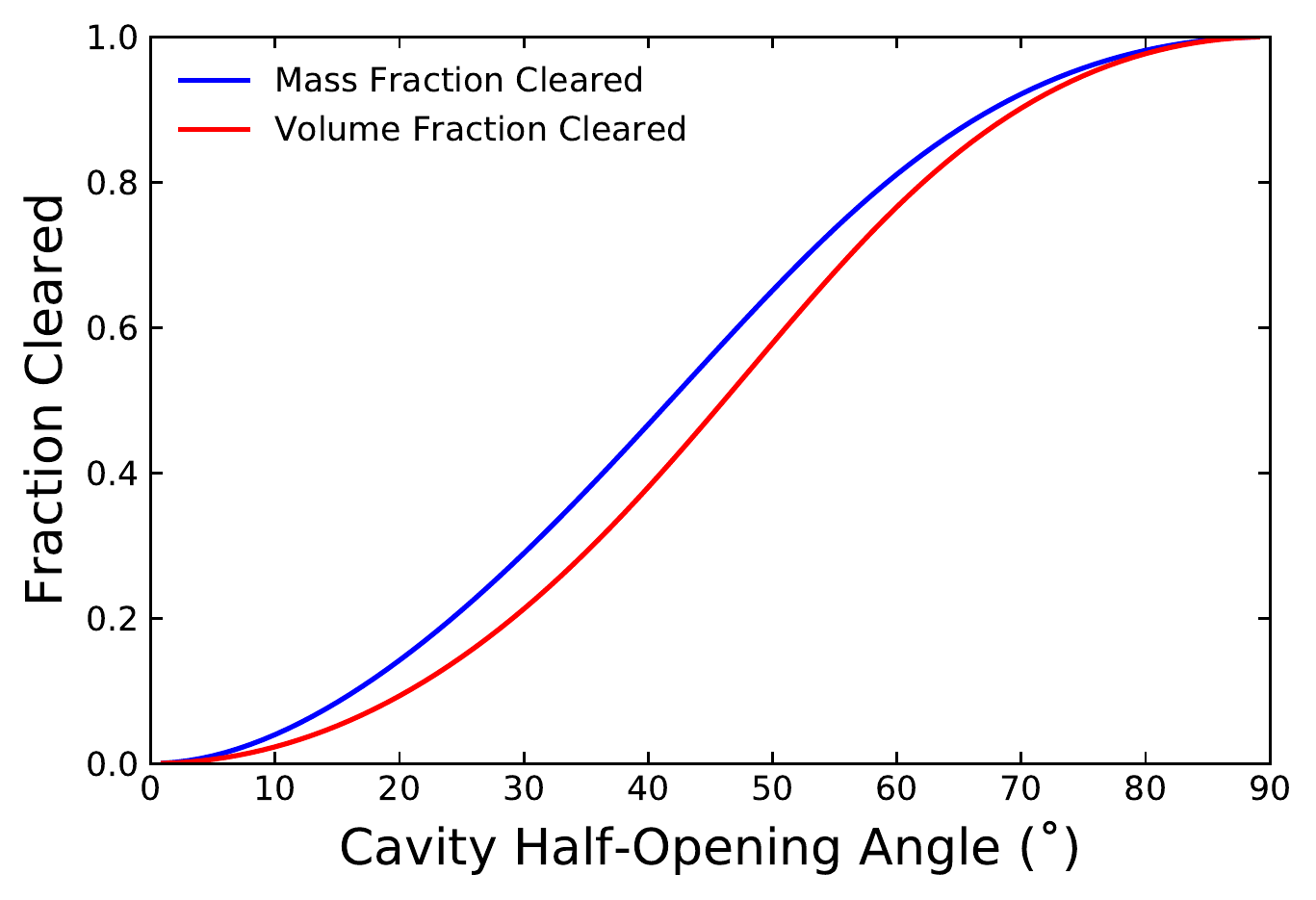}
	\caption{\label{fig:mass-vs-volume}A comparison between the volume fraction cleared used in this work and the mass fraction cleared for parabolic ($n = 2$) cavities. To obtain the mass cleared, we assumed a spherical envelope with an outer radius of \SI{8000}{\astronomicalunit} and a radial density profile of $\rho^{-1.5}$.
	The blue solid curve gives the fraction of the envelope mass evacuated by the presence of a cavity, while the red curve shows the fraction of the volume subtended by a cavity.}
\end{figure}

An alternative metric for characterizing cavity sizes is the fraction of the envelope mass cleared by the outflows, i.e. the fraction of mass that would be found in an initially spherical symmetric core with a $\rho^{-1.5}$ density law and an outer radius of 8000 AU.
We compare the volume and mass fraction cleared in \autoref{fig:mass-vs-volume}. We find the fraction of the mass cleared can be up to \SI{9}{\percent} more than the volume cleared, and that the volume cleared is a lower limit to the mass cleared. We note that this is an instantaneous mass fraction of the current envelope,\deleted{that would fill the volume of the cavity if they were not cleared}and it will differ from the total fraction of the envelope mass entrained and ejected by the outflow over the history of a protostellar collapse. Furthermore, it does not include the mass launched and ejected from the system by disk winds, X-winds, or accretion-driven stellar winds \citep[e.g.][]{watson_evolution_2016}.

\section{Results}\label{sec:results}
In this section, we classify the protostars on the basis of their \SI{1.60}{\micro\meter} morphologies in the \textit{HST} images. We then examine how the morphologies depend on the properties derived from the model fits to their SEDs. For protostars with detected outflow cavities, we develop an algorithm to measure the shape of the outflow cavity, and we calibrate this approach using the radiative transfer models in our grid.

We exclude the images with the F205W filter from this analysis as only the 83 objects successfully observed with NICMOS have these data. Furthermore, with a small number of exceptions, the morphologies are identical in the two NICMOS bands.

\subsection{Protostellar Morphologies}\label{sec:morphologies}
\begin{figure}\centering
	\includegraphics[scale=.425]{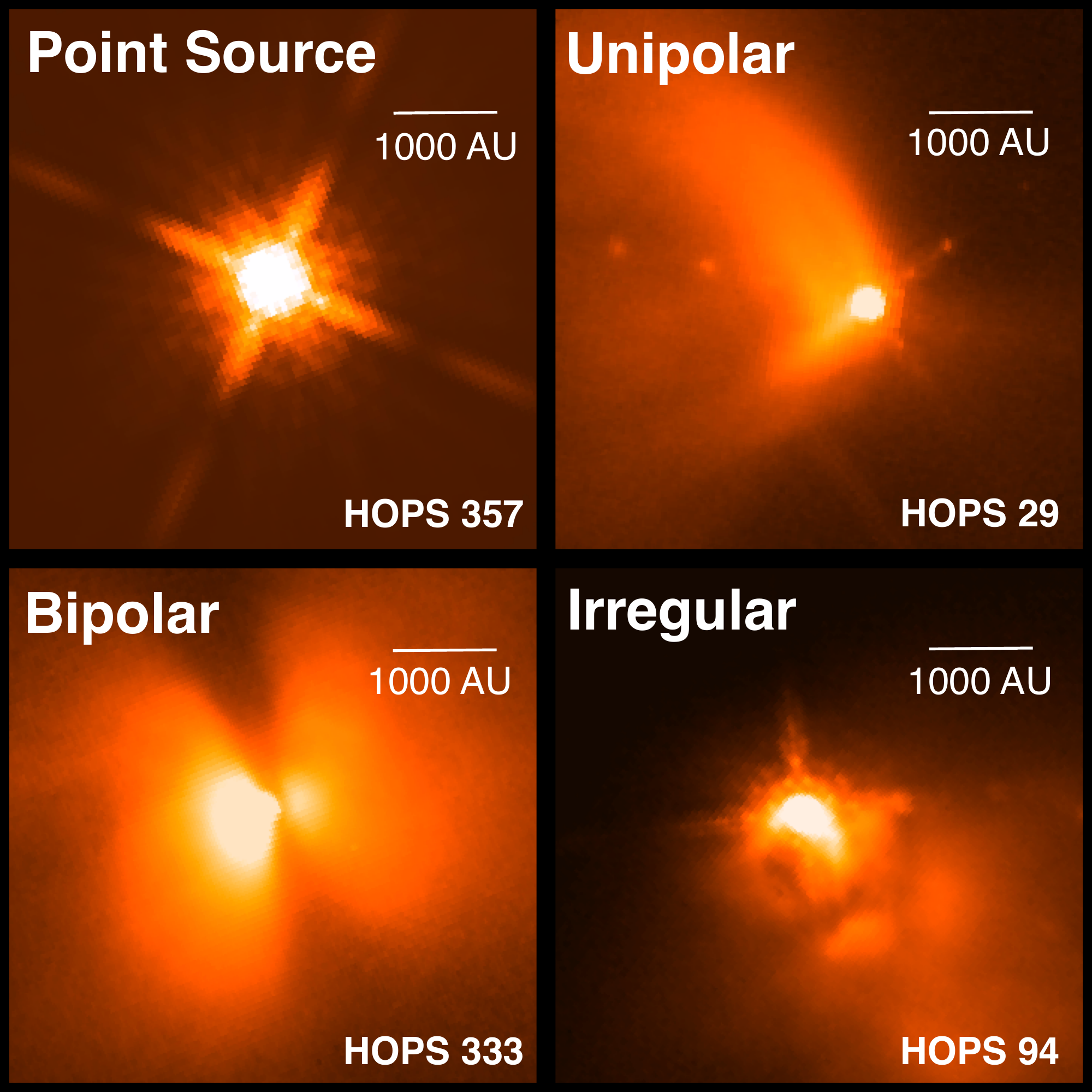}
	\caption{\label{fig:morph}Our four morphological types resolved in \textit{Hubble} WFC3 and NICMOS images as exemplified by HOPS 357, 29, 333 and 94. All images are squares with \SI{12}{\arcsecond} (\SI{5000}{\astronomicalunit}) on a side. \added{Note that HOPS 29 shows evidence of an a jet interior to its cavity and HOPS 333 shows a dark lane commonly seen in bipolar sources.}}
\end{figure}

The \textit{HST} images resolve protostars at various stages of evolution, different inclinations and differing amounts of envelope material. In these images, light primarily from the photospheres of the central protostars is scattered by  dust grains in the envelopes, delineating structures present in the envelopes. In many of the images, the structures are similar to those caused by the outflow cavities in our model grid.

As a first step in our analysis, we divide all protostars into five morphological categories (\autoref{fig:morph}). The presence of a bipolar nebula, such as two scattered-light lobes separated by a dark lane or  two outflow cavities, define the bipolar category. Sources with only one cavity visible make up the unipolar category. Unresolved protostars without detectable nebulosity are defined as point sources.  Sources too deeply embedded to detect in the F160W band are considered non-detections (not shown in \autoref{fig:morph}). The final category comprises irregular protostars; these may result from background contamination (e.g., coincidence with a more extended reflection nebula), or true inhomogeneities in the structure of the protostellar envelope.  For bipolar, unipolar and irregular categories, the presence of an unresolved point source in the nebula is noted; these are likely to be the central protostar or light scattering off of structures within 80 AU of the protostar, which is the smallest scale we can resolve in our images. 

In total, 141 HOPS objects exhibit extended structures in scattered light. The classification of all protostars are found in \autoref{tab:bigtable}, and their breakdown is summarized in \autoref{tab:counts}. Of these, 
thirty-one show a bipolar structure indicative of an edge-on inclination, although some cases show the point source of the central protostar near or offset from the midplane of the dark lane, implying that they are not exactly edge-on. One bipolar source was serendipitously observed in the same field as HOPS 334. This source was first identified as a candidate protostar by \citet{stutz_herschel_2013}; based on their values for $T_\text{bol}$ and  $L_\text{bol}$, it is determined by the criteria in \citet{furlan_herschel_2016} to be a Class 0 protostar. In this paper, we introduce this source into the HOPS catalog as HOPS 410 (\autoref{tab:bigtable}). 
Fifty-nine objects show nebulosity appearing to be a cavity on one side, with 
36 of those having detected point sources near the base of the cavity. Fifty-one remaining protostars are classified as irregular. Images of sources with unipolar, bipolar, irregular and point-like morphology are shown in \autoref{sec:images}. Two additional Class II sources with nebulosity that were serendipitously discovered in our observation are shown in \autoref{sec:extCII}.
\explain{Update counts in text and table}

\begin{table}\centering
	\caption{\label{tab:counts} Breakdown of F160W morphologies}
	\begin{tabular}{lccr}
		\firsthline
		& \textbf{Point} & \textbf{No Point} & \textbf{Total} \\
		& \textbf{Source} & \textbf{Source} & \\
		\hline
		\textbf{Non-detections} & - & 60+3 & 60+3\footnote{One of these three sources is likely an extragalactic source and two are of uncertain nature \citep{furlan_herschel_2016}.} \\
		\textbf{Point Source\footnote{Sources without associated nebulosity}} & 93+7 & - & 93+7\footnote{Six of these seven sources are likely extragalactic sources and one is of uncertain nature.}\\
		\textbf{Irregular} &  39 & 12 & 51 \\
		\textbf{Unipolar} & 36 & 23 & 59 \\
		\textbf{Bipolar} & 16 & 15 & 31 \\
		\hline
		\textbf{Total} & 184+7 & 110+3  & 294+10\footnote{Includes seven likely extragalactic sources and three of uncertain nature.} \\
		\lasthline
	\end{tabular}
\end{table}

Approximately half of our sample, 163 objects, have no resolvable nebulosity in these observations.\footnote{This includes objects identified as of uncertain nature or potential extragalactic contaminants by \citet{furlan_herschel_2016}.} One hundred of these are detected as one or more isolated point sources; these have been analyzed to determine the  companion  fractions throughout the Orion Molecular Clouds \citep{kounkel_hst_2016}. We refer to these as point sources without associated nebulosity.  In these cases, any nebulosity around the source appears to be part of an extended nebula that is illuminated by other stars in the region or is very faint and tenuous and does not delineate a clear structure around the point source. As we will discuss in Sec.~4.4, the scattered light from cavities and envelopes illuminated by these sources are likely too faint to detect against the PSF of the point source.   The remainder of the sources are non-detections.

Emission along jets, most likely dominated by the [FeII] line at \SI{1.66}{\micro\meter}, is observed in thirteen protostars, with three additional tentative detections. These are the bipolar protostars HOPS 133, 150, 186 and 216; the unipolar sources HOPS 29, (shown in \autoref{fig:morph}), HOPS 164 and 310; the irregular protostars HOPS 98, 188, 234 and 386; the point source 279 and the protostar HOPS 152, which although not detected directly at \SI{1.60}{\micro\meter}, is situated at a location that is an apparent source of jet emission. Tentative detections of jets are found toward the point source protostars HOPS 3, 344 and 345.

\begin{figure}
	\includegraphics[width = 0.45 \textwidth]{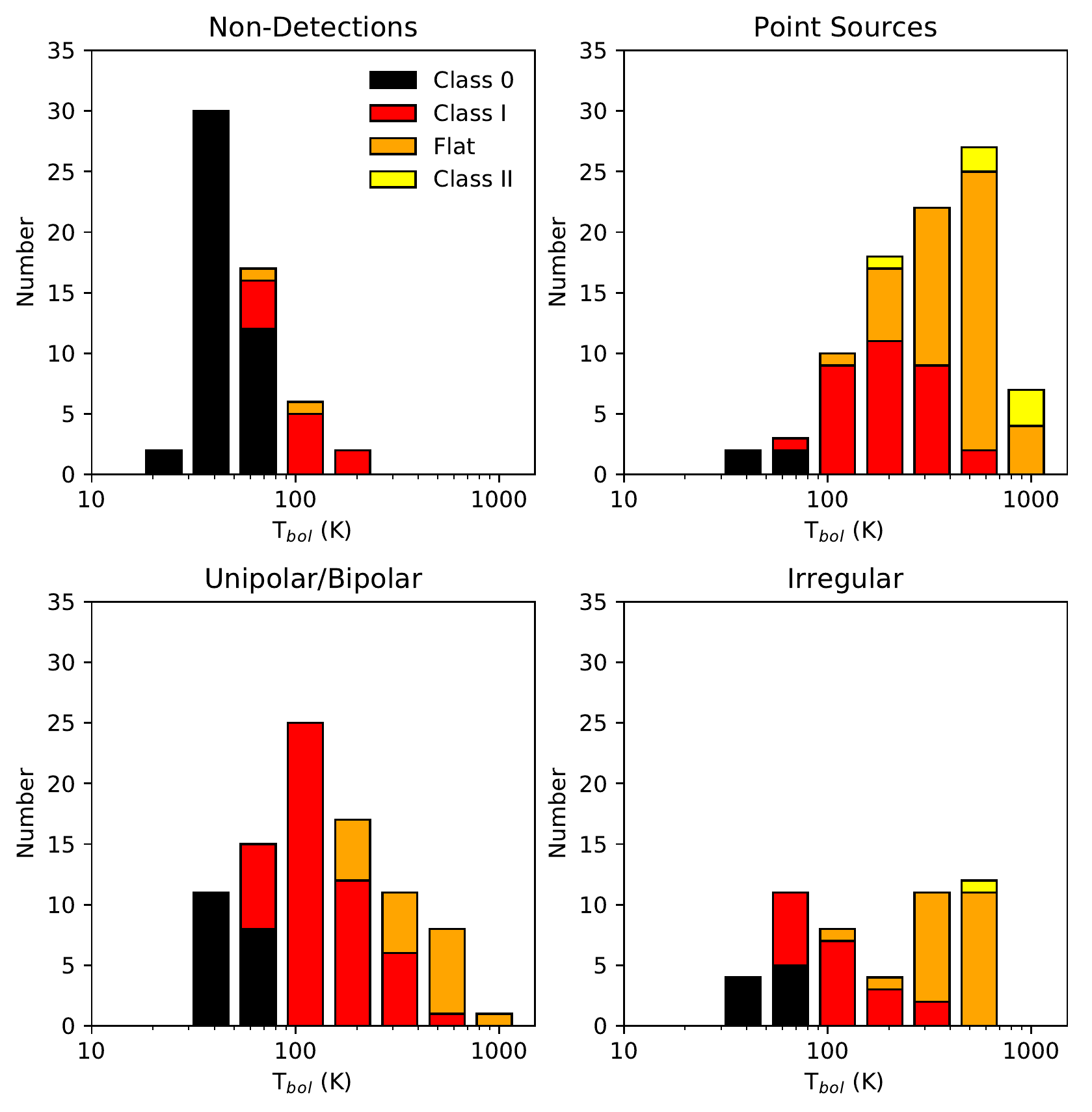}
	\caption{\label{fig:tbol-histograms}The histograms of bolometric temperatures of our sample for the different morphological classifications. The colors give the classification according to the criteria from \citet{furlan_herschel_2016}.}
\end{figure}

In \autoref{fig:tbol-histograms}, we plot the number of protostars vs bolometric temperature for four morphological groups: non-detections, point sources, unipolar or bipolar sources, and protostars with irregular morphologies. The bolometric temperature is a measure of the evolutionary stage of the protostar, although it also has some dependence on inclination \citep{ladd_c18o_1998,furlan_herschel_2016}. We also include the standard evolutionary classes, as determined with the criteria in \citet{furlan_herschel_2016}. These figures demonstrate the strong dependence of detectability and morphology in the near-IR with on the class of a protostar. The least evolved protostars (Class 0) are predominantly not detected due to the greater optical depths in their envelopes. In comparison, the most evolved sources (i.e., flat-spectrum protostars and Class II pre-main sequence stars) are dominated by unresolved point sources due to the low density of dust (and therefore low scattering probability) in their sparsely filled or non-existent envelopes.\footnote{\citet{furlan_herschel_2016} show that flat-spectrum protostars are a combination of protostars with higher density envelopes seen at low inclinations and protostars with lower envelope densities seen at any inclination. The first possibility is less common since it requires a limited range of inclinations.}
Protostars with unipolar and bipolar cavities show a broad range of $T_\text{bol}$, but peak in the  Class I phase ($T_\text{bol} \sim \SI{100}{\kelvin}$) and contain a significant fraction of Class 0 objects. Finally, the irregular protostars consist largely of Class I and flat-spectrum sources.

\begin{figure}\centering
	\includegraphics[width = 0.45 \textwidth]{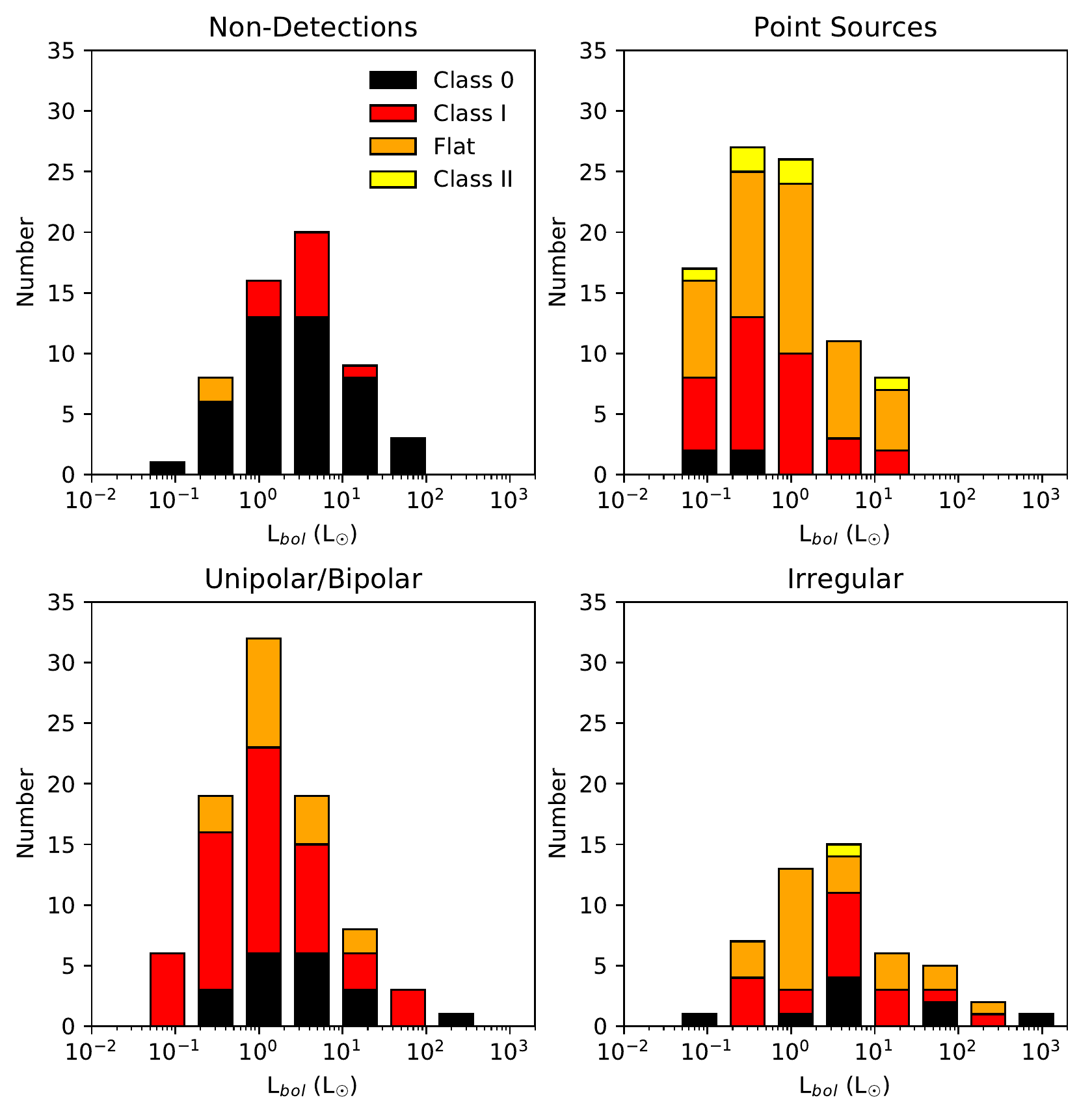}
	\caption{\label{fig:lbol-histograms}The histograms of bolometric luminosities of our sample for the different morphological classifications. The color scheme is identical to \autoref{fig:tbol-histograms}.}
\end{figure}

We show the distribution of bolometric luminosities for each morphological class in \autoref{fig:lbol-histograms}. The luminosity distributions for the three non-irregular classes display a shift in median luminosity, with point sources, unipolar/bipolar protostars and non-detections having median $L_\text{bol}$ of \SIlist{0.5;1.4;2,7}{\solarluminosity}, \added{respectively}. This change is small compared to the full range of bolometric luminosities probed, from 0.05 to \SI{480}{\solarluminosity}\replaced{, and is likely a byproduct of the difference luminosities in Class 0, I, and flat-spectrum objects}{. It is likely due to a decline in the luminosity with increasing age, as found by \citet{fischer_herschel_2017}}.

\subsection{Direct Measurements of Cavity Sizes}\label{sec:cavity-measure}
\begin{figure}\centering
	\includegraphics[angle=0,scale=1.1]{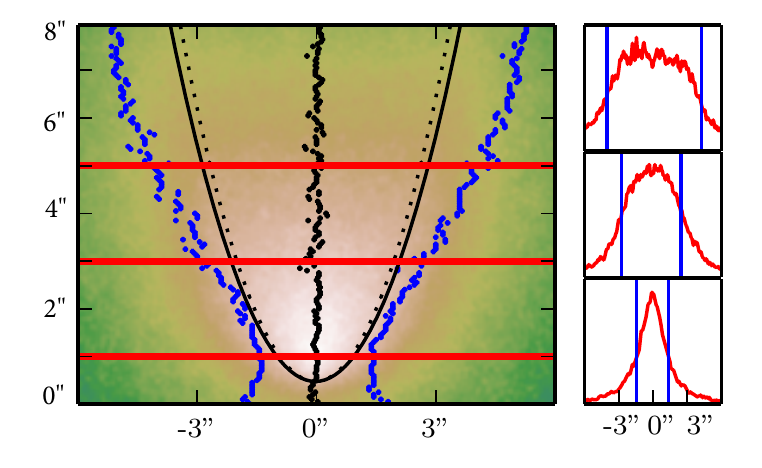}
	\caption{\label{fig:model-edge}An example of the edge detection technique applied to find the left and right edges (in blue), as well as the midpoint (in black scatter points) for a model with an inclination of \ang{60} and cavity \replaced{opening}{half-opening} angle of \ang{15}. Also shown is the analytical shape of the cavity wall with the solid black line, and the cavity wall as observed for an edge-on inclination in the \replaced{dashed}{dotted} black line. At the location of the three red lines, the three plots on the right show an intensity cut along with the location of the detected edges in blue.}
\end{figure}

For protostars with unipolar or bipolar morphologies, we fit a power-law to the shape of the cavities to estimate the amount of the envelope which was cleared by the outflows. This analysis relies on a custom edge detection routine developed to locate the outer contours of the cavities in the images. The methodology is illustrated in \autoref{fig:model-edge}. It is similar to the Sobel filter described in \citet{danielsson_generalized_1990}, constrained to the dimension perpendicular to the axis of the cavity. The image is first rotated such that the cavity is aligned with the positive $y$ axis in an $x-y$ Cartesian plane; this defines our adopted axis for the cavity.  
In three bipolar cases, those of HOPS 136, 280 and 333, we were able to measure the shape of both cavities.  For each image, a 1D Gaussian smoothing kernel was chosen by eye to account for noise and applied to every slice of constant $y$. The width of the smoothing kernel is between 2 and 4 pixels, approximately \SI{0.3}{\arcsecond}. 

We then calculate the second order finite difference along the slice (i.e.~parallel to the $x$-axis) using the equation

\begin{equation}
\frac{d^2 I}{dx^2} = I(x+2) - 2 I(x+1) + I(x),
\label{eqn:2nd}
\end{equation}

\noindent
\added{where $I(x)$ is the F160W intensity at pixel $x$. The second order finite difference, as an approximation to the second derivative, is zero at the inflection points of the slice.} \replaced{The second order discrete difference of this slice is zero at inflection points of intensity along this slice; we take this as the definition of the}{The inflection points allow us to define an} ``edge'' of the cavity. The width of the smoothing kernel is increased to obtain a consistent edge, as a small smoothing kernel can produce a discontinuous edge. Inflection points are inspected to ensure that only those tracing the cavity (as opposed to structure within or outside the outflow cavity) are retained. \added{We use this definition of an edge as it is bounded by the peak of the intensity and the background.} More sophisticated techniques  \citep[e.g.][]{canny_computational_1986} have a limited advantage due to the presence of unrelated structures in the line of sight that cannot be treated as random noise.

To determine the half-width of the cavity, $x$, at a given position along the cavity axis, $y$, we measure the full width of the cavity between the two walls and then divide by two.  Thus, the central axis of the outflow is defined by a curve tracing the midpoint of the two walls. Note that the $y$ position is the distance along a {\it straight} line that starts at the base of cavity and extends along the adopted cavity axis (\Autoref{fig:model-edge}).

In order to relate the detected edge to the physical cavity in the envelope, we ran the edge detection routine on our model grid. We compared the edges measured for the models as a function of the observed inclination to the shape of the projected cavity wall for the same model as observed from an \textit{edge-on inclination}; this allows us to correct for the effect of inclination of the shape of the outflow. The location of the projected wall is given by the analytic equation $y = A|x|^n$ (\autoref{fig:cavity-schematic})\added{, where $A$ and $n$ are determined by the parameters of our model that are described in more detail below}. \explain{Add paragraph break}

For most models, the detected edges of the cavities differ systematically from those of the projected wall (\autoref{fig:model-edge}); this is due to the combined effects of inclination, the penetration of the light from the central protostars past the cavity wall into the envelope and systematic biases of the edge detection routine. The inclination alone will broaden the cavity by \SIrange{7}{25}{\percent} for a \SIrange{40}{60}{\degree} inclination assuming a parabolic cavity. \explain{Add paragraph break}

\autoref{fig:model-edge} shows our edge fitting routine applied to a model image of a protostar with an inclination of \ang{60} and a cavity \replaced{opening}{half-opening} angle of \ang{15}. The black solid line indicate the projected cavity wall of this model, as observed from this inclination. The black dashed line indicates where the cavity wall would be for the same analytical shape, but observed at an edge-on inclination --- almost negligible even for a \ang{60} inclination. The detected edges (in blue) are characteristically wider than the known cavity wall.

To quantify the difference between the observed and actual edge, we determined the ratio of the distance to the ``detected'' edge in the model to that of the known, projected distance to the wall. For a given model, this ratio was found to be approximately constant as a function of the distance along the outflow axis. 
Thus, a single ratio can describe the difference between the observed and actual outflow cavity for a given source. Using the grid of models described in \autoref{sec:modeling}, the ratio was measured as a function of the cavity \replaced{opening}{half-opening} angle and inclination.\explain{Add paragraph break}

\begin{figure}
	\includegraphics[angle=0,scale=0.6]{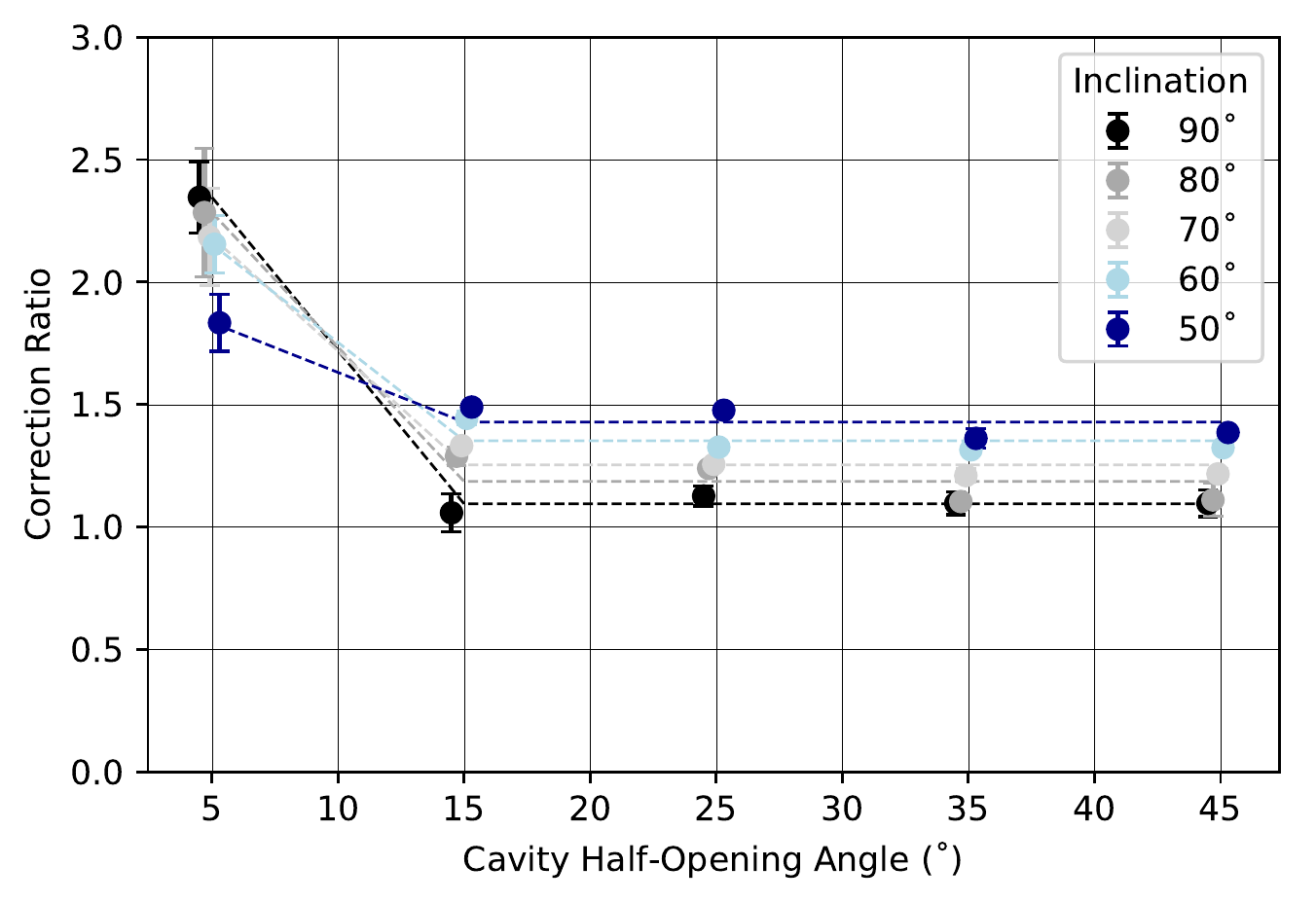}
	\caption{\label{fig:ratios-dists}Ratio of the half-width of the detected edges to the distance expected for the cavity wall of the model at an edge-on inclination. The inclination $i$ of the models used is given in the insert. The error bars give the standard deviation among models of differing envelope densities and disk properties. Note that the corrections are largely constant with cavity angle, except at the small openings.}
\end{figure}

At lower inclinations, the line of sight toward the central protostar is more likely to be directly into the cavity or to pass through less envelope material, thus significantly increasing the probability to observe the protostar as a point source (see \autoref{sec:completeness}).
In these cases, the cavity walls are difficult to detect against the PSF. Additionally, because there exist more possible lines of sight toward edge-on or near edge-on orientations than pole-on or near pole-on orientations, the probability that a protostar will be observed at a given inclination decreases as inclination decreases, assuming that the cavity may face any direction randomly.
For these reasons, we averaged the ratios determined from all models for each cavity size, considering only inclination angles from \ang{90} to \ang{50}.
The ratios are displayed in \autoref{fig:ratios-dists}, which shows that they are predominantly constant as a function of \replaced{opening}{half-opening} angle except at the smallest opening angles and that they have a weak dependence on inclination. 
The standard deviation over all parameters aside from cavity size and inclination are shown as the error bars in this figure.

\begin{figure}
	\includegraphics[angle=0,scale=0.6]{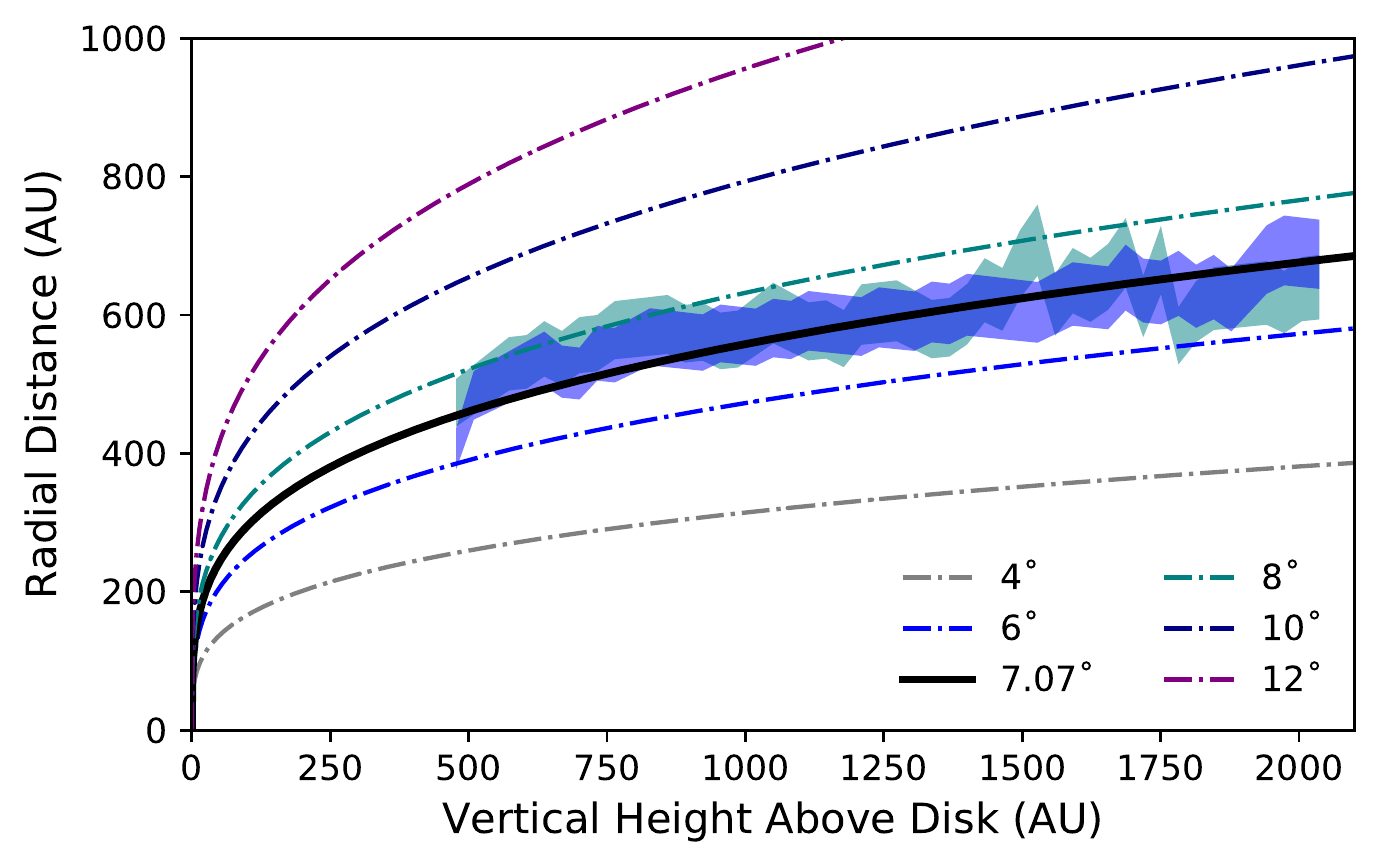}
	\caption{\label{fig:hops136-dist}The detected edges of the northern cavity of HOPS 136 and several power-law curves corresponding to a range of opening angles. Both the northwestern and northeastern cavity edges (in blue and teal respectively) were measured in this protostar. The detected half-widths of the cavity have been corrected by the model derived ratios shown in \autoref{fig:ratios-dists}.  The filled regions display the uncertainty in the location of the cavity edges due to this correction. The half-widths of the northeastern and northwestern edges were folded together and fit with the power-law curve of exponent $n = 3.61$ half-opening angle $\theta = 7.07$ shown in black. Five power-law curves of exponent $n = 3.61$ and various opening angles are shown in comparison.}
\end{figure}

\begin{figure}\centering
	\includegraphics[angle=0,scale=1.1]{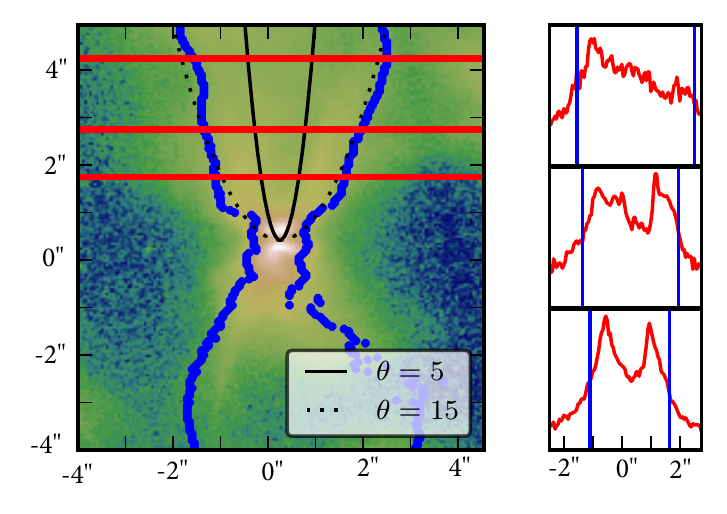}
	\caption{\label{fig:obs-edge}
	~An example of the edge detection technique applied to find the left and right edges (in blue), for the bipolar source HOPS 136. For comparison, two parabolic cavities, with half-opening angles of \ang{5} and \ang{15}, are shown in solid and dotted black lines respectively. At the location of the three red lines, the three plots on the right show an intensity cut along with the location of the detected edges in blue.}
\end{figure}

These ratios shown in \autoref{fig:ratios-dists} were applied to the measured half-width of the cavities
from the \textit{HST} images. Generally, we initially divided the half-width by $1.3$, which is the approximate average ratio for  \ang{50}-\ang{80} inclination cavities with a half-opening angle greater than \ang{15}. For cavities of these sizes that were also bipolar and thus presumably near \ang{90} in inclination, we restricted our initial ratios to $1.1$. For cavities with narrower opening angles and unipolar and bipolar morphologies, we chose initial ratios of $2.0$ and $2.3$ respectively.  We then iterated when necessary, modifying the correction ratio until the combination of the ratio and recovered half-opening angle was consistent with combinations observed in our models, shown in \autoref{fig:ratios-dists}. 

We fit the cavity width as a function of the distance along the outflow axis to the function
\begin{equation}\label{eqn:cavity}
	y = A |x-x_0|^n + y_0.
\end{equation}
to both the model grid discussed in \autoref{sec:modeling} and observed images (\autoref{fig:hops136-dist}). Here, $(x_0, y_0)$ identify the location of the protostar, and are fixed to the center of our model images. 
In the observed data, the parameters $(x_0, y_0)$ are manually centered on the central protostar when apparent from a point source or an area of maximum flux along the profile of the cavity or were placed along the disk absorption lane in the case of some bipolar sources.
The midpoints of the cavity, as shown in  \Autoref{fig:obs-edge}, were used to fit a center line which in turn allowed us to perform a final small-angle rotation correction. Our two detected edges were then considered for fitting in three ways: both the left and right edges were independently fit with a power-law profile, and, after folding over the now vertical center line, both detected edges were simultaneously fitted. This allowed us to counter minor asymmetries in detected cavity edges as well as outlying points biasing our fitting regime on a single edge. The recorded parameters were those given by the single-edge fit with an exponent $n$  greater than 1, or in cases where both edges met this criteria, the parameter from the folded fit was recorded.
The exponent $n$, which is referred to as the cavity exponent, gives a measure of the collimation of the outflow cavity and may be indicative of the physical mechanism behind the outflow creation. For example, \citet{shu_star_1991} show how a shell of molecular gas composed of the outflow and swept up material has a shape dependent on the angular distribution of the outflow. 

By allowing $n$ to be an \replaced{unfixed}{unconstrained free parameter} in our fitting (with the caveat that it be greater than 1), we allow for conical cavities ($n = 1$) as well as parabolic cavities ($n = 2$). The amplitude $A$ parameterizes the size of the cavity. For the model used by the \texttt{HO-CHUNK} code, this relates the radius of the envelope ($R_\text{max}$) and the cavity \replaced{opening}{half-opening} angle $\theta$ by:
\begin{equation}\label{eqn:Avalue}
	A = R_\text{max}^{1-n} \cot^n\theta.
\end{equation}

\noindent
The value of $\theta$ is only dependent on $A$ for conical cavities ($n = 1$), but for other values of $n$, the value of $\theta$ depends on our choice of $R_\text{max}$, which we set to \SI{8000}{\astronomicalunit}.
Error analyses for functions of the fitted values are discussed in \autoref{sec:errors}.

For three protostars with bipolar morphologies, we were able to measure parameters for both cavities. In all other bipolar cases, the cavity appearing brighter was fitted. From our monochromatic model grid, we can see that inclination is responsible for variations in brightness between the two cavities. We expect the closer cavity to have a stronger signal due to a smaller extinction along the line of sight, although inhomogeneous envelopes may also be responsible for differences in cavity brightness.

An example of our fitting technique applied to the protostar HOPS 136 can be seen in \autoref{fig:obs-edge}. \cite{fischer_hops_2014} determined that this protostar is a late stage object with a \SI{10(2)}{\degree} half-opening angle for a $R_\text{max}=\SI{10000}{\astronomicalunit}$ envelope. The detected edges of its northern cavity are compared with power-law curves as given by \autoref{eqn:cavity} in \autoref{fig:hops136-dist}, revealing the northern cavity of this protostar is best fit by a \ang{8.7} half-opening angle,\footnote{For the bipolar protostar HOPS 136, measurement of both the northern and southern cavity edges was possible. In \autoref{tab:bigtable}, the average of both sets of parameters are reported.} in close agreement with \citet{fischer_hops_2014}.

For thirty of the ninety protostars in our sample with unipolar or bipolar morphologies, we use this technique for measuring the cavity shape, and tabulate the values of $n$ in \autoref{tab:bigtable} along with the half-opening angle. 
The median uncertainty for $n$, as obtained from the least squares fitting of \autoref{eqn:cavity}, is $\delta n \sim 0.14$. 
We find from \autoref{eqn:theta_error} that uncertainties in half-opening angle measurements are on average $\delta \theta < \ang{0.3}$. We also include in \autoref{tab:bigtable} the volume fraction of the envelope cleared by the cavity as in \autoref{sec:modeling}. We calculate this assuming a spherical envelope and a cavity volume given by the profile in \autoref{eqn:cavity}. For simplicity, we calculate the uncertainty in this measurement for the case of a conical cavity of the same half-opening angle. 
Uncertainties in this metric are $\sim \SI{5}{\percent}$ of the measured volume fraction (\autoref{sec:errors}). We note that these uncertainties describe the accuracy of our fits after carefully determining the axis of the outflow cavity and selecting regions for fitting over which our edge detection was well behaved and not confused by errant nebulosity, background stars, or the PSF of the central protostars. These determinations and selections may introduce systematic uncertainties that are not accounted for in the formal uncertainties.

The remaining sixty unipolar and bipolar protostars are excluded due to the inability to accurately and fully trace the cavity with the \textit{HST} images. For several sources we see a  morphology indicative of an edge-on or nearly edge-on disk but do not see evidence for a cavity (e.g. HOPS 65 and HOPS 200); these may be pre-main sequence stars with disks. Other protostars have cavities that are too faint to reliably trace (e.g.~HOPS~220 and HOPS~235), show only one edge of a cavity wall -  due to either a non-uniform extinction or an irregularly shaped envelope (e.g.~HOPS 18 and HOPS 310), or are coincident with nebulosity - making it impossible to disentangle the cavity from larger scale structures (e.g. HOPS 387 and HOPS 384). Finally, some cavities exhibit morphologies inconsistent with a power-law cavity (e.g.~HOPS 8 and HOPS 232). In general, the factors that prevented automated fitting appeared incidental and not obviously correlated with apparent cavity size. Future efforts will focus on expanding the range of brightness levels and morphologies analyzed as well as understanding the nature of cavities with only one apparent wall.

\begin{figure}[t]\centering
\includegraphics[angle=0,scale=0.6]{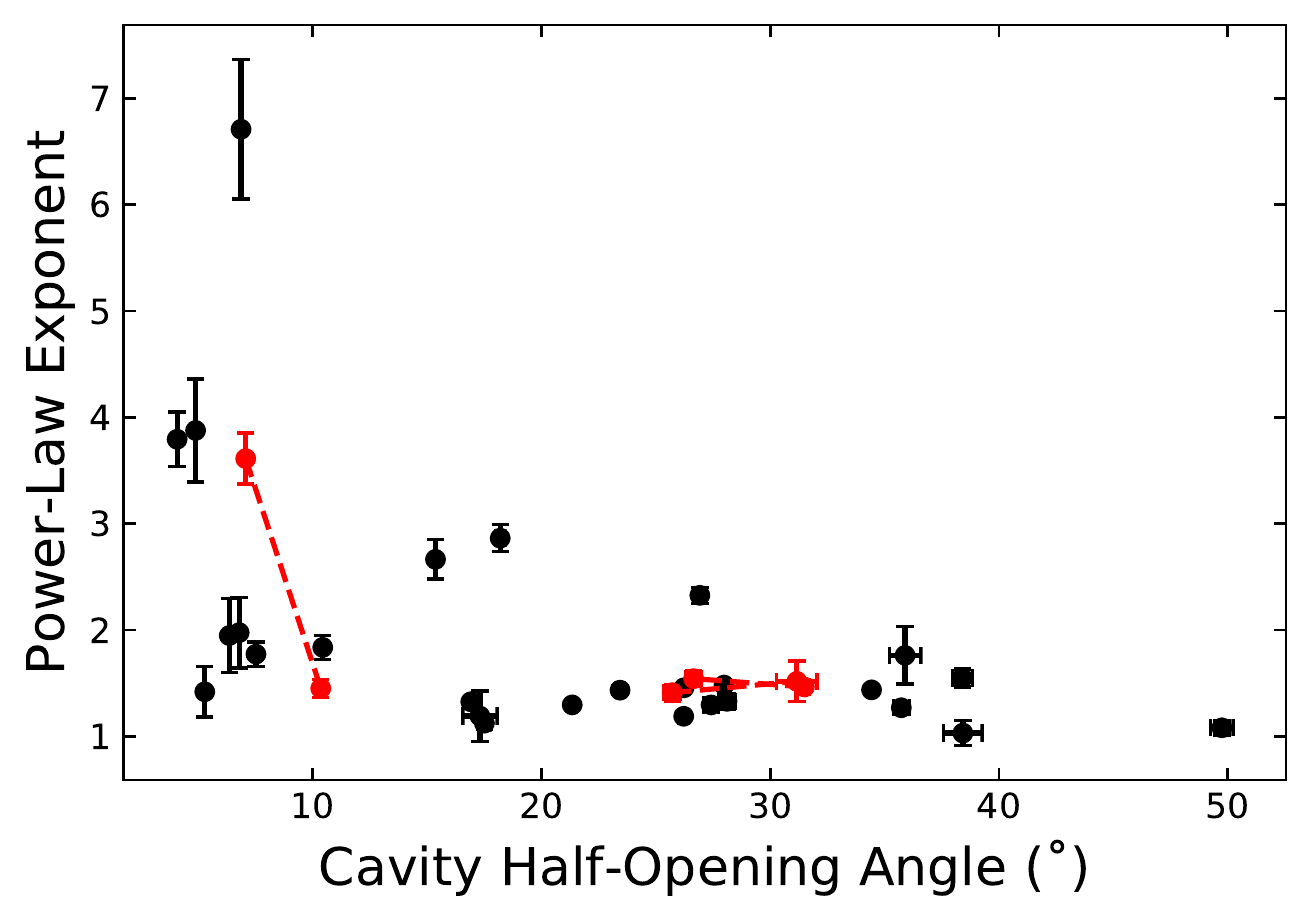}
	\caption{\label{fig:exponent-vs-angle}The exponents ($n$) and cavity half-opening angles  ($\theta$) from the fits to the detected cavity wall edges of 30 measured protostars. The three bipolar sources where both cavities were able to be fitted are connected with red lines. Typical uncertainties are  $\delta \theta \sim\ang{0.3}$ and $\delta n \sim 0.14$. (See \autoref{sec:errors}.)}
\end{figure}

\begin{figure}[t]\centering
	\includegraphics[angle=0,scale=.6]{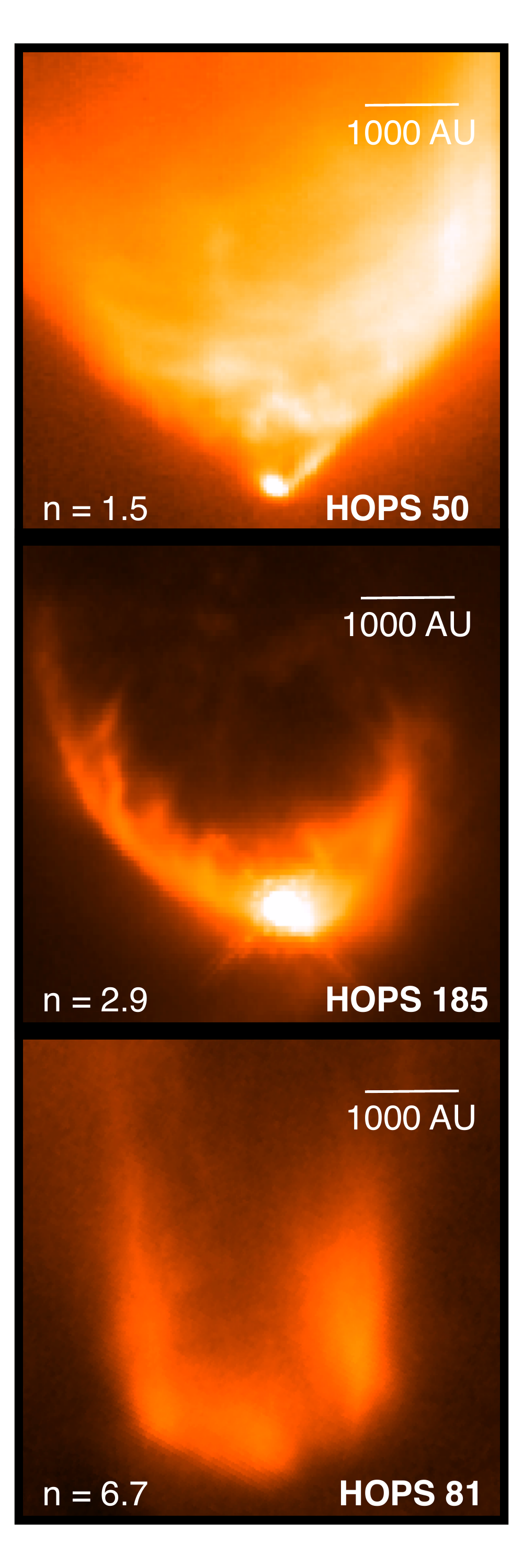}
	\caption{\label{fig:exponent-examples}Examples of protostars with cavities 
	\replaced{growing with a $>2$ exponent (HOPS 66), a roughly quadratic cavity (HOPS 185), and a streamline cavity (HOPS 93)}{with an assortment of cavity power-law exponents: HOPS 50 ($n = 1.5$), HOPS 185 ($n=2.9$) and HOPS 81 ($n = 6.7$)}.}
\end{figure}

\autoref{fig:exponent-vs-angle} shows the range of fitted exponents $n$ and cavity half-opening angles measured in this work. We find the mean and median of the cavity exponents are $1.9$ and $1.5$ respectively, indicating that parabolic cavities are a reasonable model assumption. Cavity exponents vary significantly between the protostellar cavities; we show examples of this variation in \autoref{fig:exponent-examples}. Finally, we show our distribution of cavity half-opening angles and volume fractions in \autoref{fig:cavity-histograms}. \explain{Move paragraph to end of subsection}

\subsection{Cavity Sizes vs. SED Derived Properties}
\label{sec:cavity-sed}

The SEDs of the protostars provide information on both their evolutionary phase as well as their total luminosity \citep[e.g.][]{whitney_two-dimensional_2003}. Correlations between the SED derived properties of protostars with the cavity sizes provide a means to probe the evolution of cavities as well as, potentially, their dependence on the final mass of the protostar \citep{fischer_herschel_2017}.  \autoref{fig:bigvsplot} shows two ways of parameterizing the cavity size, half-opening angle and volume fraction cleared, against an assortment of evolutionary indicators derived from the \SIrange{1.6}{870}{\micro\meter} SED \citep{furlan_herschel_2016}. We calculate the half-opening angles with \autoref{eqn:Avalue} using the values of $A$ and $n$ 
and an envelope radius of \SI{8000}{\astronomicalunit}. 

\begin{figure*}\centering
	\includegraphics[angle=0,scale=.85]{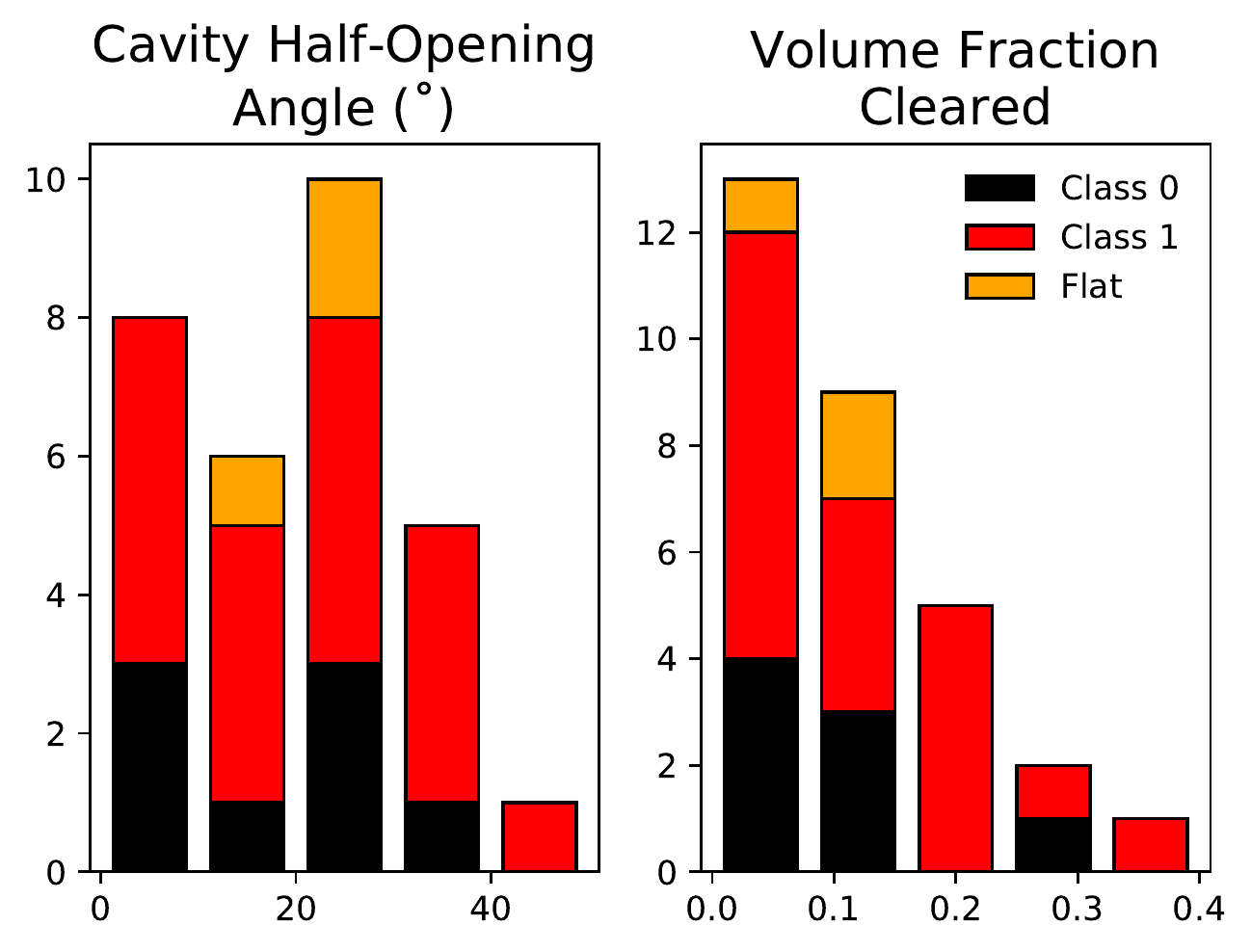}
	\caption{\label{fig:cavity-histograms}Histograms of cavity opening angles and volume fraction cleared for our sample of protostars with detected cavities in the \textit{HST} images. Color scheme is identical to \autoref{fig:tbol-histograms}.}
\end{figure*}

\begin{figure*}\centering
	\includegraphics[angle=0,scale=0.55]{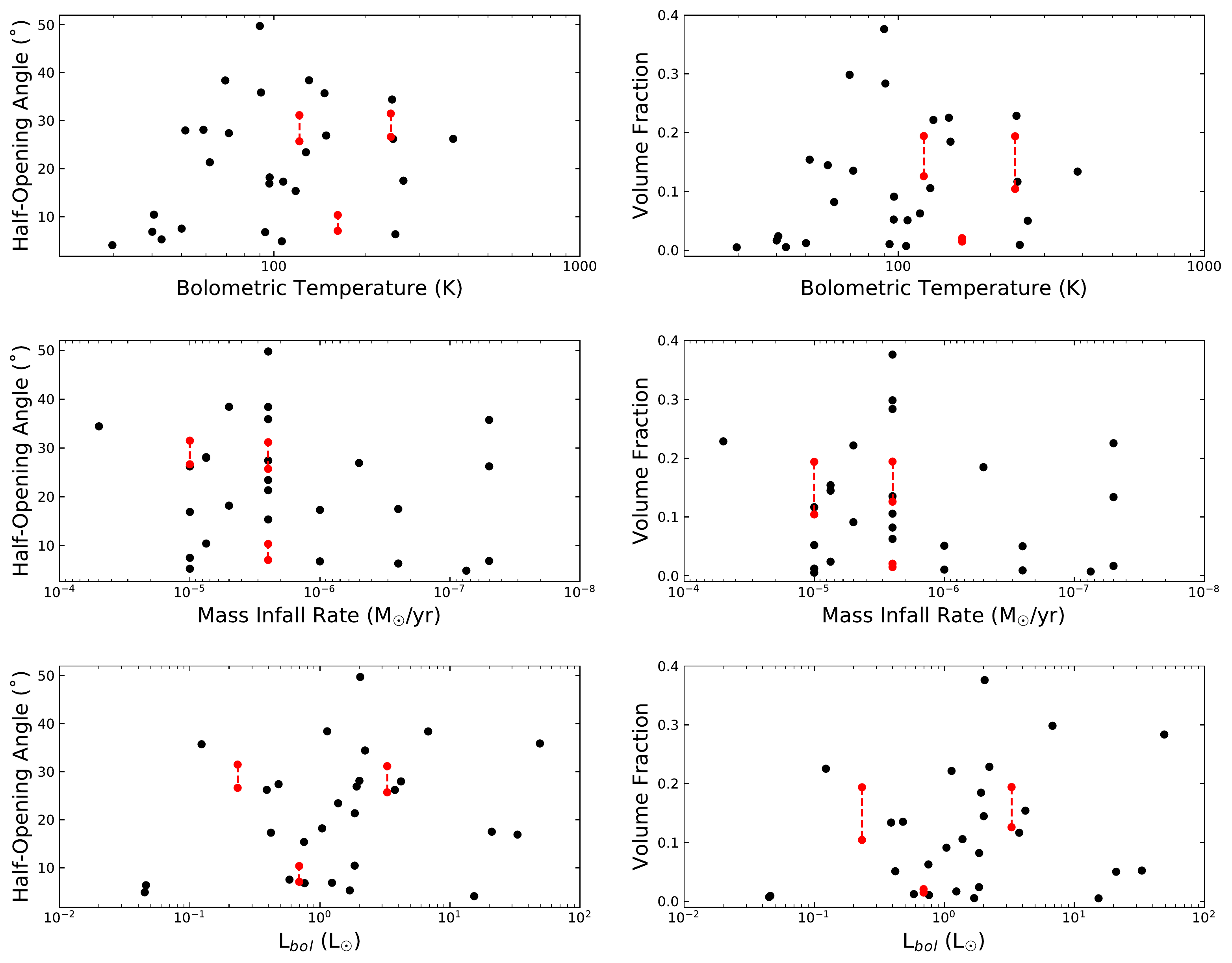}
	\caption{\label{fig:bigvsplot}
	Cavity size diagnostics, half-opening angle (\textbf{left}) and volume fraction cleared (\textbf{right}), against evolutionary indicators.
	The mass infall rates (from SED model fitting\added{, assuming a \SI{0.5}{\solarmass} stellar mass}) and bolometric temperatures and luminosities are found in \citet{furlan_herschel_2016}. Data from bipolar sources with both cavities fitted are connected with red lines.}
\end{figure*}

We quantify the degree of correlation by finding the Spearman Rank Correlation Coefficient $r$, a measure of the monotonic correlation between two variables in our thirty measured protostars. A correlation coefficient of 1 or -1 implies a strictly monotonic correlation. The Spearman coefficients and the two-sided $p$-value for a hypothesis test are given in \autoref{tab:spearmanr} for each of the diagnostic indicators and methods of parameterizing the cavity size. The hypothesis test uses a null hypothesis of no correlation; therefore, low $p$-values indicate evidence of a correlation and evolutionary trend. For the three bipolar sources with both cavities measured, the found parameters of the two cavities were averaged before computing Spearman Coefficients and $p$-values.

We do not find statistically significant  correlation between cavity size and $T_{bol}$ or mass infall rate, (which should be considered a proxy for envelope density, as discussed in \autoref{sec:modeling}).  
As shown in \autoref{fig:cavity-histograms}, the sample of protostars is dominated by Class I sources; at \SI{1.60}{\micro\meter}, many Class 0 protostars are not detected, while flat-spectrum sources are often point sources or have irregular nebulosity (see \autoref{fig:tbol-histograms}).
Hence, these results can be primarily interpreted as a lack of evidence for an evolution in cavity properties across the Class~I phase. The wide scatter in cavity sizes does not appear to be the result of evolution, but must depend on other environmental or intrinsic factors.

A higher correlation coefficient is found between cavity size and luminosity, with more luminous objects tending to have larger cavities; however, the p-values show that we cannot rule out the null hypothesis.

\begin{deluxetable}{lcccc}
	\tablecaption{\label{tab:spearmanr}Spearman Coefficients and $p$-values.}
	\tablehead{%
		\colhead{Evolutionary} & \multicolumn{2}{c}{vs Half-Opening Angle} & \multicolumn{2}{c}{vs Volume Fraction} \\
		\colhead{Diagnostic} & \colhead{$r$} & \colhead{$p$-value} & \colhead{$r$} & \colhead{$p$-value}
	}
	\startdata
	$T_\text{bol}$ & 0.24 & 0.20 & 0.23 & 0.21 \\
	$\dot{M}_\text{infall}$ & 0.16 & 0.41 & 0.12 & 0.52 \\
	$L_\text{bol}$ & 0.26 & 0.16 & 0.29 & 0.12 \\
	\enddata
\end{deluxetable}

\subsection{The Prevalance of Point Sources}\label{sec:completeness}

\begin{figure}
	\includegraphics[angle=0,scale=0.65]{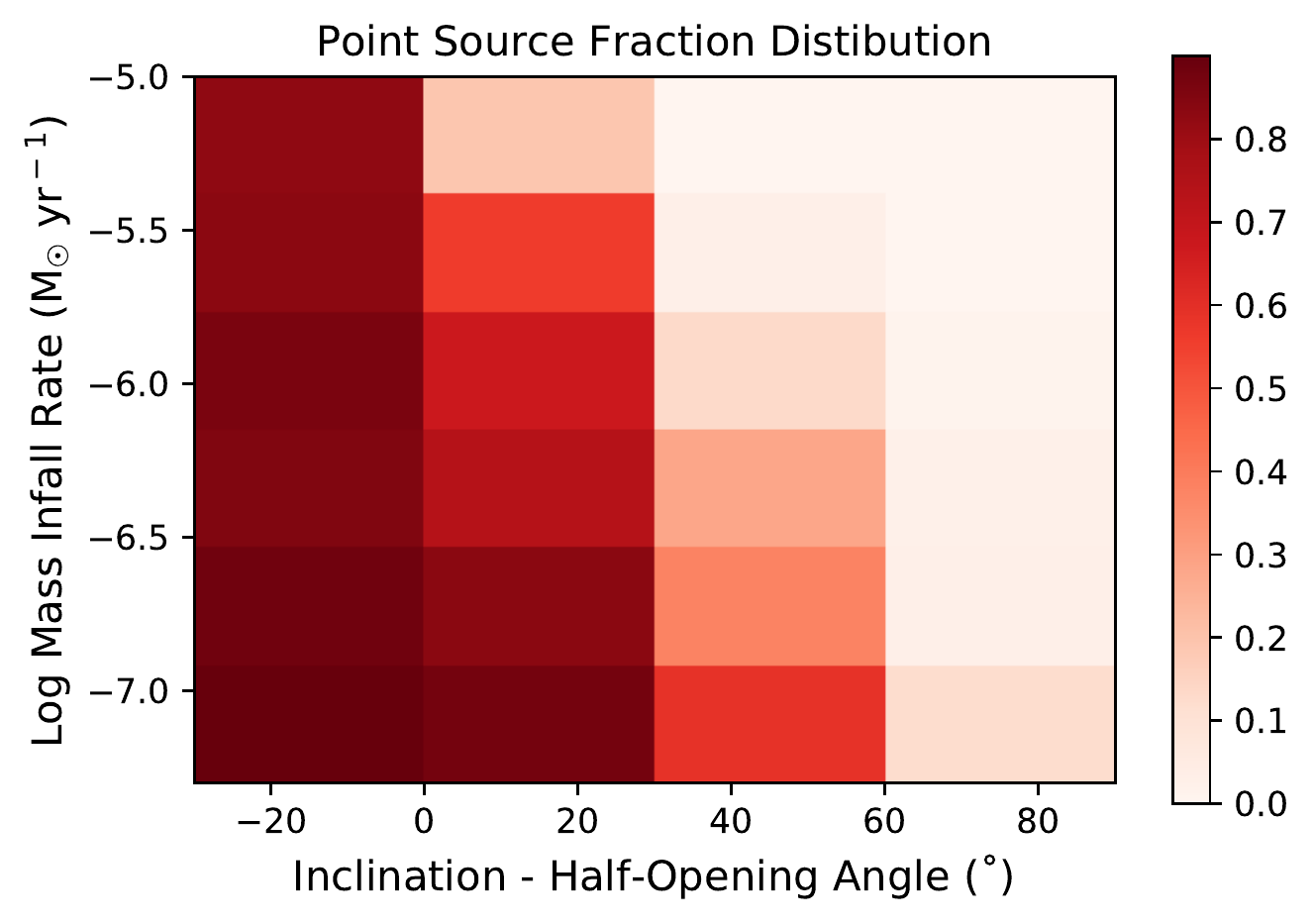}
	\caption{\label{fig:point_papameters}Fraction of the model protostars observed as point sources in our simulation as a function of parameter space. Darker colors indicate a higher fraction of point sources. In models where the inclination minus the half-opening angle is less than zero, the line of sight toward the central protostar is directly into the cavity, not passing through the infalling envelope. }  
\end{figure}

\begin{figure}
	\includegraphics[angle=0,scale=0.55]{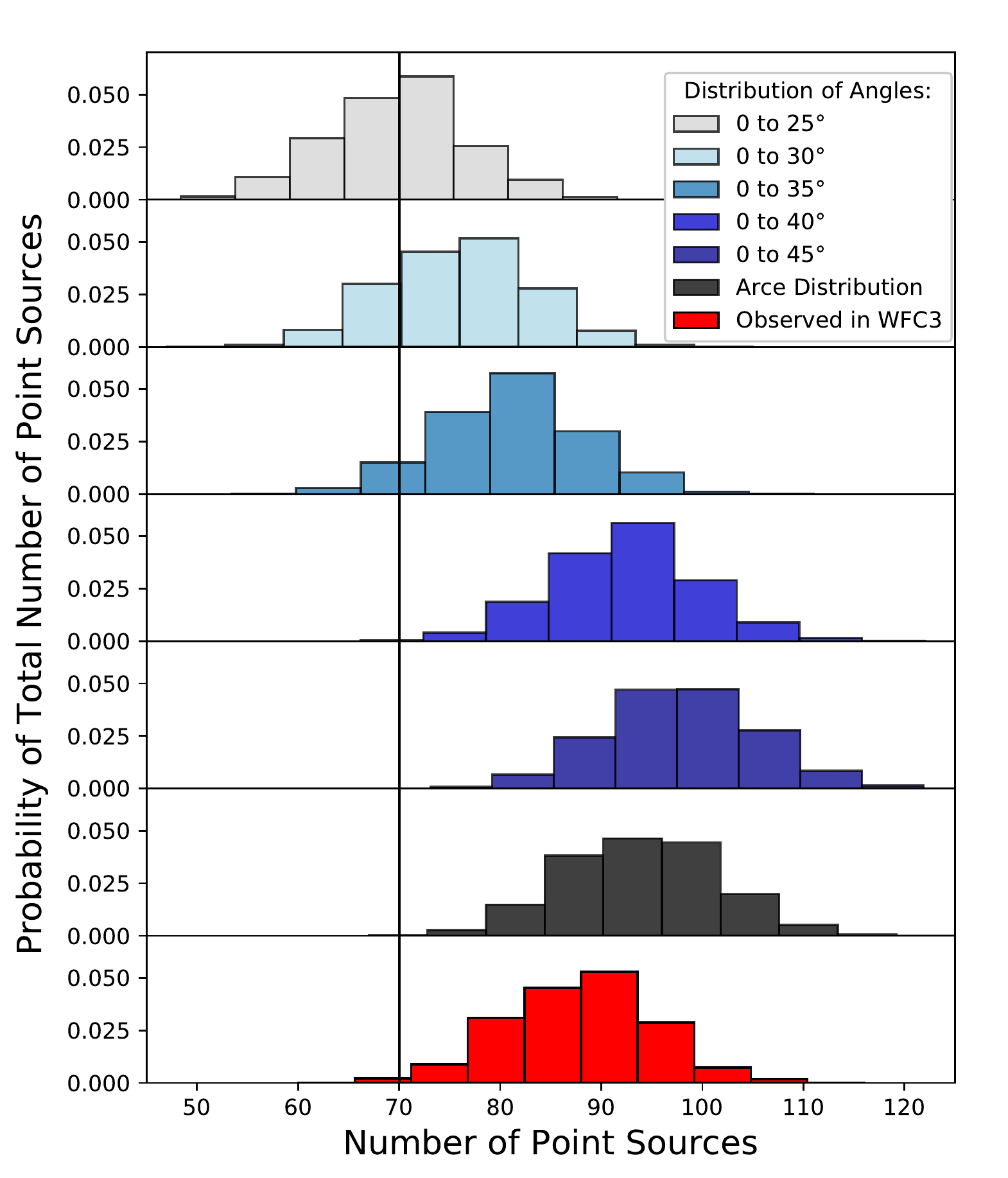}
	\caption{\label{fig:acre-mc}Histograms of the number of sources detected as point sources  when simulating observations of 230 protostars. The top five histograms examine five even distributions of opening angles over ranges indicated in the legend.  The "Arce" model, in grey,  assumes \citet{arce_evolution_2006}'s dependence between $T_\text{bol}$ and cavity \replaced{opening}{half-opening} angle to derive a distribution of opening angles using the bolometric temperature distribution of our sample. A distribution of opening angles drawn from our 30 measured protostars produces the histogram in red. The horizontal line marks the 70 point sources observed among the 230 protostar sub-sample of our WFC3 observations.}
\end{figure}

We detect cavities towards {90 (30\%)} of our sample, while {100 (33\%)} are observed as point sources without nebulosity. Since protostars are surrounded by envelopes that scatter light, the substantial number of point sources without detected scattered-light nebulosity is surprising. In this section, we examine why the point source morphology is common and test whether the  number of point sources implies an observational bias in our cavity size distribution.

Protostars may be observed as point sources without detected cavities in two primary cases. First, the central protostar is observed along a line of sight directly into the cavity.   In this case the brightness of the PSF from the central protostar, which will not be attenuated by the envelope, can be significantly stronger than scattered light from surrounding cavity walls, which may only contribute a diffuse scattering around the bright protostar (\autoref{fig:model-grid}).  Even if the line of sight grazes the cavity wall, the bright PSF can dominate over the nebulosity. Second,  a low density envelope leads to a more diffuse, lower surface brightness cavity wall and a brighter point source; once again, the cavity may not be visible against the PSF. In both of these cases, the extended nebulosity often found in the Orion clouds can also hide the scattered light from the cavities. 

The first case may lead to a bias against detecting large cavities. For envelopes with large cavities, the central protostar can be directly observed over a larger range of inclinations. Furthermore, since the walls of the cavity are further from the  protostar, they will have systematically lower densities than narrower cavities and they will intercept less flux from the central star; consequently, the  walls will be fainter and harder to detect for large cavities (\autoref{fig:model-grid}).

To determine the combinations of inclinations, cavity sizes and envelope densities that lead to the point source morphology, and to ascertain potential biases in our observed cavity size distribution, we use a Monte Carlo simulation that combines the model grid in \autoref{sec:modeling}, the envelope densities from the SED model fitting of \citet{furlan_herschel_2016} and several adopted cavity size distributions. The steps of the simulation are as follows. 

We first determined for each model in our grid whether a cavity would be detected by the WFC3 observations.
To determine  whether a cavity is detectable, two criteria were applied to each model. Non-detections of cavities were noted when no distinct edge that delineates a cavity is found in the image using the technique of \autoref{sec:cavity-measure} or when the signal in a cut taken across the cavity \SI{2000}{\astronomicalunit} from the central protostar has a peak value below the typical RMS of an image.  At  \SI{2000}{\astronomicalunit}, every protostar with a detected cavity shows nebulosity; if the signal from the nebulosity in the models is below the typical RMS values in the WFC3 images, then it is unlikely that the cavity would be detected.  The typical RMS was obtained from $\SI{30}{\arcsecond}\times\SI{30}{\arcsecond}$ off-source patches  chosen to avoid point sources or outflow cavities. These patches commonly included extended, diffuse nebulosity that is common in the Orion Molecular Clouds.  

We then performed a Monte Carlo simulation to predict the number of point sources without cavities we would detect for different assumed cavity half-opening angle distributions. We sampled the models drawing randomly for four parameters: infall rate (i.e. envelope density), inclination, inner F160W flux, and cavity half-opening angle. The distribution of infall rates for our sample of protostars (including those in the ``irregular'' category) were the best fit values in \citet{furlan_herschel_2016}. Inclinations were drawn assuming the outflow axes were randomly oriented. The maximum disk radius, the mass of the disk and the radial exponent in the disk density law were left as free parameters to be randomly drawn from those in \autoref{tab:fixedpars}. The brightness in the inner 0.2" region for each model was determined by scaling the image flux to correspond to \SI{1.60}{\micro\meter} magnitudes randomly drawn from the distribution of F160W magnitudes in the tabulation of  \citet{kounkel_hst_2016}. Finally, we sampled the cavity half-opening angles from several different distributions discussed below.

We plot the fraction of models resulting in point sources as a function of infall rate (i.e. envelope density), inclination and cavity half-opening angle in \autoref{fig:point_papameters}. Here we subtract the cavity half-opening angle of a source from the inclination to measure the angle of the line of sight with respect to that of the cavity. Where this value is below zero, the line of sight is directly into the cavity and does not pass through the envelope. We find in \autoref{fig:point_papameters} a strong preference for point source morphologies in models observed at such inclinations and opening angles.  When the inclination minus half-opening angle is positive and near zero, then the line of sight toward the central protostar  intersects the lower density, outer regions of the envelope.  In this case the incidence of  a point source morphology increases with decreasing infall rate.  Finally, if the infall rate is low, point source morphologies can be detected at every inclination and cavity half-opening angle combination, although the incidence increases at lower inclinations.  As expected, point source morphologies arise when either the protostar is observed through its outflow cavity or when the envelope is thin.  This is consistent with the point source morphology being dominated by protostars with flat-spectrum SEDs (\autoref{fig:tbol-histograms}); the flat SEDs are expected for protostars observed at low inclinations or with low envelope densities \citep{calvet1994,furlan_herschel_2016}.

Each iteration \footnote{For each distribution of opening angles, we performed 30 thousand iterations.} of the Monte-Carlo simulation returns the number of point sources without detected nebulosity.  We compare this to the number of point sources in our data. Before comparing, we removed from our sample those sources identified by their SEDs as possible extragalactic contaminants or of an uncertain nature and those without complete SEDs \citep{furlan_herschel_2016}, except for sources where \text{HST} imaging has revealed a unipolar or bipolar morphology, confirming their protostellar nature.  
Seventeen sources observed with WFC3 and classified as either non-detections or point sources are removed based on these criteria.
Finally, we choose only the point sources observed with WFC3, in order to account for differences in sensitivity. This reduces our sample down to 230 protostars, with 70 point sources.

In \autoref{fig:acre-mc}, we show normalized histograms of the number of point sources observed for various models of the cavity half-opening angle distributions. In red, we show the simulation results when the cavity sizes are randomly drawn from the values in \autoref{tab:bigtable}.
The observed number of point sources is marked with a vertical line. Realizations of 230 protostars with this simulation attain 70 point source detections or less of at  rate of {1.02\%}. We note that our exclusion criteria, described above, reject 11 objects with point source morphologies from our sample.
These objects could not be determined morphologically to be extragalactic contaminants, however, and were removed due to their SEDs. We note, however, that protostars may have extragalactic-like SEDs. HOPS 48, 67 and 301 were classified by \citet{furlan_herschel_2016} as extragalactic contaminants based on potential emission features in their Spitzer IRS spectra. In the case of these three sources, however, the features appear to originate in contamination from reflection nebulae or HII regions, and we observe cavities clearly associated with all three with \textit{HST} WFC3. Thus we consider our observed number of 70 point sources to be a lower limit. 

We also compare to fiducial models assuming a uniform distributions of cavities from 0 to 25, 30, 35, 40 and 45 degrees. The distributions extending beyond \ang{35} include enough large cavities to overpredict the number of point sources. These results indicate that our observations are not significantly biased against the detection of large cavity openings. 

Finally, we examined the consequences of outflow cavities that grow with time. We first adopt the relationship between cavity half-opening angle and $T_\text{bol}$ found by \citet{arce_evolution_2006}. We used this relationship and the observed distribution of bolometric temperatures of our protostars to derive the half-opening angle distribution we entitle ``Arce Model.'' We used a linear fit between the infall rate and $T_\text{bol}$ to pick a model in our grid on each iteration of the Monte Carlo. This model overpredicts the number of point sources, as it does not take into account the highly evolved protostars with low cavity half-opening angles found in our sample \citep[e.g.,][]{fischer_hops_2014}. 

In summary, we find that the histogram produced from the observed distribution overlaps with the observed number of sources, although with a 1\% probability of predicting the predicted number of protostars or less. If some of our excluded contaminant sources are in fact  protostellar in nature, the observed distribution may provide a better match.  Importantly, the result here is that our observed cavity angle distribution is largely consistent with our observed number of point sources. Uniform distributions of half-opening angles extending to \ang{45} overpredict the number of point sources, and we do not find evidence that our observations fail to detect larger cavities.  We also find that the uniform distributions with angles $< \ang{35}$ better reproduce the observed point sources  than our observed distribution.  This suggests that we may be missing small cavities that can be hard to detect due to higher extinction from their envelopes.

\section{Discussion: Consequences for Protostellar Evolution}\label{sec:discussion}

The  goal of this study is to assess the impact of  jets and winds on protostellar envelopes.  This is an essential step toward both understanding how feedback lowers the efficiency of star formation and determining the importance of feedback in halting mass infall and setting the final masses of protostars.  Feedback can lower efficiency and mass infall in three ways: by ejecting mass that would have otherwise been accreted, by clearing the envelope, and by entraining gas in the envelope and surrounding cloud into an outflow \citep[e.g.][]{watson_evolution_2016,zhang_alma_2016}.  This paper aims to quantify the role of cavity clearing.  

One way outflows may halt infall is by the progressive clearing of the envelope as the protostar evolves \citep{arce_alma_2013}. This may be driven, for example, by successive bursts of a wide angle wind \citep{zhang_entrainment_2019}.  The signature of this clearing would be a correlation between cavity size and the evolution of the protostellar SEDs. We find no significant trend between cavity half-opening angle with either $T_{bol}$ or model inferred $\dot M$, both indicators of envelope evolution (\autoref{sec:cavity-sed}). Instead, we find that there is a range of cavity half-opening angles extending from \SIrange{5}{50}{\degree} present across the observed range of $T_{bol}$ and the range of $\dot M$ values inferred from model fits \citep{furlan_herschel_2016}. This implies that the evolution from dense to thin envelopes is not driven by the progressive growth of the outflow cavities.

\begin{figure*}\centering
	\includegraphics[scale=.8]{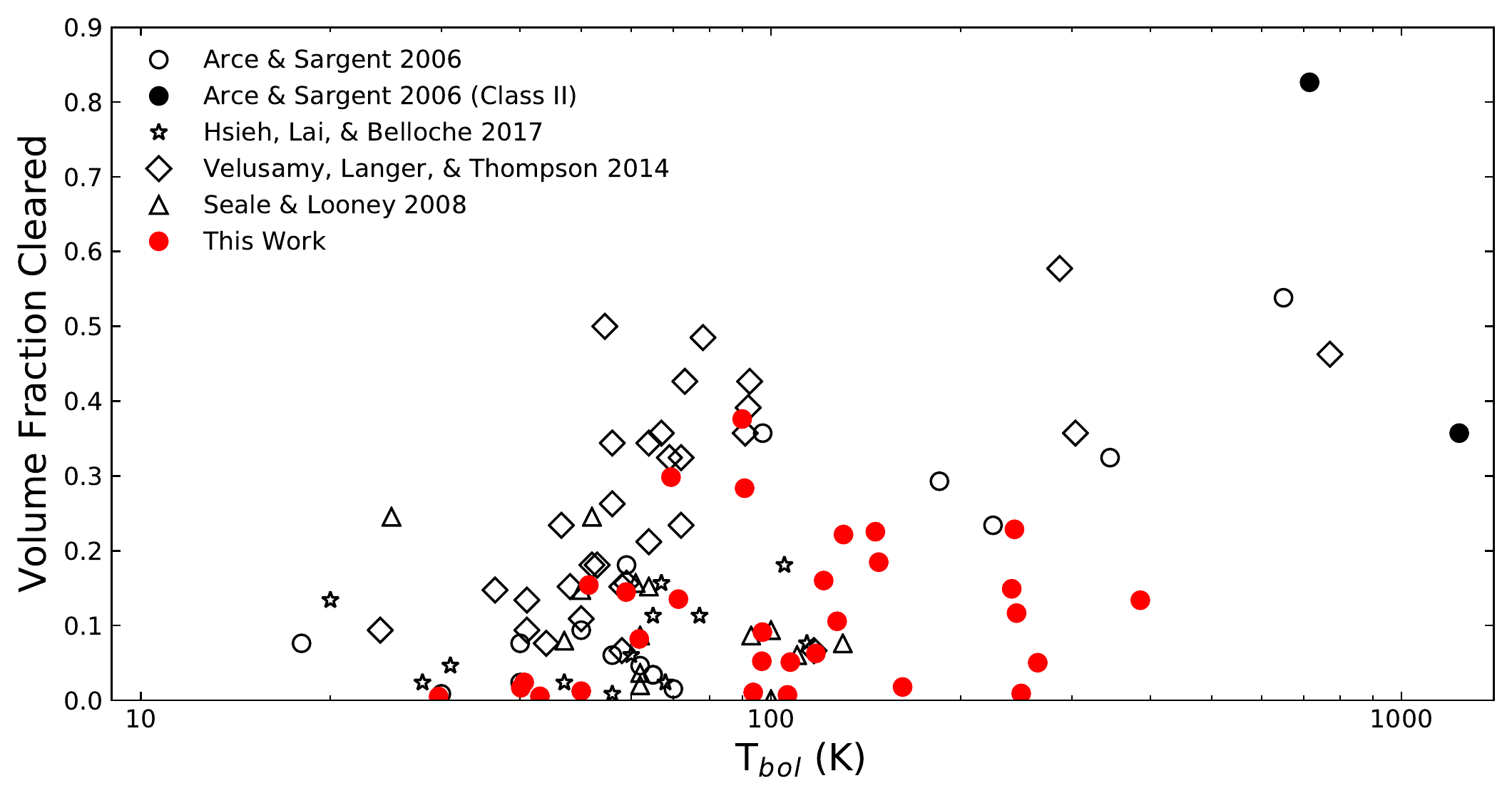}
	\caption{\label{fig:tbol-vs-frac-vs-acre}The fraction of the volume cleared by the outflow cavity, as described in \autoref{sec:modeling}. Our observations, using the procedure described in \autoref{sec:cavity-measure}, \replaced{is}{are} shown in filled red circles. Black circles are measurements of the outflow cavity angles found by \citet{arce_evolution_2006} and by the references therein. The two filled black circles indicate Class II sources in this sample. Black triangles are the measurements by \citet{seale_morphological_2008} using a different \replaced{edge detection algorithm}{technique} on \textit{Spitzer} IRAC images. Bolometric temperatures are from \citet{dunham_luminosities} or computed where possible with Spitzer photometry from Gutermuth et al. (in prep) and PACS photometry from Pokhrel et al. (in prep).
	Black diamonds are the opening angles measured by \citet{velusamy_hires_2014} using the HiRes deconvolution algorithm on IRAC images.\@ Finally, black stars are fits by \citet{hsieh_widening_2017} of WIRCam and IRAC images to synthetic images from a model grid generated by the same \citet{whitney_two-dimensional_2003} code.}
\end{figure*}

To extend this result, we compare our cavity sizes with volume fractions calculated from millimeter and lower resolution IR studies in  \autoref{fig:tbol-vs-frac-vs-acre}. We use the tabulated outflow cavity angles and assumed conical cavity shapes to calculate the volume fractions. Our scattered-light measurements extend these by providing a relatively large sample at a common distance observed with a uniform spatial resolution, which eliminates possible biases due to distance, and by detecting a significant number of protostars with relatively high $T_\text{bol}$ ($> \SI{100}{\kelvin}$) and smaller cavities ($<\SI{20}{\percent}$ of the envelope cleared). The range of volume fractions (and hence, cavity half-opening angles) tabulated in the literature are consistent with those measured from our data, and there is no evidence for large systematic differences between the data sets, despite the different types of observations and methods used to measure the cavity sizes. 

\citet{arce_evolution_2006} use millimeter line emission in the blue and red lobes identified in \ce{CO} maps to measure the cavity angle, assuming a conical outflow geometry. Although we do not share sources (so a direct comparison between the different methods cannot be made), we find that both the size scales probed and the range of observed volume fractions are similar, indicating that there are not large, systematic differences between the two techniques. \citet{arce_evolution_2006} suggest a correlation between an age diagnostic based on $T_\text{bol}$ and the cavity size. This correlation, however, is driven significantly by the Class II objects in their sample, shown 
in \autoref{fig:tbol-vs-frac-vs-acre}. For instance, the Spearman Rank Correlation Coefficient of $T_\text{bol}$ and volume fraction decreases in their sample from 0.7 to 0.6 ($p=0.0011$ to $p=0.015$) without the Class II objects. By definition, these objects only contain small remnants of their protostellar envelopes. They lack well-defined defined cavities, and the lack of envelopes may not necessarily be the result of clearing by progressively expanding outflows cavities.

\citet{seale_morphological_2008} also measure the opening angles of envelope cavities in scattered light detected by \textit{Spitzer} IRAC\@. Although this technique has a lower angular resolution than our study and encompasses a sample of objects spanning a much broader range of distances, it has the advantage of being able to detect outflow cavities from Class 0 objects which are apparent in the \textit{Spitzer} \SI{3.6}{\micro\meter} band. They also find \replaced{no statistically significant correlation}{correlation coefficients that indicate no or weak statistically significant correlation} between cavity size and age indicators, \replaced{with the exception of a correlation between IRAC color and cavity size}{with two exceptions}. \replaced{This indicator}{The first age indicator that is correlated with cavity size is the IRAC color (Figures~11\textit{b}--\textit{d} of that work), an indicator that} may also depend on cavity size since larger cavities allow more radiation to escape at the wavelengths probed by IRAC\@. The second correlation is with the age parameter from the \citet{robitaille_interpreting_2007} model grid. 
This age is used to set the sampling of cavity angles assuming cavity growth and thus \replaced{should}{could} have induced a correlation. \added{Futhermore, the correlation with the age parameter is relatively weak, with the Pearson product moment at a significance of $\alpha=\SI{4}{\percent}$, and no evidence for a correlation using Kendall's Tau rank correlation coefficient.} \replaced{The five sources from their search with bolometric temperatures from the c2d survey (Evans et al.\ 2003) are plotted as squares in Figure~18.}{Sources from \citet{seale_morphological_2008} with bolometric temperatures in the literature are plotted as triangles in \autoref{fig:tbol-vs-frac-vs-acre}.}

Other works have found evidence for cavity growth during the Class 0 phase, as suggested by \citet{arce_evolution_2006}. \citet{velusamy_hires_2014} measure the full opening angle near the base of the cavity using the HiRes reduction of \textit{Spitzer} IRAC images. They find a broken power-law growth showing a clear increase in the sizes of cavities with increasing $T_\text{bol}$ from protostars with $T_\text{bol} < \SI{100}{\kelvin}$, but do not reproduce the growth for more evolved objects.\footnote{The \citet{velusamy_hires_2014}  power-law breaks at an age of \SI{8000}{\year }, as determined from $T_{bol}$ using the empirical relation of \citet{ladd_c18o_1998}. This corresponds to a $\log{(T_\text{bol}/\si{\kelvin})}$ of $2 \pm 0.25$.} \citet{hsieh_widening_2017} also present a survey of low luminosity protostars using IRAC images, in addition to \textit{CFHT} WIRCam \textit{Ks}-band observations. These authors use the same radiative transfer modeling code described in \autoref{sec:modeling}, but they use a direct least-squares fit of their model grid to their images to determine the cavity parameters. They find evidence for a similar growth during the Class 0 phase.  Although we do not find a similar correlation in our data, the Class 0 phase is dominated by non-detection in our \SI{1.60}{\micro\meter} imaging and the smallest cavities will be harder to detect (\autoref{fig:tbol-histograms}). Thus, we do not rule out the growth of cavities during the Class 0 phase.

\citet{furlan_herschel_2016}, using the SEDs and modeling described in \autoref{sec:cavity-sed}, find that the envelopes decrease in density by a factor of 50 as protostars transition from the Class 0 to the flat-spectrum phase. \footnote{It is well recognized that the SED depends on both the inclination and the evolutionary stage, and the SED classes encompass a mixture of evolutionary stages. The SED classes, however, provide an approximate indicator of the evolution suitable for this analysis and has the advantage that they are not model dependent. See \citet{robitaille_interpreting_2007} and \cite{furlan_herschel_2016} for further discussion.} By the end of the Class I phase, it is thought that most of the stellar mass has been accreted. We should therefore expect the processes that reduce the mass and density of the protostellar envelope to continue through the Class I phase after starting in the Class 0 phase. 

The lack of a correlation between the fraction of the volume cleared and the evolutionary indicators, in a sample preferentially probing Class I objects, implies that the evolution of the envelope during the Class I phase is not driven by growth of the outflow cavities. Although envelope clearing contributes up to a \SI{40}{\percent} reduction, the more than an order of magnitude drop in the envelope density cannot be explained by this clearing alone. Of particular importance are the number of protostars with $<\SI{15}{\percent}$ of the envelope cleared throughout the entire range of $T_\text{bol}$ covered. One of the best examples is the protostar HOPS 136, which has a volume cleared of \SI{1.8}{\percent}. \citet{fischer_hops_2014} found that this protostar was in the late stages of stellar formation. The envelope mass of \SI{0.06}{\solarmass} was much smaller than the estimated stellar mass of \SIrange{0.4}{0.5}{\solarmass}, showing that most of the stellar mass has been accreted. A relatively low density envelope is inferred from both the SED and the detection of scattered light in the envelope in the \textit{HST} images (which implies a low optical depth at \SI{1.60}{\micro\meter}). The presence of such protostars with a low density, low mass envelope, and narrow outflow cavities at the late stages of stellar formation are clear examples where the clearing of the envelopes by outflows cannot explain the observed low envelope densities.

Our results also put limits on the ability of feedback from outflows to explain the \replaced{lack of efficiency in star formation}{low star formation efficiency}. Comparisons of the Core Mass Function and Initial Mass Function suggest that \SIrange{60}{70}{\percent} of the core mass will not \replaced{form into}{accrete onto} stars \citep{alves_mass_2007,konyves_census_2015}, and previous authors have invoked outflows as partially responsible for this effect \citep[e.g.,][]{alves_mass_2007}. 
Assuming the growth of the cavities is monotonic in time, the volume fraction cleared provides a lower limit on the mass fraction cleared by the outflows. From our \textit{HST} data, the Class I protostars have cleared at most \SI{40}{\percent} of their volume (\autoref{fig:tbol-vs-frac-vs-acre}).  Recalling that the mass fraction cleared from a cavity may be as much as \SI{9}{\percent} higher than the volume cleared (\autoref{fig:mass-vs-volume}), the maximum fraction of mass cleared is \SI{50}{\percent}. Most of the protostars have cleared a much smaller mass fraction, even those toward the end of their protostellar phase; the median volume fraction cleared for the HST sample is only \SI{10}{\percent}.
These results suggest that the feedback via clearing is not sufficient to explain the small star formation efficiency inferred for dense cores, and other mechanisms should be investigated.

There are \replaced{several other possibilities}{other possible ways outflows may reduce star formation efficiency}. The gas launched by the star-disk system in a jet or wind can escape the protostar and its envelope.
Using estimates of the mass loss rates of 84 protostars, \citet{watson_evolution_2016} found that the median fraction of gas launched is $0.09$ of the gas accreted (although with a wide dispersion); this may decrease the star formation efficiency by up to an additional \SI{10}{\percent}.
We find a median star formation efficiency of $\sim$ \SI{70}{\percent} given a \SI{10}{\percent} median volume fraction cleared, a \SI{9}{\percent} increase for mass fraction cleared, and an additional \SI{10}{\percent} for mass directly launched and ejected by the central protostar. Only for the largest cavities, which clear up to $\sim$ \SI{40}{\percent} of their envelopes, can the efficiency be as low as $\sim$ \SI{40}{\percent}.

Secondly, the size of the cavity seen in scattered light may not measure the entire volume of the gas entrained in the outflow. In support of this, \citet{seale_morphological_2008} noticed a possible discrepancy between their scattered-light outflow cavity sizes and the extent of the outflowing gas traced by millimeter line data. This outflowing gas may be slower moving, denser gas entrained into the outflow that is located  outside of the  cavities. 

In the case of the HH46/47 outflow, \citet{zhang_alma_2016} used ALMA data in the $^{12}$CO, $^{13}$CO, and C$^{18}$O lines to measure the mass  in the outflow, including the slower, denser, entrained gas. They  find that the gas mass in the outflow with velocities exceeding the escape velocity is $\sim 3$~times the current stellar mass.  If this instantaneous efficiency persists throughout the protostellar collapse, then the entrainment of gas in the outflow may account for the observed inefficiency. Simulations of collapsing cores with turbulence have also been able to achieve star formation efficiencies of 40\% \citep{offner_investigations_2014}. Radiative transfer models based on these simulations are needed to predict the evolution of cavities and compare them to the cavities measured in this work.

Finally, if outflows are not sufficient to reduce star formation efficiencies to the observed levels or to slow/halt accretion, then other mechanisms must be identified. For example, the collapse of a finite Bonner-Ebert core leads to an exponential tapering in the infall rate \citep{vorobyov_lifetime_2010}; however, this does not explain the inferred low star formation efficiency of cores. Furthermore, protostellar cores embedded in molecular clouds can draw gas from their surroundings and may not be limited by the mass in the surrounding core \citep{myers_distribution_2009}.  Oscillating molecular filaments, as suggested by \citet{stutz_evolution_2016} and \citet{stutz_evolution_2018} may eject protostars. Alternatively dynamical interactions in small non-hierarchical systems or clusters may also eject protostars \citep{reipurth2010,bate_stellar_2012}.  Identifying this mechanism should be considered a key problem in star formation since it plays an important role in determining both the masses of stars and the efficiency of star formation.

\section{Summary}
We present WFC3 \SI{1.60}{\micro\meter} and NICMOS \SI{1.60}{\micro\meter} and \SI{2.05}{\micro\meter} images of 304 protostars and pre-main sequence stars in the Orion Molecular Clouds. All of these objects were studied as part of the Herschel Orion Protostar Survey (HOPS) and are well characterized by their \SIrange{1.6}{870}{\micro\meter} SEDs \citep{furlan_herschel_2016}. In this work, we use the images to resolve light from the central protostar scattered by dust in the envelopes surrounding the protostars, allowing us to probe structures with approximately \SI{80}{\astronomicalunit} spatial resolution. The specific results are as follows:
\begin{itemize}
	\item We divide the sample into five distinct morphological classes. These morphological classes are non-detections (63), point sources without nebulosity (100), protostars with unipolar cavities (59), protostars with bipolar cavities (31), and irregular protostars (51). Thirteen of these protostars have jets appearing to originate from the protostars, and an additional three have tentative detections of jets. The relative incidence of each morphology depends on SED class: non-detections are dominated by Class 0 objects, protostars with cavities are dominated by Class I objects and the point sources are primarily composed of flat-spectrum and Class I protostars. The irregular morphological class contains a relatively even mixture of Class 0, Class I and flat-spectrum protostars. We find that non-detections have the highest bolometric luminosities while point-sources have the lowest.
	\item For the protostars with observed cavities, we developed an edge detection routine to find the structure of the cavity walls. From this, we fit a power-law to the cavity shape and find the best fit shape (e.g., conical, parabolic, etc.) for 30 protostars in our sample with unipolar or bipolar morphologies. We calibrated this technique against our large model grid to reliably measure the opening of cavities. We find a distribution of cavity \added{half-opening} angles ranging from \SIrange{4.1}{49.7}{\degree}, while the power-law exponent varies from $1.1$ to $6.7$ with a median of $1.5$. We note that these cavity angles are not correlated with the SED derived angles of \citet{furlan_herschel_2016}, demonstrating that fitting radiative transfer models to SEDs does not provide reliable constraints on cavity sizes (\autoref{sec:sed-compare}).
	\item Using the well characterized SEDs  of \cite{furlan_herschel_2016}, we look for correlations between the observed cavity \replaced{opening}{half-opening} angle and evolutionary diagnostics such as SED class and bolometric temperature. Our data show no evidence for a dependence of outflow \replaced{opening}{half-opening} angle and volume fraction cleared with any of the evolutionary indicators. Furthermore, several evolved protostars with relatively small cavity sizes are identified. We conclude that there is no systematic growth of the cavity \replaced{opening}{half-opening} angle during the Class I phase.
	\item We find that the incidence of point sources is consistent with both the observed cavity angle distribution and the distribution of envelope densities from \citet{furlan_herschel_2016}. This implies that the point sources are protostars observed through a line of sight passing through the outflow cavity (hence seeing the protostar directly) or protostars with lower envelope density (as are typical of flat-spectrum protostars). Furthermore, we show that the number of point sources is inconsistent with a significant population of large cavities missed by our survey. Instead, our sensitivity to detecting cavities may decrease toward the smallest opening angles. As a whole, this is evidence that the cavity size distribution we obtain is reasonably complete and representative of the true distribution.
	\item Our findings indicate that outflow clearing is not the primary mechanism for the dissipation of the envelope during the Class I phase. It further suggests that  clearing alone cannot explain the $\sim\SIrange{30}{40}{\percent}$ star formation efficiencies inferred from core mass functions. Current measurements of the amount of mass directly launched by protostar in winds or jets suggest that this additional factor is not sufficient. 
	Measurements of the molecular gas with millimeter interferometry are needed to determine whether slower, higher density flows  entrained by the outflows are responsible for the halting of infall/accretion and the $\sim\SIrange{30}{40}{\percent}$ star formation efficiencies.  If they are not,  mechanisms other than feedback may be required. 
\end{itemize}

\acknowledgments
\added{This work was supported by NASA Origins of Solar Systems grant 13-OSS13-0094.}

Based on observations made with the NASA/ESA Hubble Space Telescope, obtained at the Space Telescope Science Institute, which is operated by the Association of Universities for Research in Astronomy, Inc., under NASA contract NAS~5-26555. These observations are associated with program \#11548.

Support for programs \#11548, \#14181 and \#14695 was provided by NASA through a grant from the Space Telescope Science Institute, which is operated by the Association of Universities for Research in Astronomy, Inc., under NASA contract NAS~5-26555.

\added{

AS gratefully acknowledges funding support through Fondecyt Regular (project code 1180350) and from the Chilean Centro de Excelencia en Astrofísica y Tecnologías Afines (CATA) BASAL grant AFB-170002.

The National Radio Astronomy Observatory is a facility of the National
Science Foundation operated under cooperative agreement by Associated
Universities, Inc.

}

\software{Astropy \citep{the_astropy_collaboration_astropy:_2013}, \texttt{DrizzlePac} \citep{drizzlepac}, \texttt{HO-CHUNK}  \citep{whitney_model_1992,whitney_model_1993},Matplotlib \citep{hunter_matplotlib:_2007}, NumPy \citep{walt_numpy_2011}, SciPy \citep{scipy}, Tiny Tim \citep{krist_20_2011}}

\bibliography{main}

\appendix
\onecolumngrid

\section{Images of All Protostars}
\label{sec:images}
The images in this appendix show the NICMOS and WFC3 images of protostars which display bipolar or unipolar morphologies, are point sources without associated nebulosity, or are classified as irregular.
\input{figures.tex}

\clearpage

\section{Table of All \textit{HST} Sources in this Text}
\input{table_HOPS_Cleaner_Format.tex}

\section{Error Analysis}
\label{sec:errors}
Uncertainties for functions of the fitted values described in \autoref{sec:cavity-measure} were computed without assuming any independence between the fitted parameters, particularly $n$ and $A$. Since $\theta$ is found as a function of these two parameters, the uncertainty $\delta \theta$ is given by
\begin{align}\label{eqn:theta_error}
	\delta \theta &\leq \left|\frac{\partial \theta}{\partial A}\right| \delta A + \left|\frac{\partial \theta}{\partial n}\right| \delta n \\
	&= \frac{R_\text{max} (A R_\text{max})^{1/n}\left( A \left|\ln\left(\frac{R_\text{max}}{A}\right)\right| \delta n + n ~\delta A \right)}{A n^2 \left(R_\text{max}^{2/n} + A^{2/n} R_\text{max}^2\right)}.
\end{align}
We note that this uncertainty is more strongly dependant on uncertainties in $n$ than on those in $A$.

We calculate the adjusted uncertainty in the power-law coefficient $A$ in \autoref{eqn:power_law} by the equation:
\begin{equation}
\delta A = \frac{1}{C^n}\delta A_{fit}+\frac{A~n}{C^{n+1}}\delta C+\frac{A}{C^n}ln(C)\delta n,
\label{eqn:delta_A}
\end{equation}
where $C$ represents the correction factor shown in \autoref{fig:ratios-dists} used to account for the effects of inclination on where cavity edges are detected, $\delta C$ is the uncertainty for a given correction value, $\delta A_{fit}$ is the non-adjusted uncertainty in parameter $A$ resulting from least squares fitting \autoref{eqn:power_law} to the location of detected cavity edges and $\delta n$ is the uncertainty in $n$ given the same fit.

By approximating conical outflow cavities of half-opening angles $\theta$, (as computed from $A$ and $n$ from the relation in \autoref{eqn:Avalue}), the estimated uncertainty in the fraction of the envelope volume subtended by our measured outflow cavities is given as 
\begin{equation}
\delta f_{vol}=|\frac{\partial f_{vol}}{\partial \theta}|\delta \theta=\sin(\theta)\delta \theta,
\label{eqn:delta_Vol_frac}
\end{equation}
where $f_{vol}$ is given by
\begin{equation}
f_{vol}=1-\cos(\theta)
\label{eqn:Vol_frac}.
\end{equation}

Values for these uncertainties are discussed in \autoref{sec:cavity-measure}.

\section{Extended Class II Objects}
\label{sec:extCII}

Approximately 200 pre-main sequence stars with disks, or Class II objects, were serependipitously in our WFC3 observations; these are tabulated in \citet{kounkel_hst_2016}. We have found that two of these objects are associated with bright, compact nebulosity similar to that found around protostars (\autoref{fig:class2}).  MGM 2742 (V2475 Ori) is a binary which is associated with a nebula with an irregular morphology.  MGM 925 (V2674 Ori) appears to be seen nearly edge-on and has a clear bipolar morphology.

\section{ Morphological Identification of Contamination}
\label{sec:vermin}

\begin{figure}\centering
	\includegraphics[scale=.425]{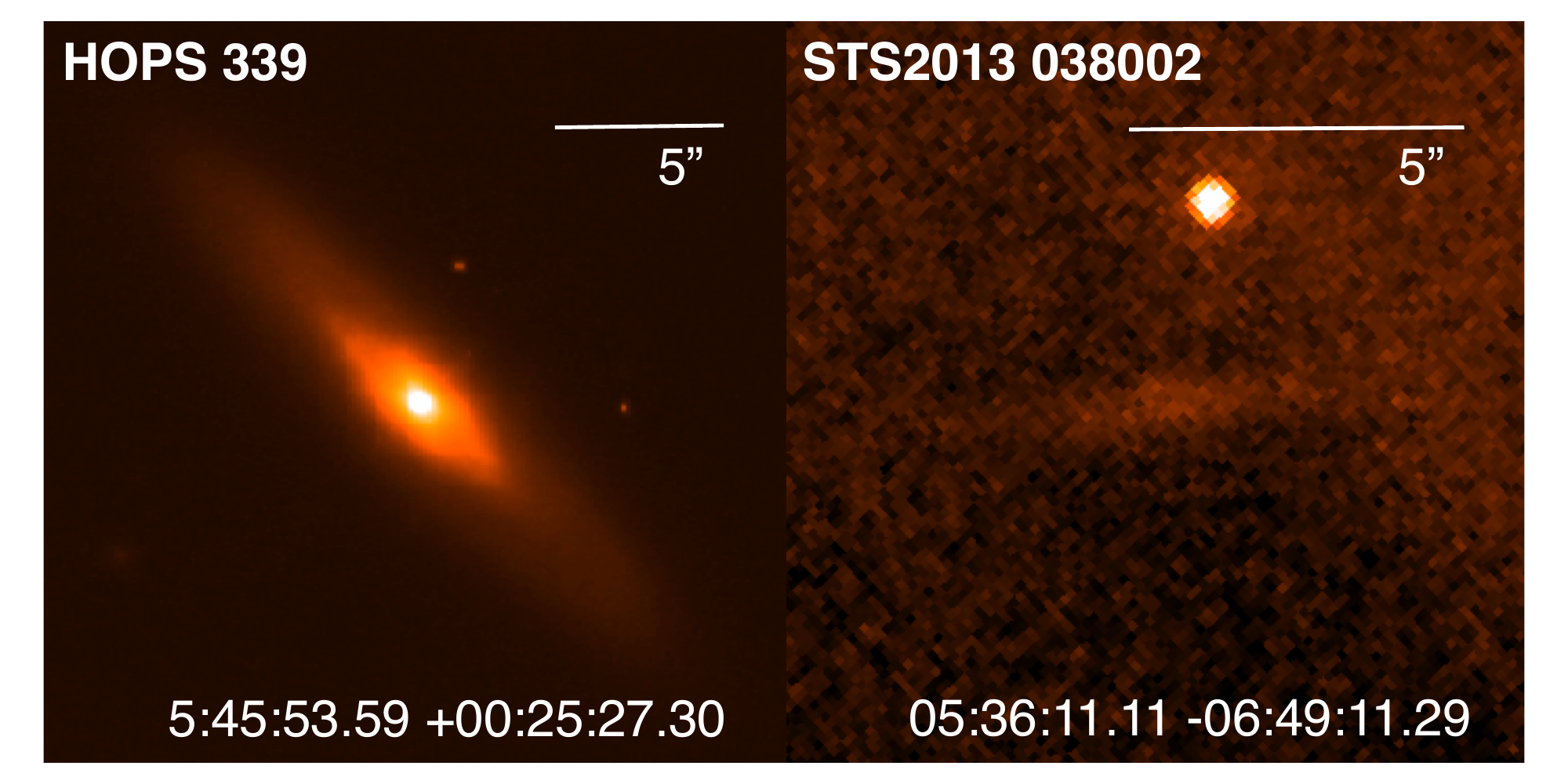}
	\caption{\label{fig:vermin}\textit{Hubble} WFC3 images of two non-protostellar sources clarified to be extragalactic.}
\end{figure}

We are able to revise our identification of three objects by their morphology, showing them to be contamination. 
HOPS 339, shown in \autoref{fig:vermin}, is determined to be a disk galaxy. {~\citet{furlan_herschel_2016}} describes its SED as mostly flat with a strong $\SI{10}{\micro\meter}$ absorption feature and notes that by SED alone it would not be flagged as a possible extragalactic contaminant. This illustrates the importance of high-resolution near-infrared observations in disentangling galactic contaminants and protostellar objects. One  protostellar candidate identified in {~\citet{stutz_herschel_2013}}, STS2013 038002 (\autoref{fig:vermin}), was targeted by the \textit{HST} program GO 14695. The source appears to be an extended streak; we suggest that it is a background galaxy observed through substantial extintion. The object STS2013 92011 \citep{stutz_herschel_2013} was also targeted by GO 14965. It is is shown by WFC3 imaging to be an outflow knot.

\begin{figure}\centering
	\includegraphics[scale=.425]{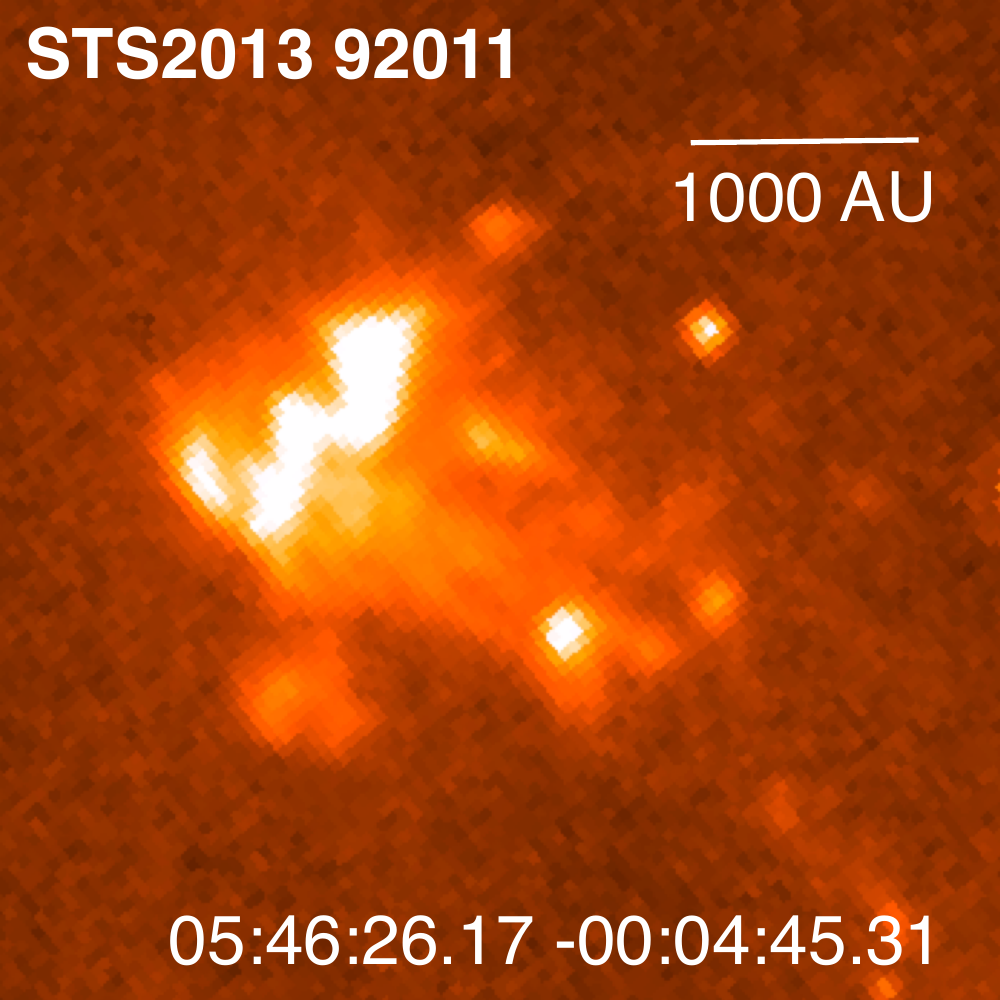}
	\caption{\label{fig:knot}One of the four objects with weak $\SI{24}{\micro\meter}$ flux targeted by \textit{HST} program 14695 is revealed in WFC3 imaging to be an outflow knot. Emission is likely dominated by the [FeII] line at \SI{1.66}{\micro\meter}}.
\end{figure}

\begin{figure}\centering
	\includegraphics[scale=.425]{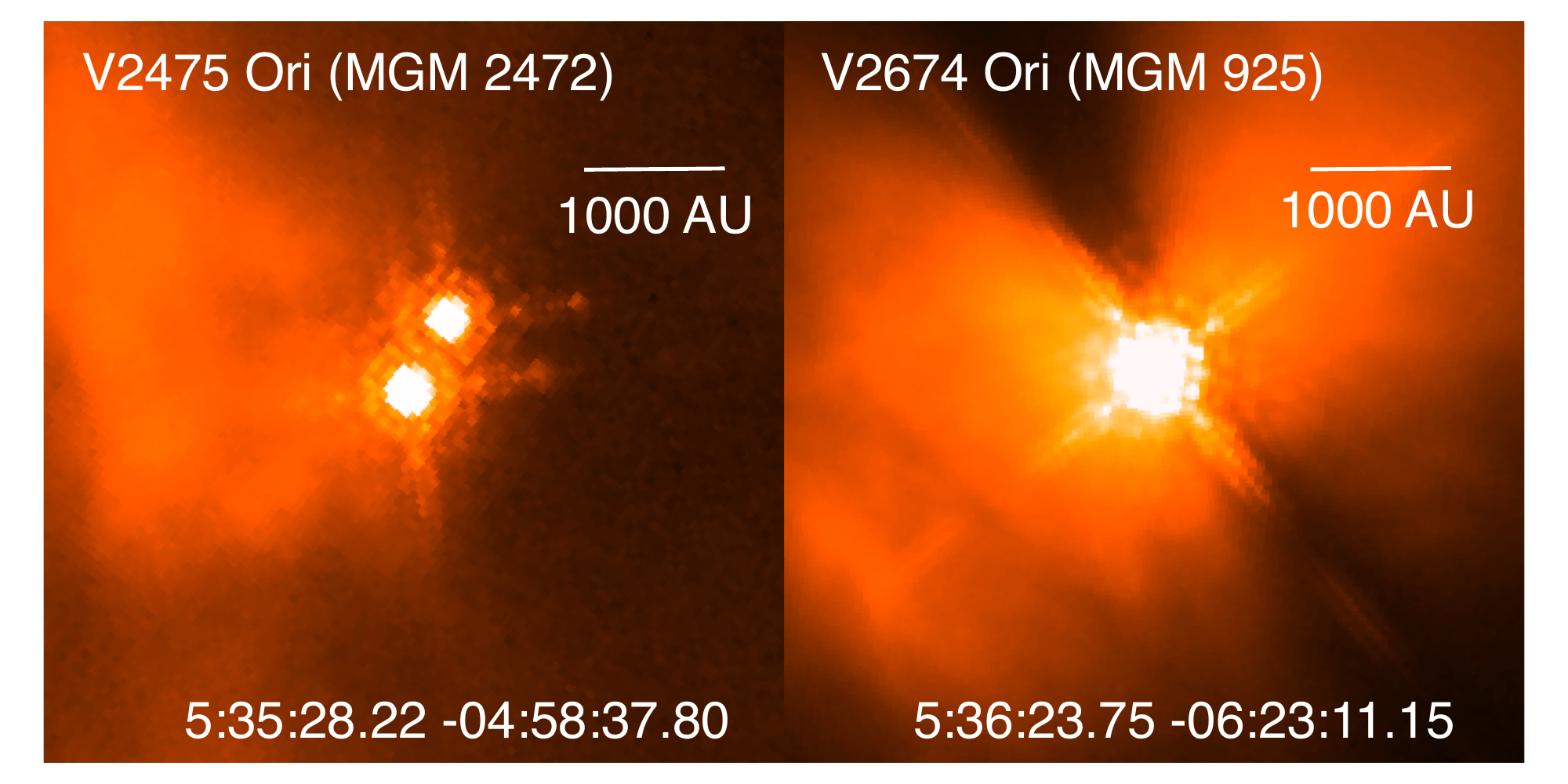}
	\caption{\label{fig:class2}\textit{Hubble} WFC3 images of two Class II objects appearing incidentally in WFC3 observations of HOPS sources.}
\end{figure}

\section{The Effect of the Adopted Dust Law on Cavity Morphology}\label{sec:dust_law_comparison}

\begin{figure}\centering
	\includegraphics[scale=.425]{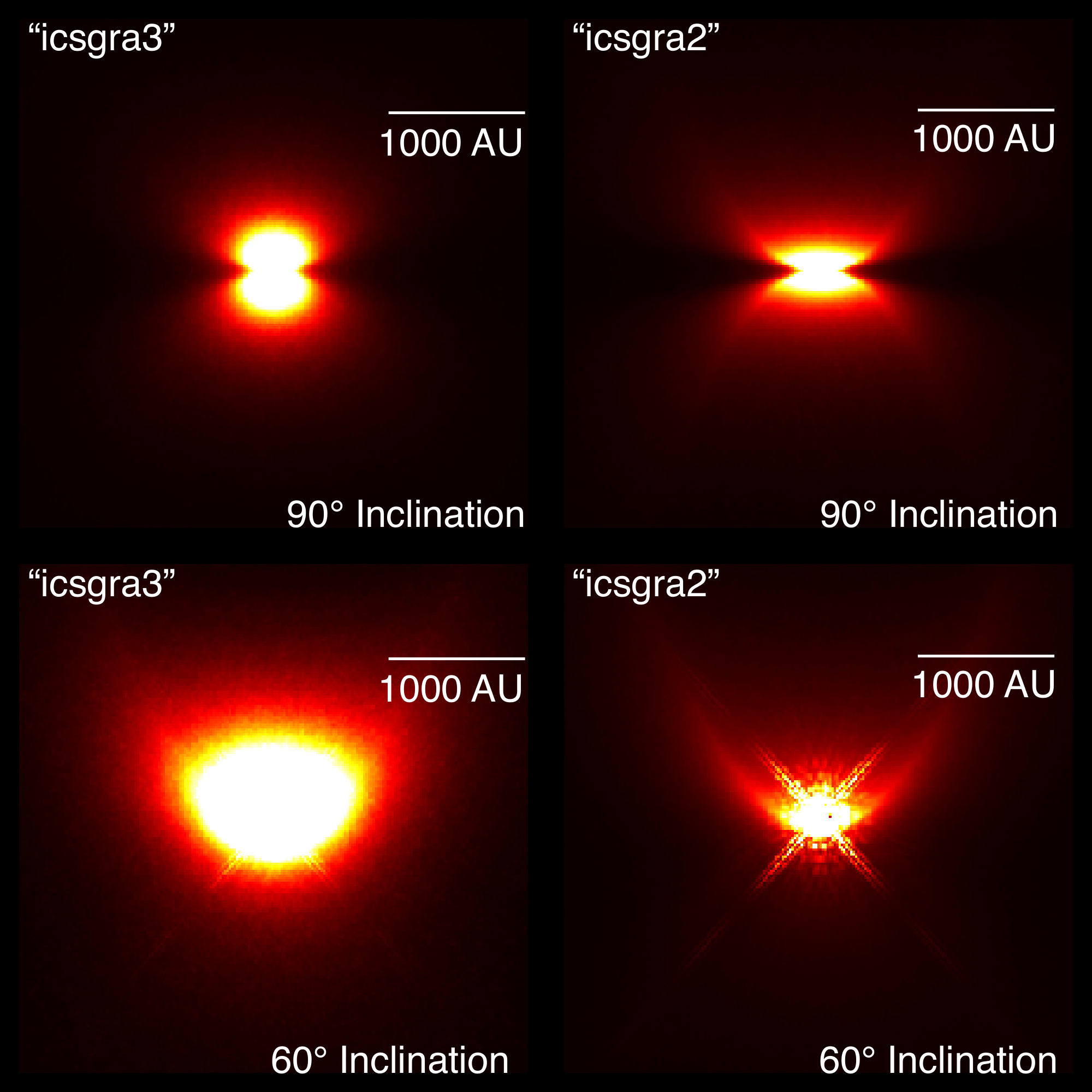}
	\caption{\label{fig:dustlaws} Two examples of a radiative transfer models created from different dust opacity models and otherwise identical parameters. On the left, the "icgras3" dust law used in this paper displays more forward scattering and less limb brightening along the cavity edge.  }
\end{figure}

In our modeling, we used the dust opacity models adopted by \citep{furlan_herschel_2016}.  To investigate the role of the assumed dust law on the observed morphology, we compared images generated with two dust opacity models from \citet{ormel_dust_2011}. The opacity model used in this paper, "icsgra3," is described in \autoref{sec:modeling} and adopts a grain coagulation time of 0.3~Myr. We compare this to the opacity model "icsgra2" which in contrasts adopts a time of 0.1~Myr with consequently more grains of smaller sizes. When comparing model protostars from our grid generated with otherwise identical parameters, we observe that those using "icsgra2" show strongly limb brightened cavity edge profiles and bright point sources (\autoref{fig:dustlaws}). In contrast, the larger grains present in  "icsgra3" result in more forward scattering where the intensity peaks toward the center of the cavity instead of the edges.  The "icsgra3" are more consistent with \textit{HST} observations which typically show the cavities filled with emission (\autoref{fig:morph}), although there are some examples that show enhanced edges (\autoref{fig:exponent-examples}). This suggests that the larger grains in "icsgra3" are more representative of our sample (\autoref{sec:images}).   
Although beyond the scope of this investigation, future studies of the observed cavity morphologies may provide new constraints on grain properties and their variations.  

\section{Comparison Between SED Modeling and Near-IR Morphologies}\label{sec:sed-compare}
\begin{figure}
	\includegraphics[scale = .5]{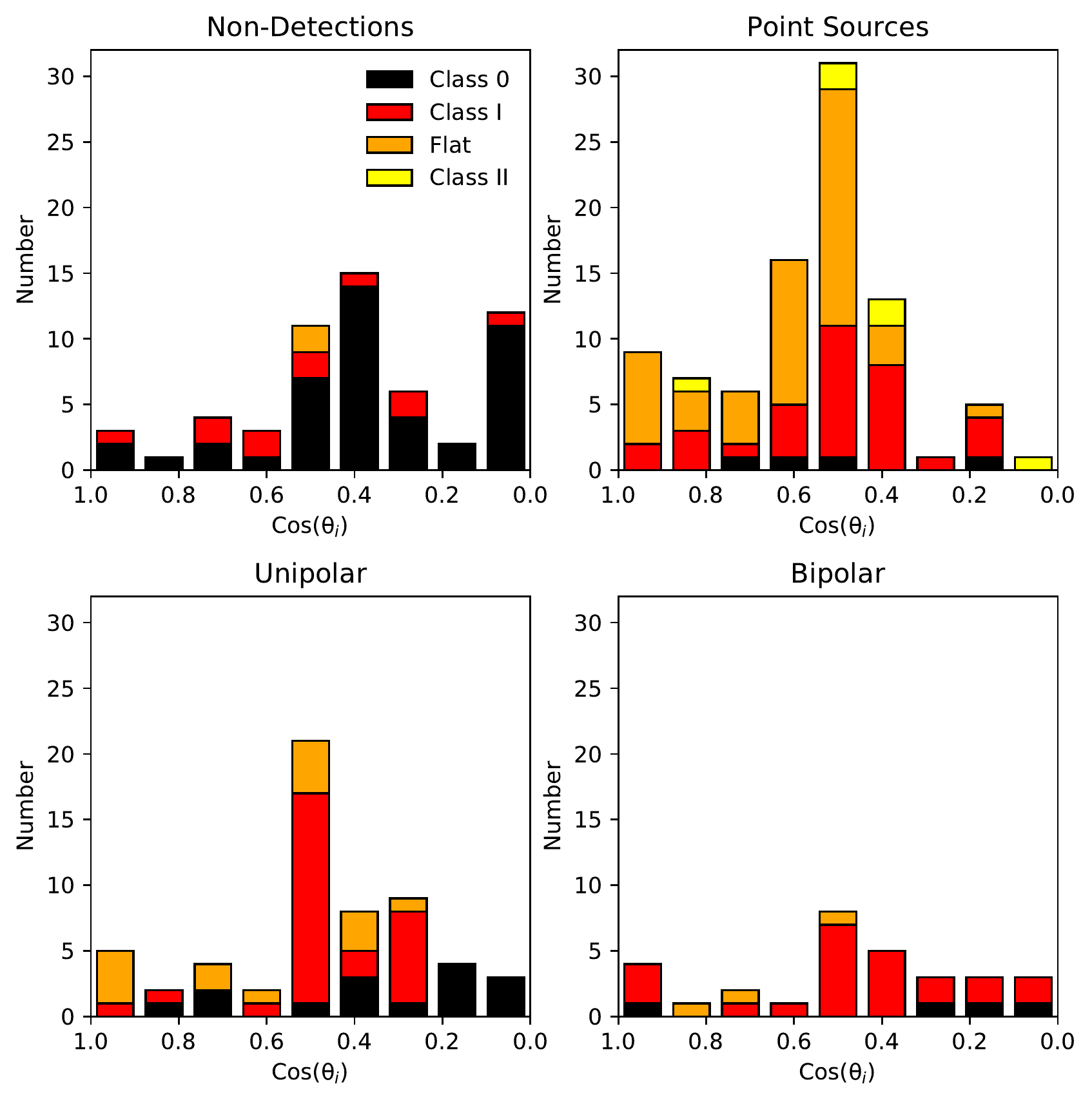} 
	\caption{\label{fig:cat_by_class}These four panels contain histograms of inclinations from the SED fitting of \citet{furlan_herschel_2016}, for a selection of four scattered-light morphologies. In each histogram, the color indicates the classification scheme used by \citet{furlan_herschel_2016}}
\end{figure}
Most of the protostars in this paper have been characterized in detail by \citet{furlan_herschel_2016} using modeling in concert with the SEDs of these objects. The models used by these authors differ from those discussed in \autoref{sec:modeling}, and are fit to the SED from \SIrange{1.6}{870}{\micro\meter}. The primarily differences that affect the \SI{1.60}{\micro\meter} emission are that our grid uses a finer sampling of high inclination protostars, \replaced{far fewer spacing}{a sparser grid} of envelope densities, and a different cavity opening angle exponent (2 vs 1.5).

\autoref{fig:cat_by_class} shows histograms of the number of protostars vs inclinations determined by SED fitting. Four different histograms are displayed, one each for bipolar, unipolar, point sources, and non-detections. (Irregular protostars are not shown). Since the bins are chosen to have equal intervals in the cosine of the inclination, a random distribution should result in an equal number of sources in each bin; however, the SED-determined inclination for the overall sample of HOPS protostars peak at \SIrange{60}{70}{\degree}, suggesting that the SED derived inclinations may have systematic biases \citep{furlan_herschel_2016}. The distribution of inclinations for point sources is similar in this respect to the overall sample \citep[see][Figure~29]{furlan_herschel_2016}, except at the highest inclinations. Obscuration by the disk likely accounts for this deficiency, both by decreasing the ability to detect a point source and by decreasing the contrast with the surrounding nebulosity. The unipolar and non-detections also peak around \SIrange{60}{70}{\degree}, although they show a deficiency of low inclination objects, this is expected since the outflow cavities cannot be detected at low inclinations the lower obscuration at these angles makes non-detections unlikely.

Finally, the bipolar protostars show a broad range of inclinations, even though their observed morphologies require a nearly edge-on perspective. \citet{furlan_herschel_2016} showed that there are large systematic uncertainties in the SED derived inclinations. This figure further demonstrates the limitations of using SED derived inclinations, particularly for edge-on protostars. In a detailed study of the \textit{HST} morphology of the HOPS~136 protostar, \citet{fischer_hops_2014} could only find agreement between the SED models and the edge-on morphology by adding a low density component of dust in the outflow cavity to increase the scattering at shorter wavelengths. This suggests that our models are incomplete and therefore under-predict the brightness of protostars in the near-IR;\@ consequently, model fits erroneously favor inclined models where the near-IR emission is less absorbed by the disk.

\begin{figure}
	\includegraphics[scale=.6]{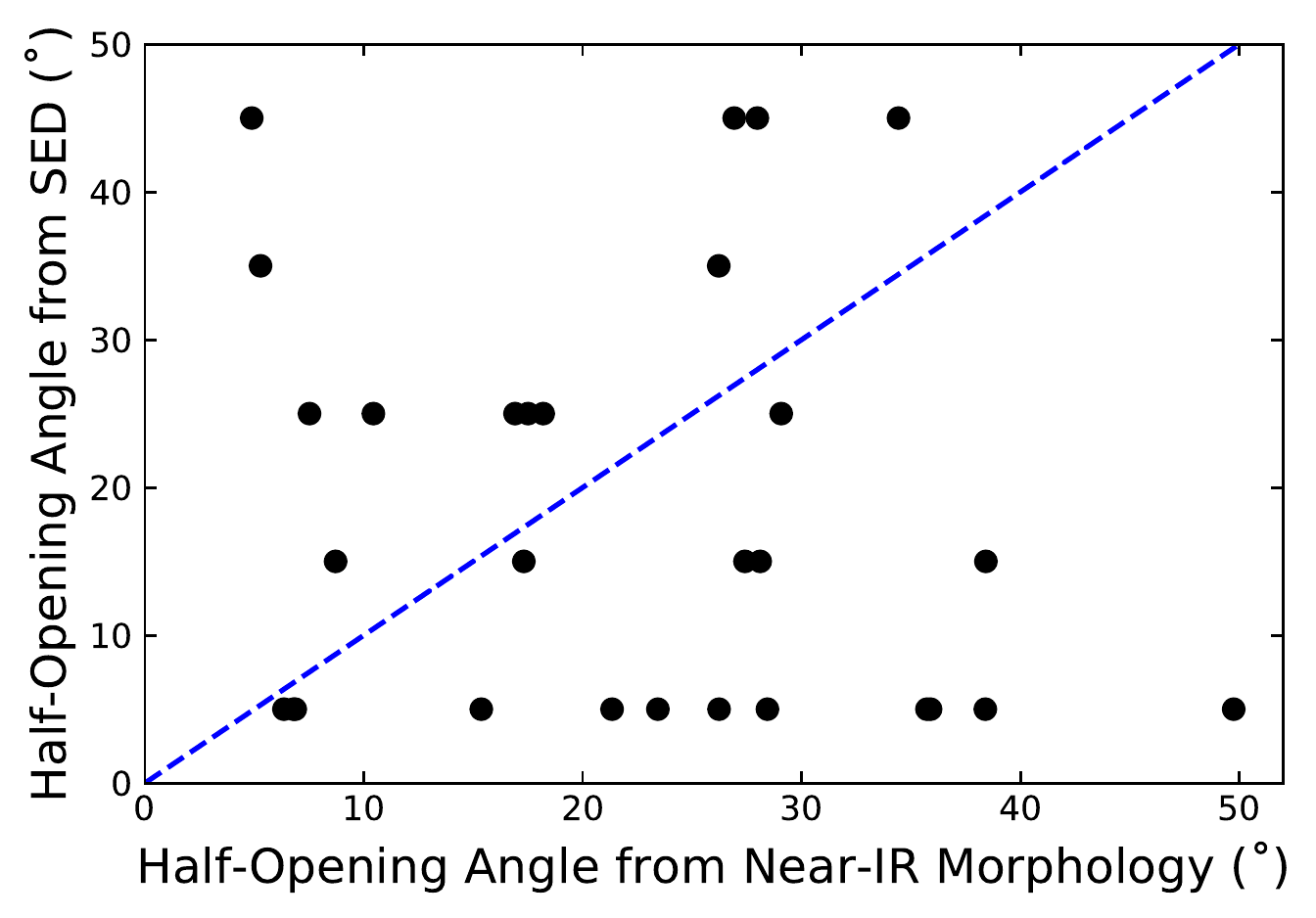}
	\caption{\label{fig:hst-vs-sed}Half-opening angles compared between this work and those in \citet{furlan_herschel_2016}. The blue dotted line indicates equality in the two methods.}
\end{figure}

Our measurements of the cavity half-opening angles can also be used to test the angles derived from SED models. \citet{furlan_herschel_2016} fitted models with discrete cavity \replaced{opening}{half-opening} angles of \SIlist{5;15;25;35;45}{\degree} for cavities with a $r^{1.5}$ power-law shape.\footnote{Note that they use the terminology ``cavity opening angle'' where we use ``cavity half-opening angle''.} The consistency of the best fit cavity angle compared to the mode for various criteria of close models are shown in Figure~46 of that work; \added{these show there are a wide range of cavity angles that can be fit to a given SED.} \replaced{however, we show another measure of this consistency in}{This is further demonstrated in} \autoref{fig:hst-vs-sed}, where we compare  half-opening angles inferred from the best SED fits  from \citet{furlan_herschel_2016} to the cavity half-opening angles derived in this work.
In spite of the different cavity shapes assumed, a monotonically increasing correlation should be seen if both are accurate measurements of the cavity size. Since we do not see the expected correlation, and since most protostars have decent quality model fits for a range of cavity angles in the SED model grid, we find the cavity opening angles are not well constrained by SED modeling. A similar result was found by \citet{seale_morphological_2008}.

\end{document}

%% file: figures.tex
\begin{figure*}[t]\centering
	\includegraphics[scale=.80]{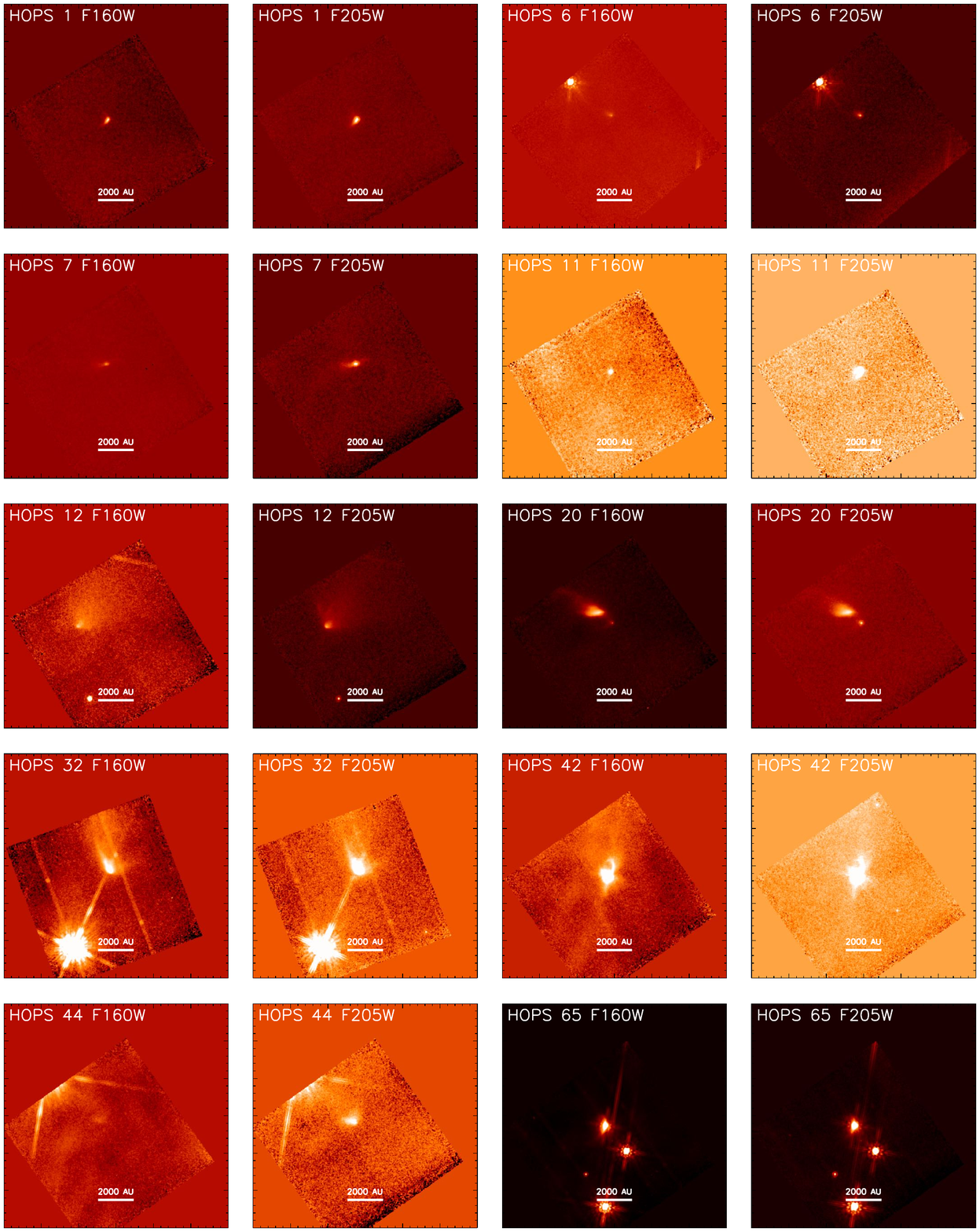}
	\caption{NICMOS F160W and F205W images of sources with unipolar or bipolar nebulosity.}
\end{figure*}

\begin{figure*}[t]\centering
	\includegraphics[scale=.80]{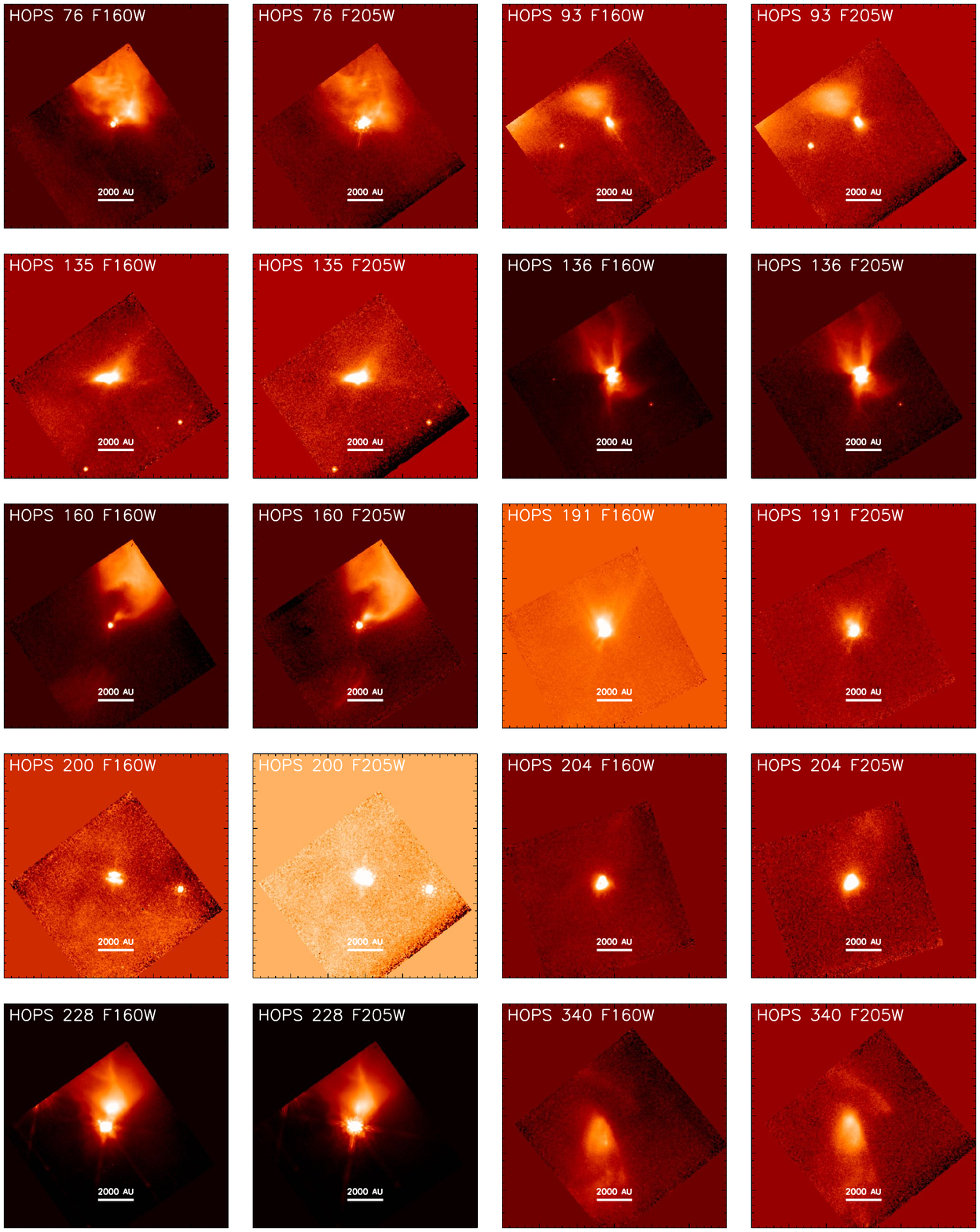}
	\caption{NICMOS F160W and F205W images of sources with unipolar or bipolar nebulosity, continued.}
\end{figure*}

\newpage

\begin{figure*}[t]\centering
	\includegraphics[scale=.80]{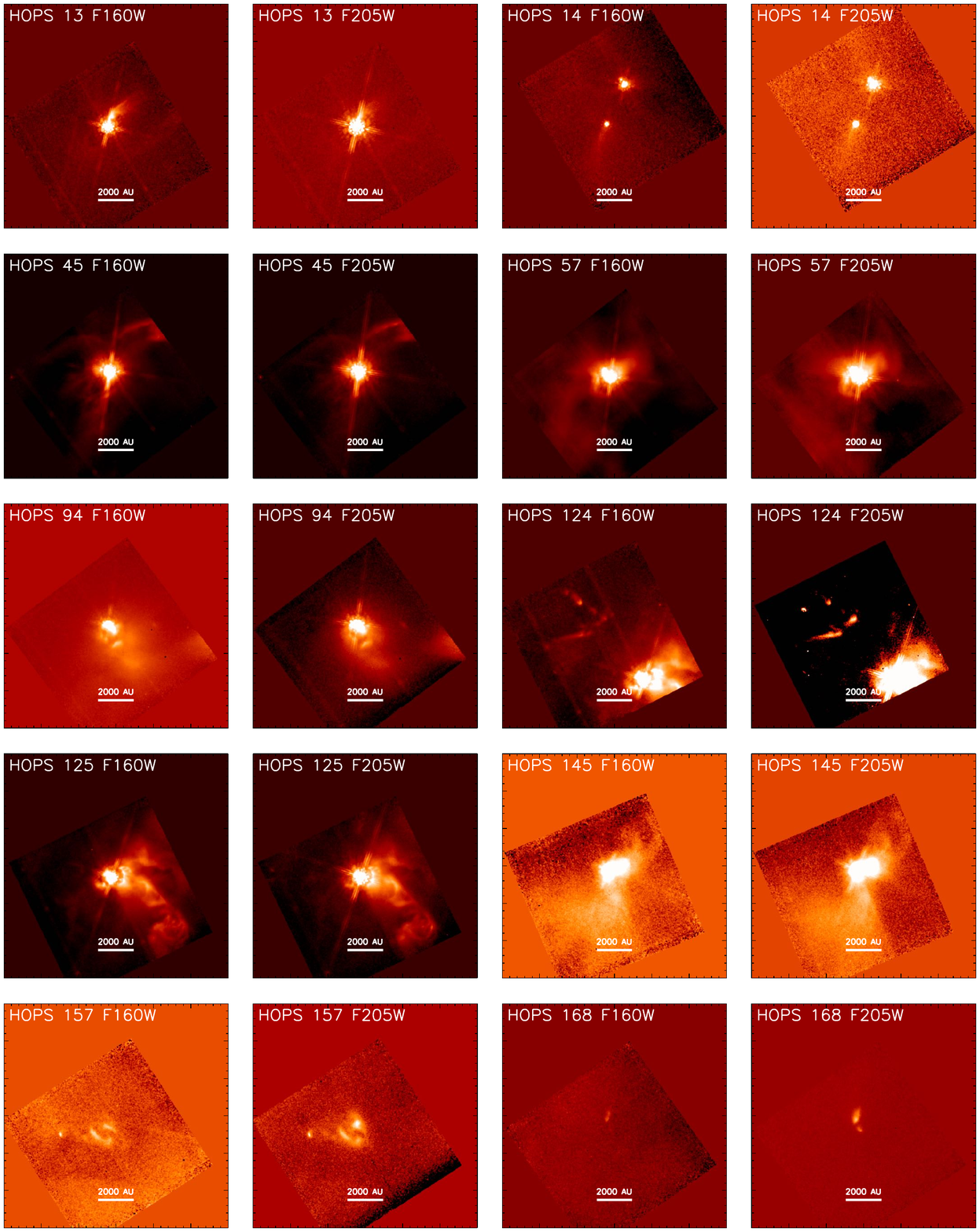}
	\caption{NICMOS F160W and F205W images of irregular sources.}
\end{figure*}

\begin{figure*}[t]\centering
	\includegraphics[scale=.80]{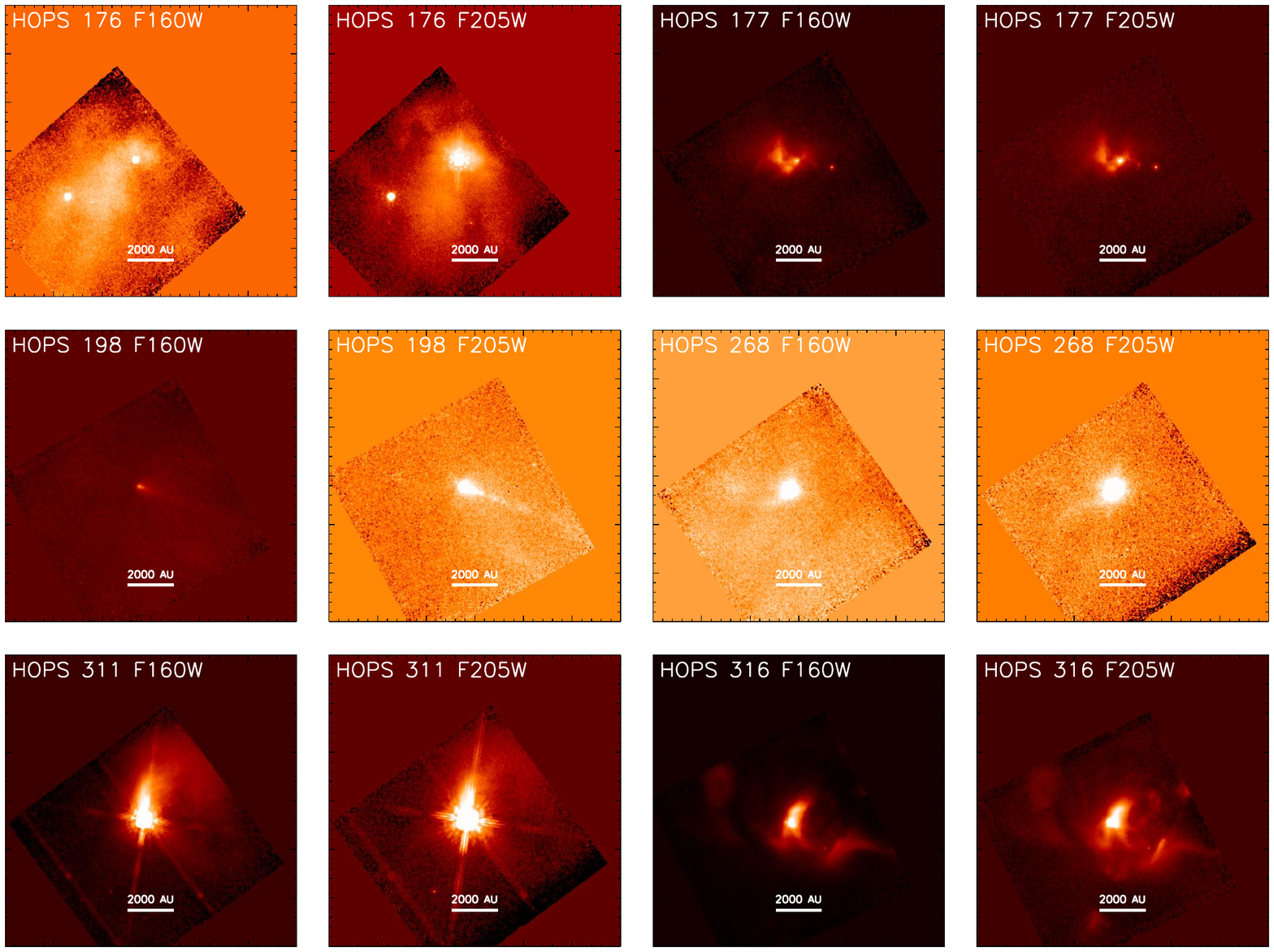}
	\caption{NICMOS F160W and F205W images of irregular sources, continued.}
\end{figure*}

\newpage

\begin{figure*}[t]\centering
	\includegraphics[scale=.80]{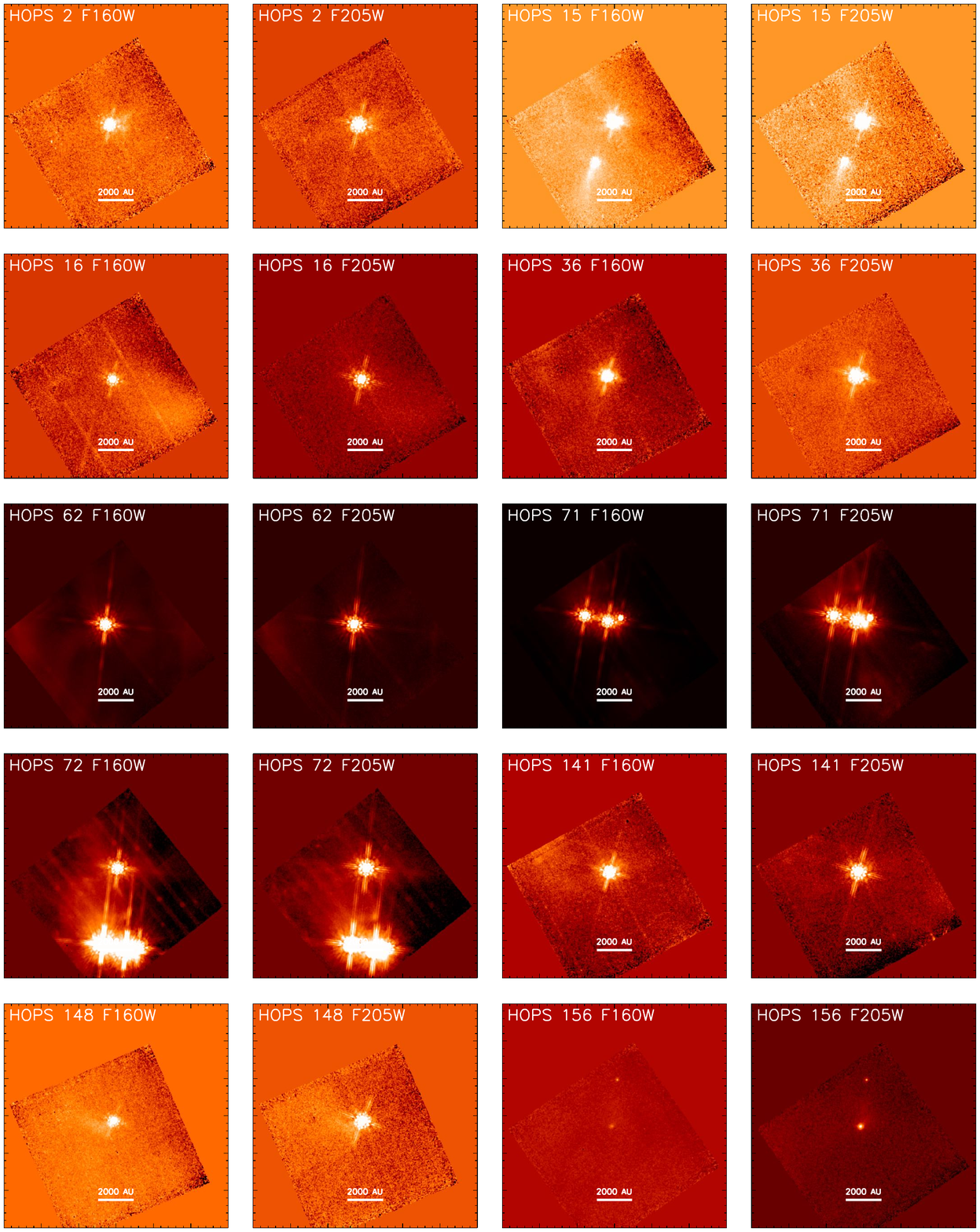}
	\caption{NICMOS F160W and F205W images of point sources without associated nebulosity.}
\end{figure*}

\begin{figure*}[t]\centering
	\includegraphics[scale=.80]{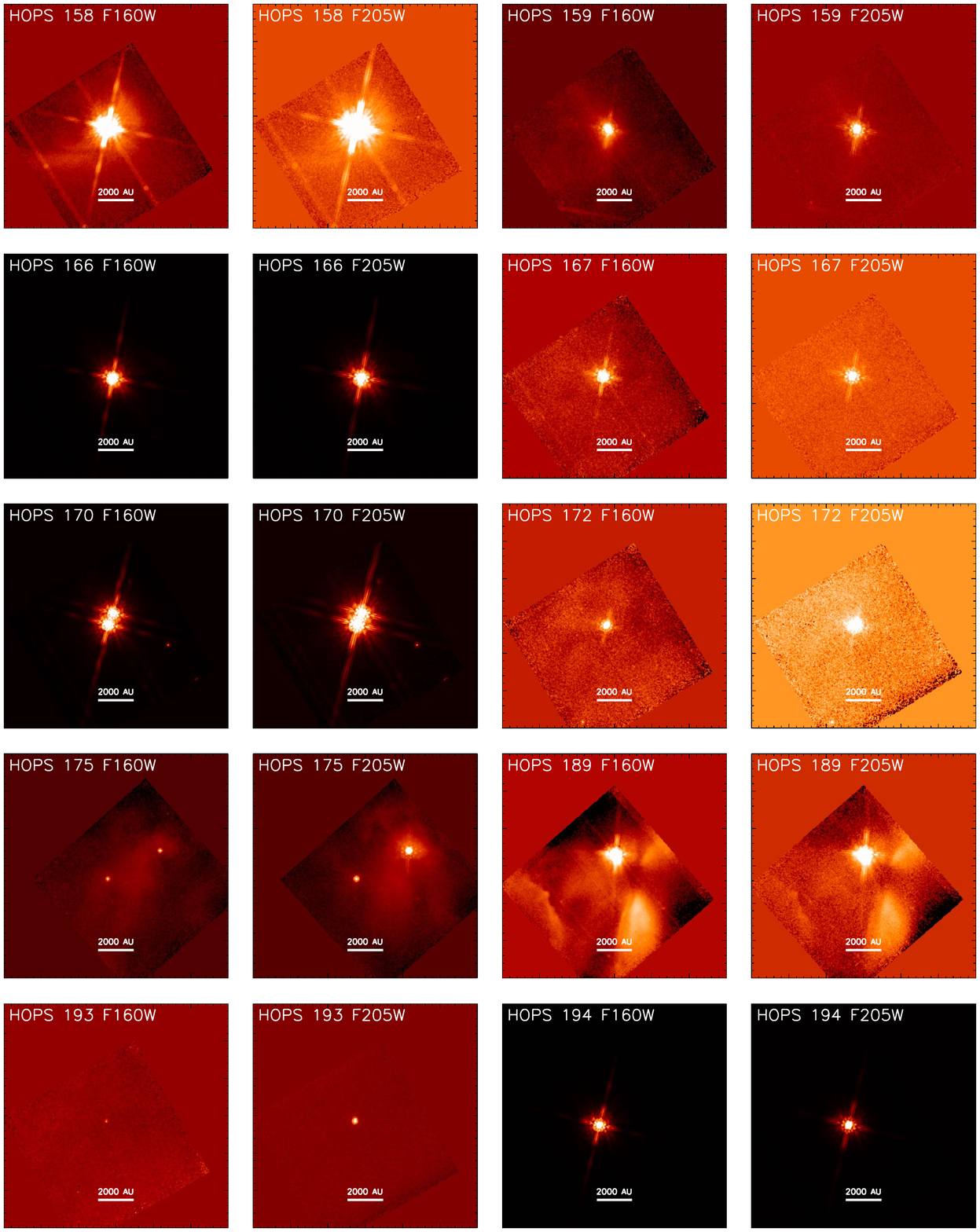}
	\caption{NICMOS F160W and F205W images of point sources without associated nebulosity, continued.}
\end{figure*}

\begin{figure*}\centering
	\includegraphics[scale=.80]{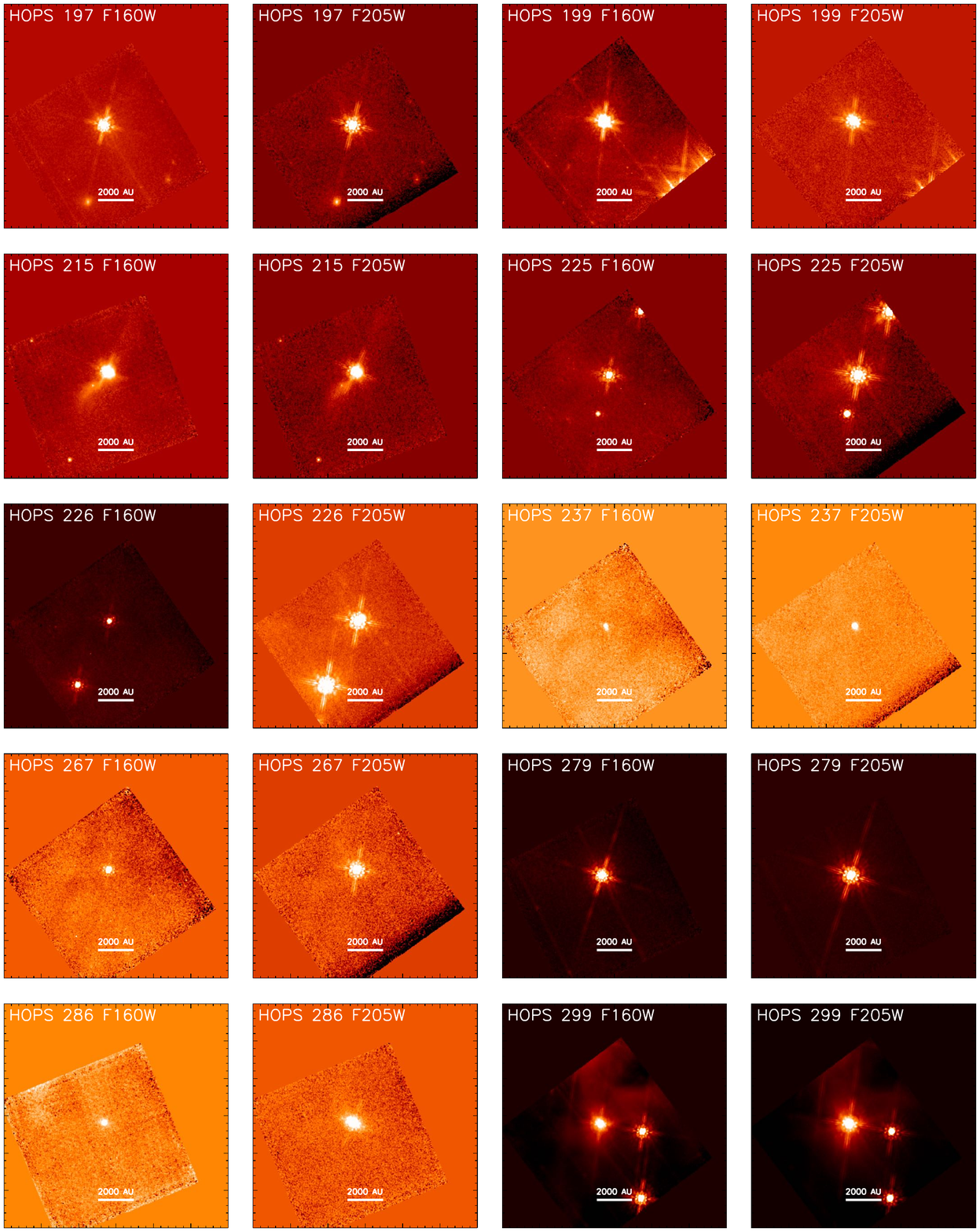}
	\caption{NICMOS F160W and F205W images of point sources without associated nebulosity, continued.}
\end{figure*}

\newpage

\begin{figure*}[t]\centering
	\includegraphics[scale = 0.8]{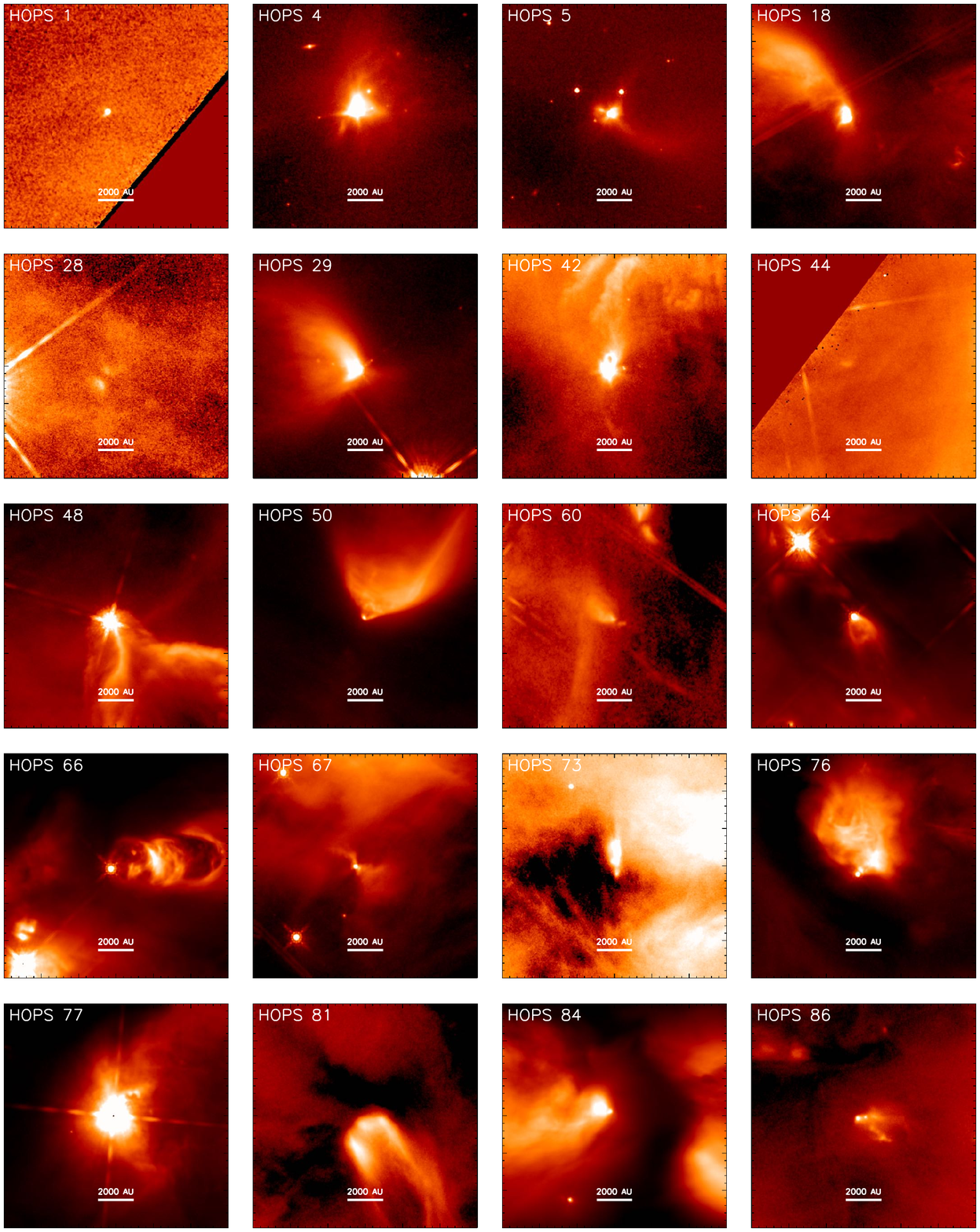}
	\caption{WFC3 F160W images of sources with unipolar or bipolar nebulosity.}
\end{figure*}

\begin{figure*}[ht!]\centering
	\includegraphics[scale = 0.8]{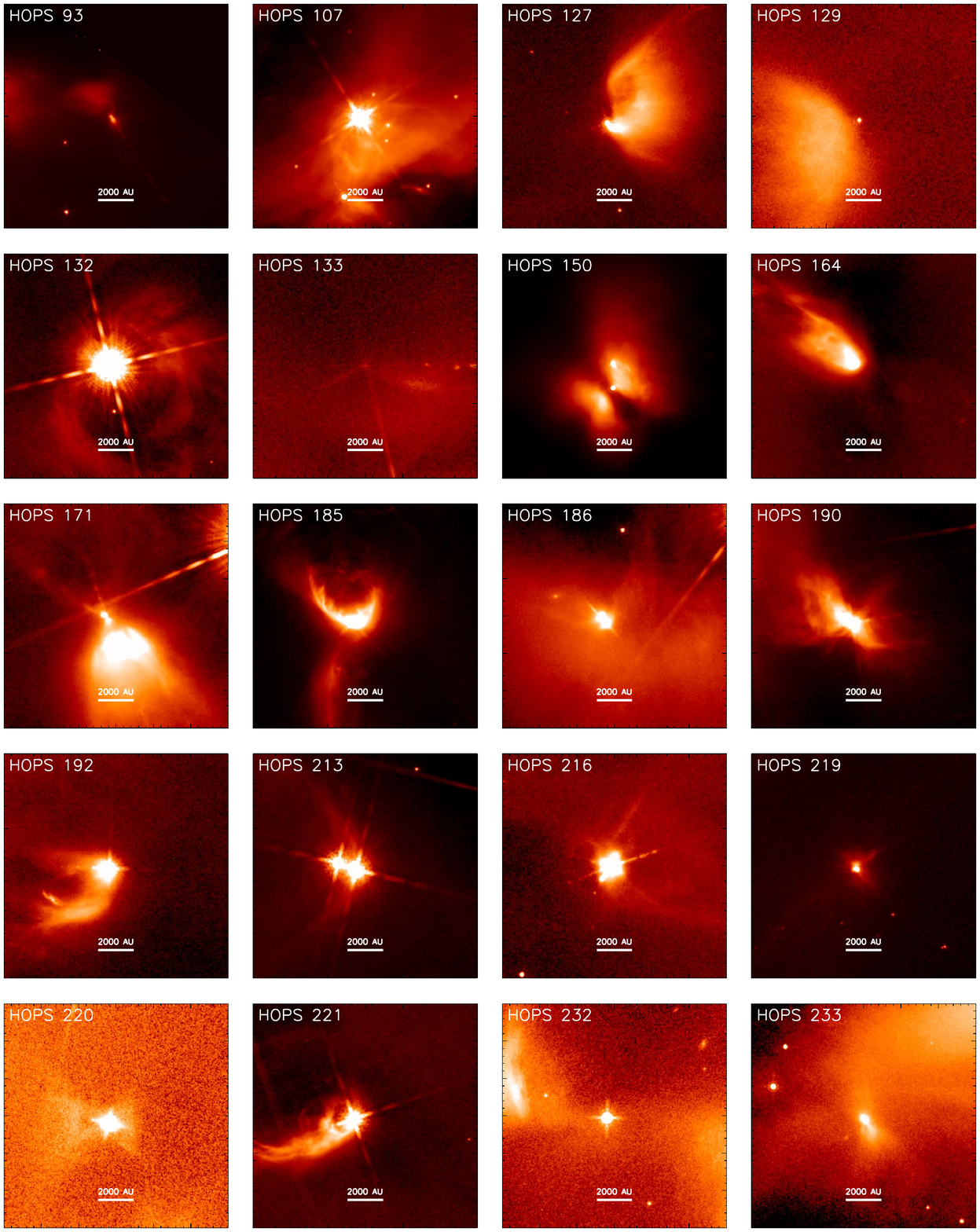}
	\caption{WFC3 F160W images of sources with unipolar or bipolar nebulosity, continued.}
\end{figure*}

\begin{figure*}[ht!]\centering
	\includegraphics[scale = 0.8]{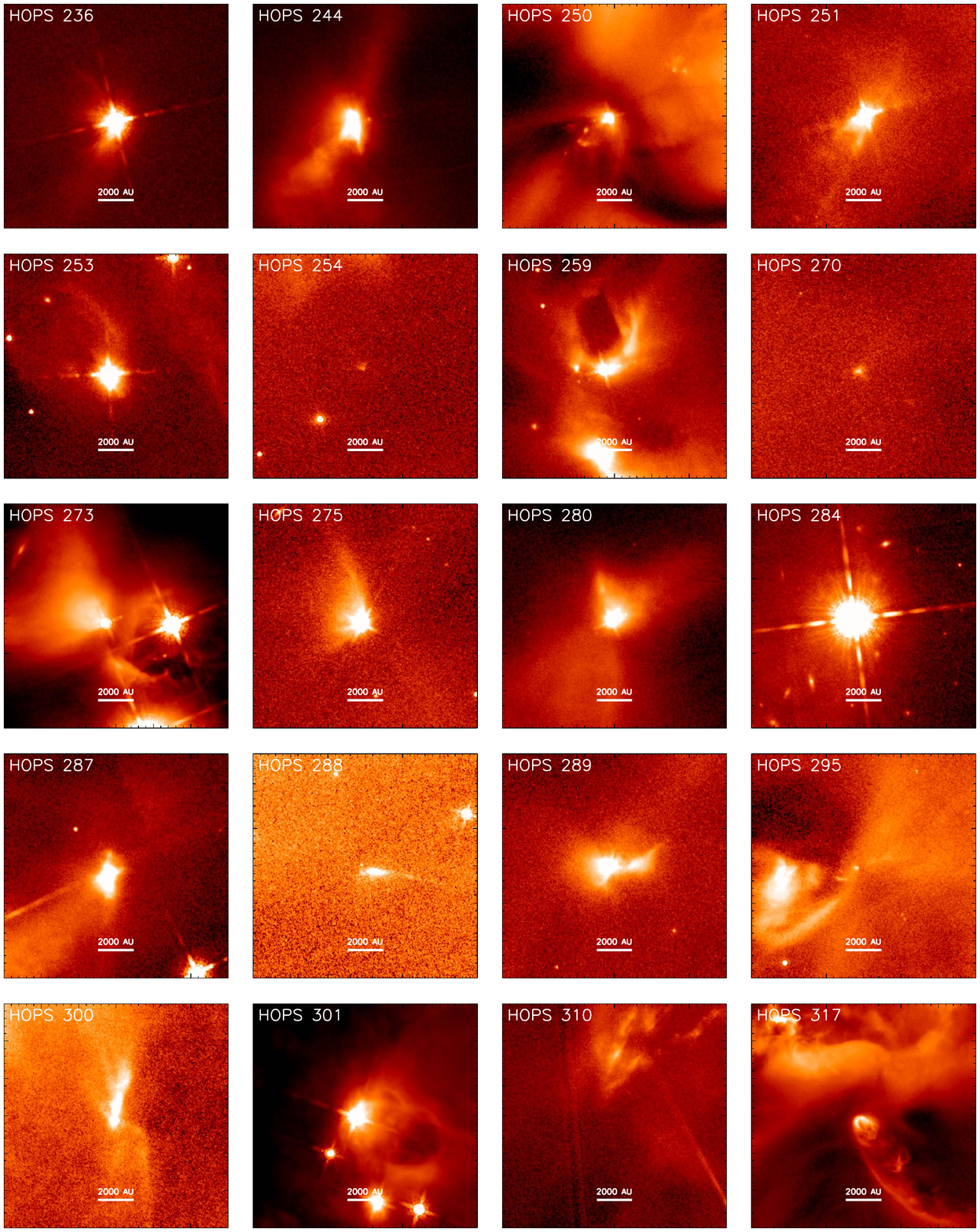}
	\caption{WFC3 F160W images of sources with unipolar or bipolar nebulosity, continued.}
\end{figure*}

\begin{figure*}[ht!]\centering
	\includegraphics[scale = 0.8]{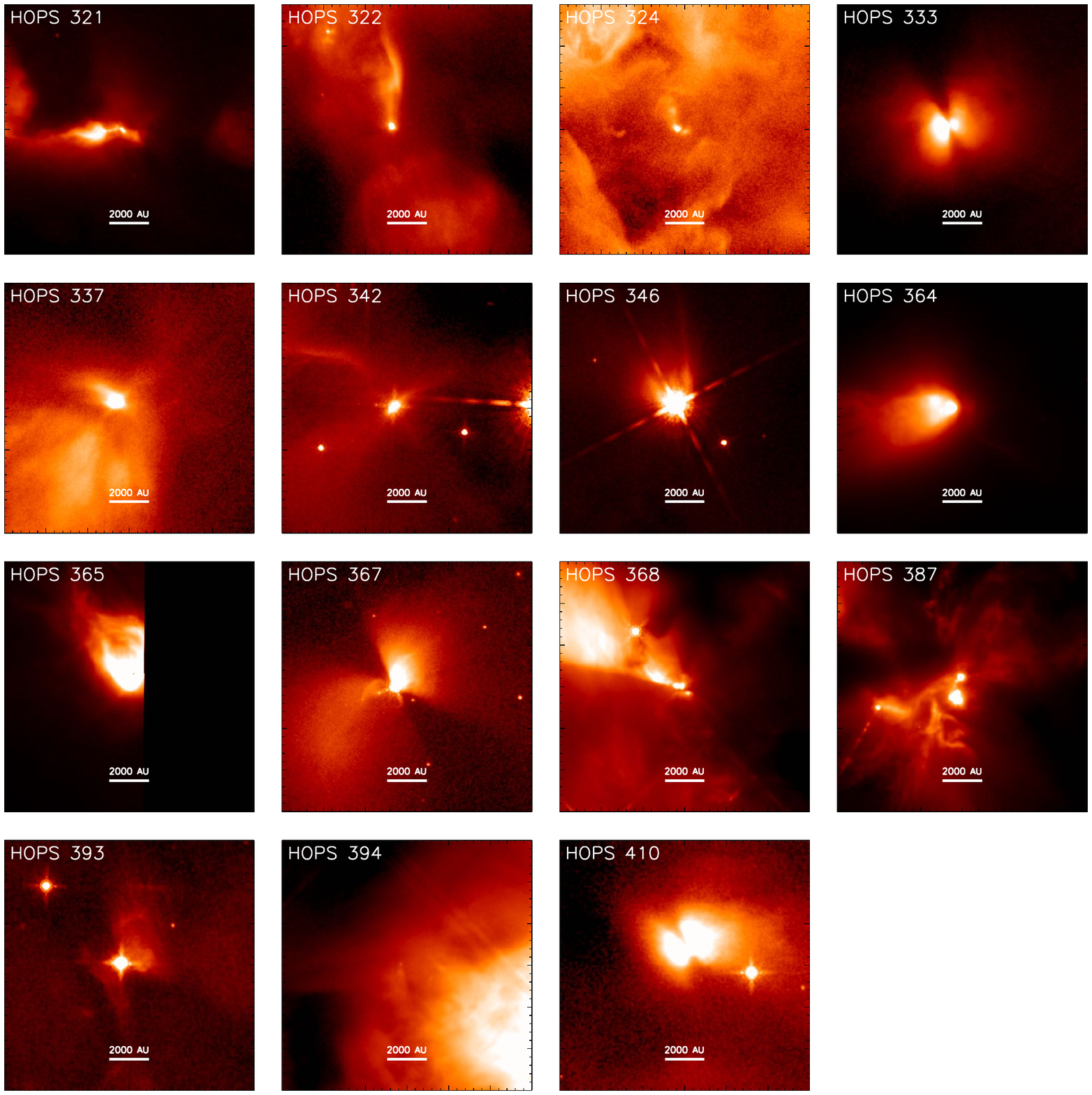}
	\caption{WFC3 F160W images of sources with unipolar or bipolar nebulosity, continued.}
\end{figure*}

\newpage

\begin{figure*}[ht!]\centering
	\includegraphics[scale = 0.8]{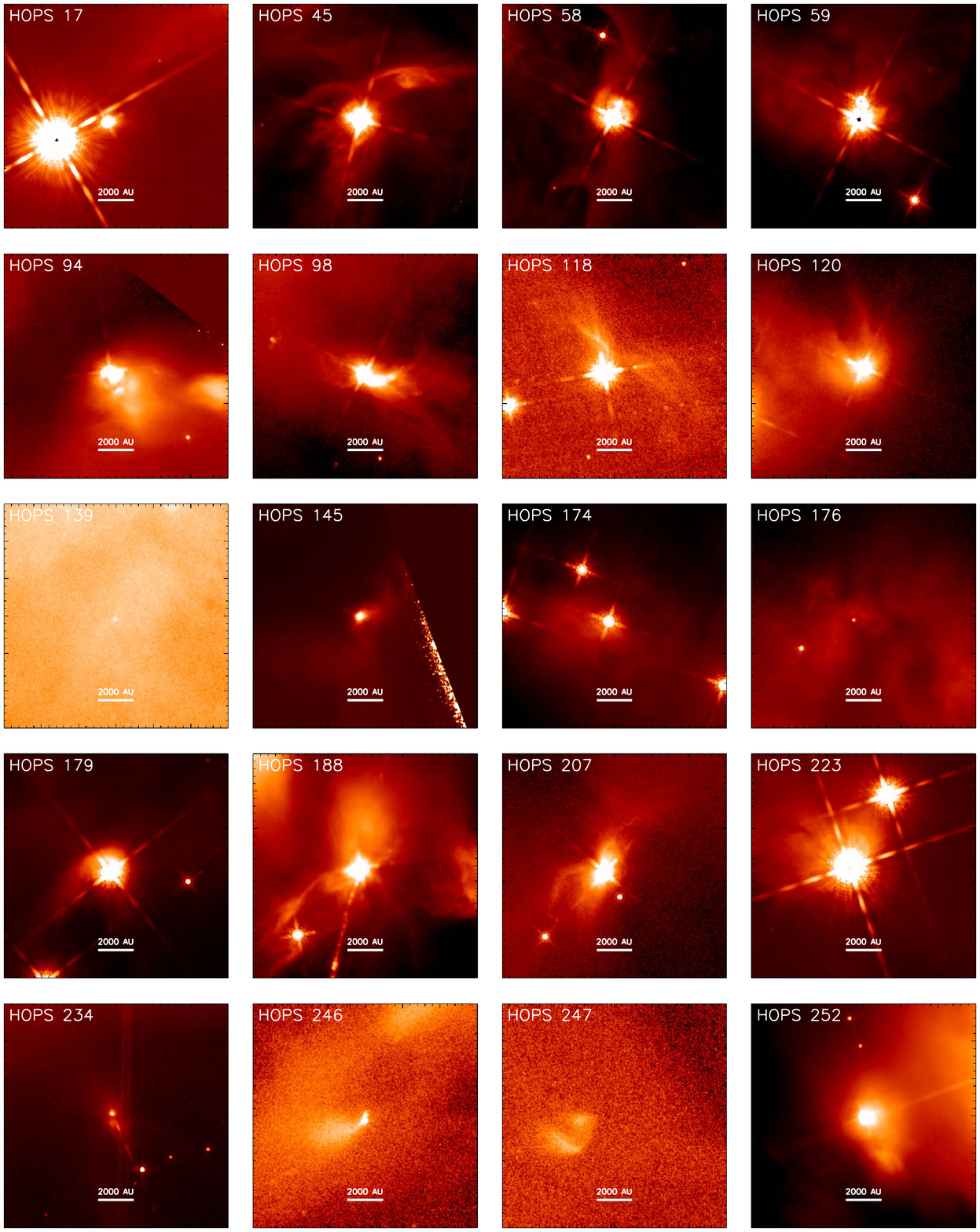}
	\caption{WFC3 F160W images of irregular sources.}
\end{figure*}

\begin{figure*}[ht!]\centering
	\includegraphics[scale = 0.8]{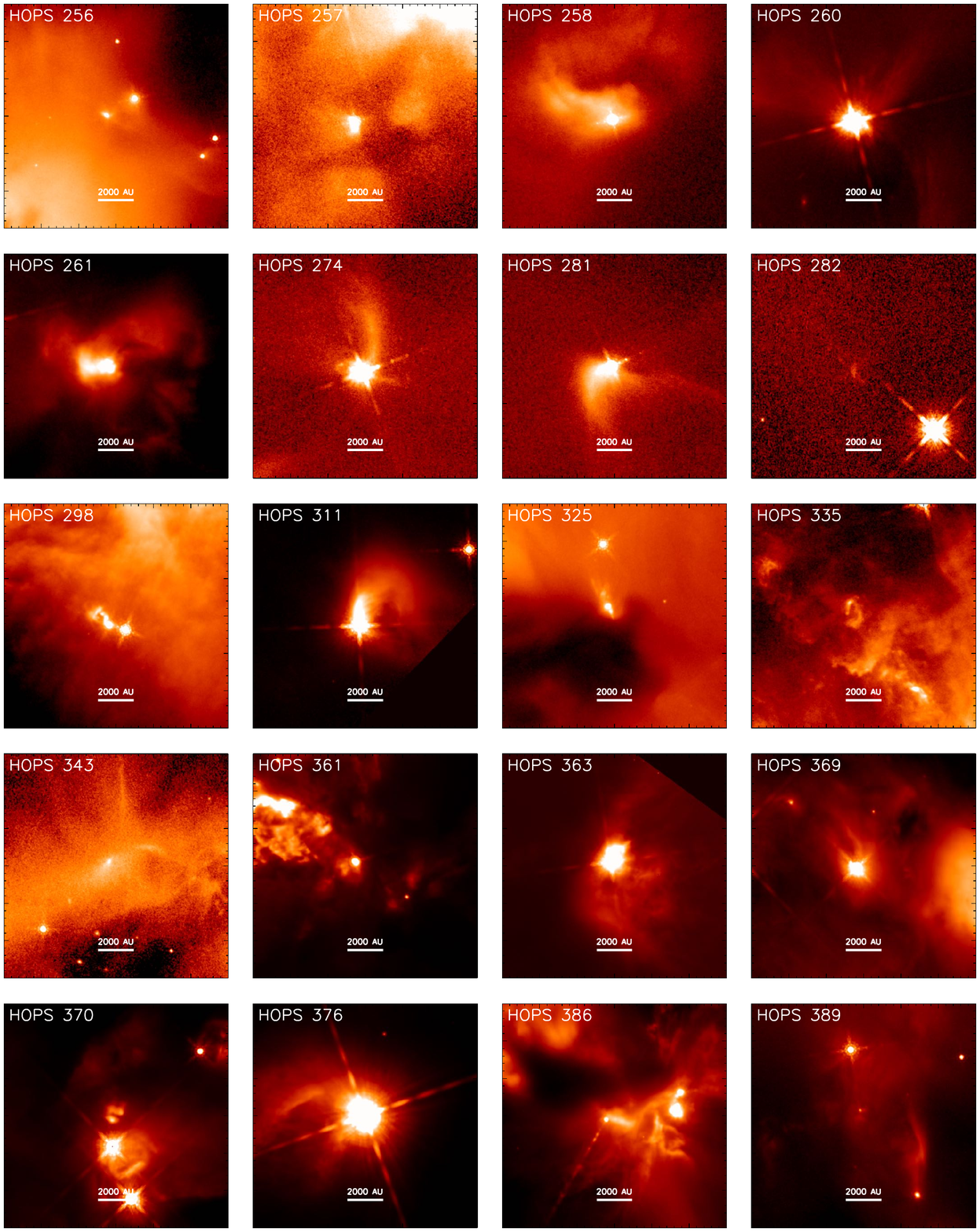}
	\caption{WFC3 F160W images of irregular sources, continued.}
\end{figure*}

\newpage

\begin{figure*}[ht!]\centering
	\includegraphics[scale = 0.8]{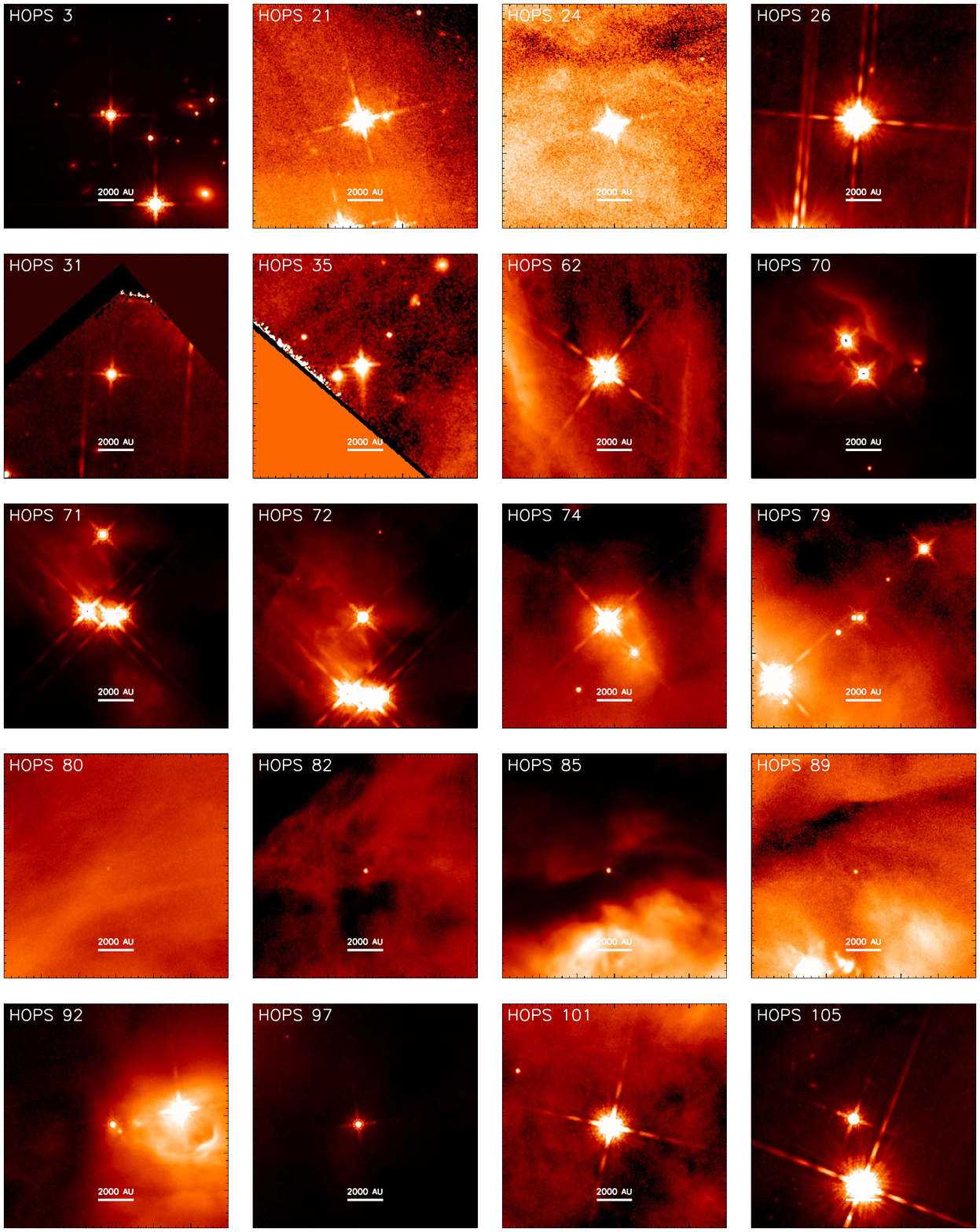}
	\caption{WFC3 F160W images of point sources without associated nebulosity.}
\end{figure*}

\begin{figure*}[ht!]\centering
	\includegraphics[scale = 0.8]{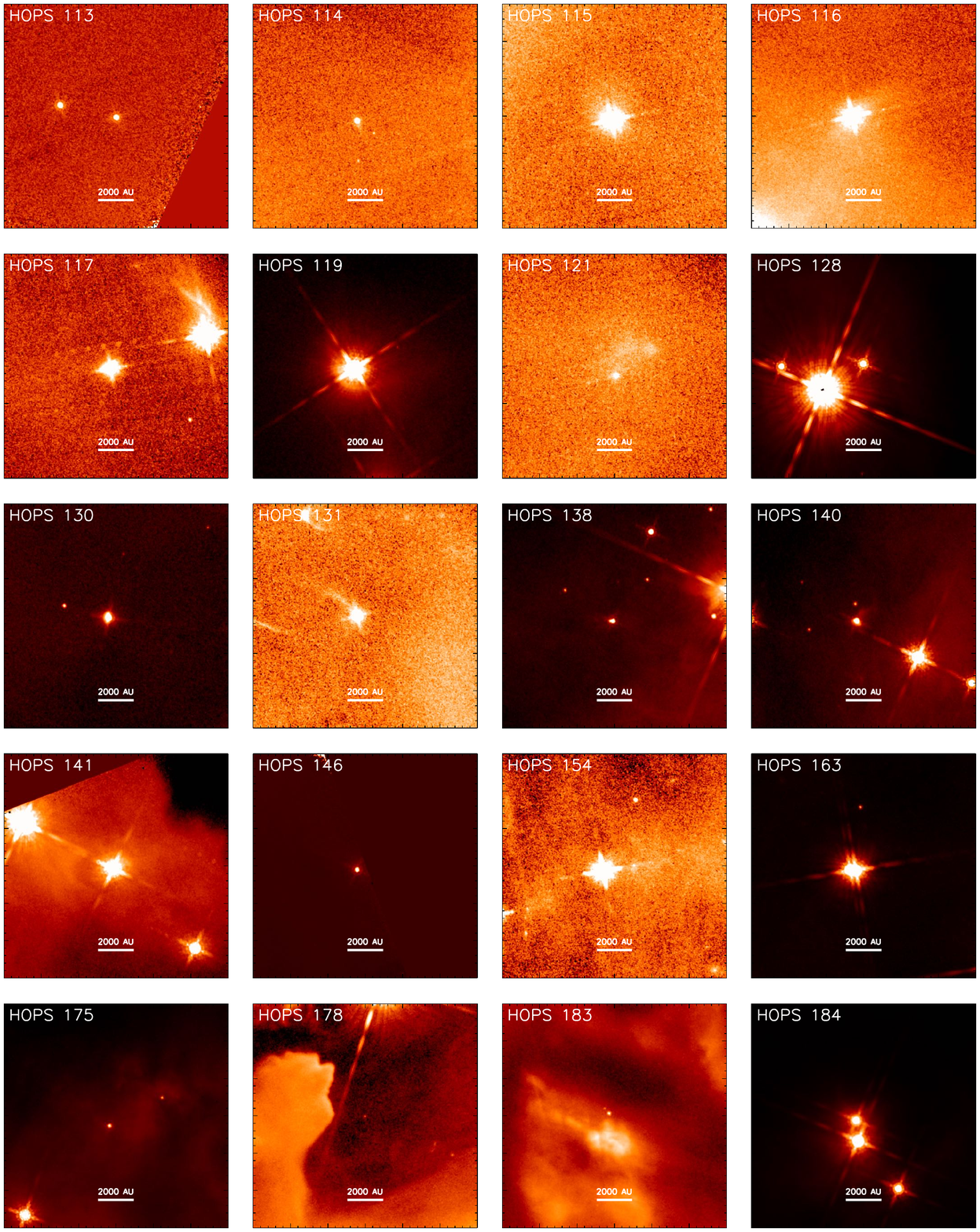}
	\caption{WFC3 F160W images of point sources without associated nebulosity, continued.}
\end{figure*}

\begin{figure*}[ht!]\centering
	\includegraphics[scale = 0.8]{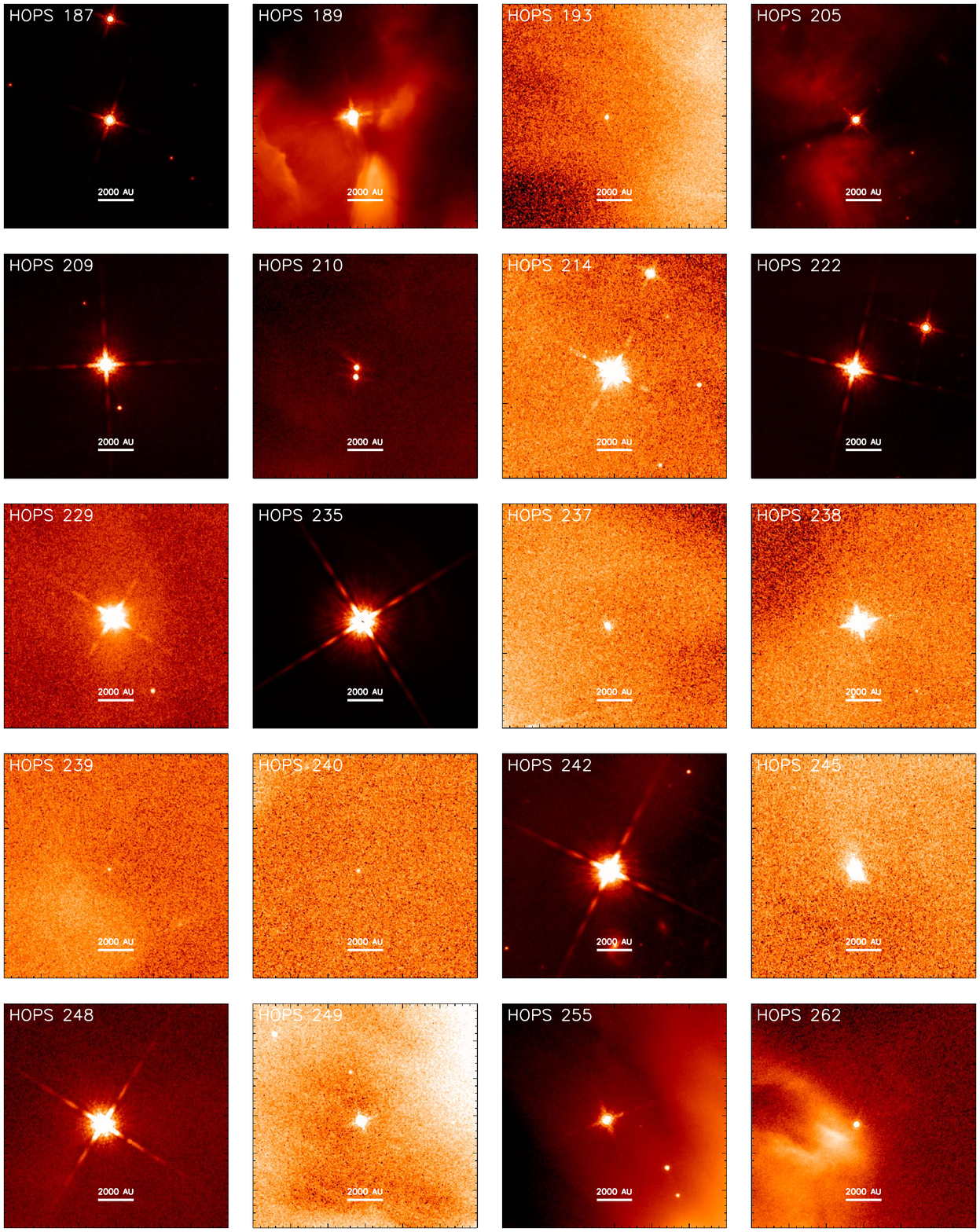}
	\caption{WFC3 F160W images of point sources without associated nebulosity, continued.}
\end{figure*}

\begin{figure*}[ht!]\centering
	\includegraphics[scale = 0.8]{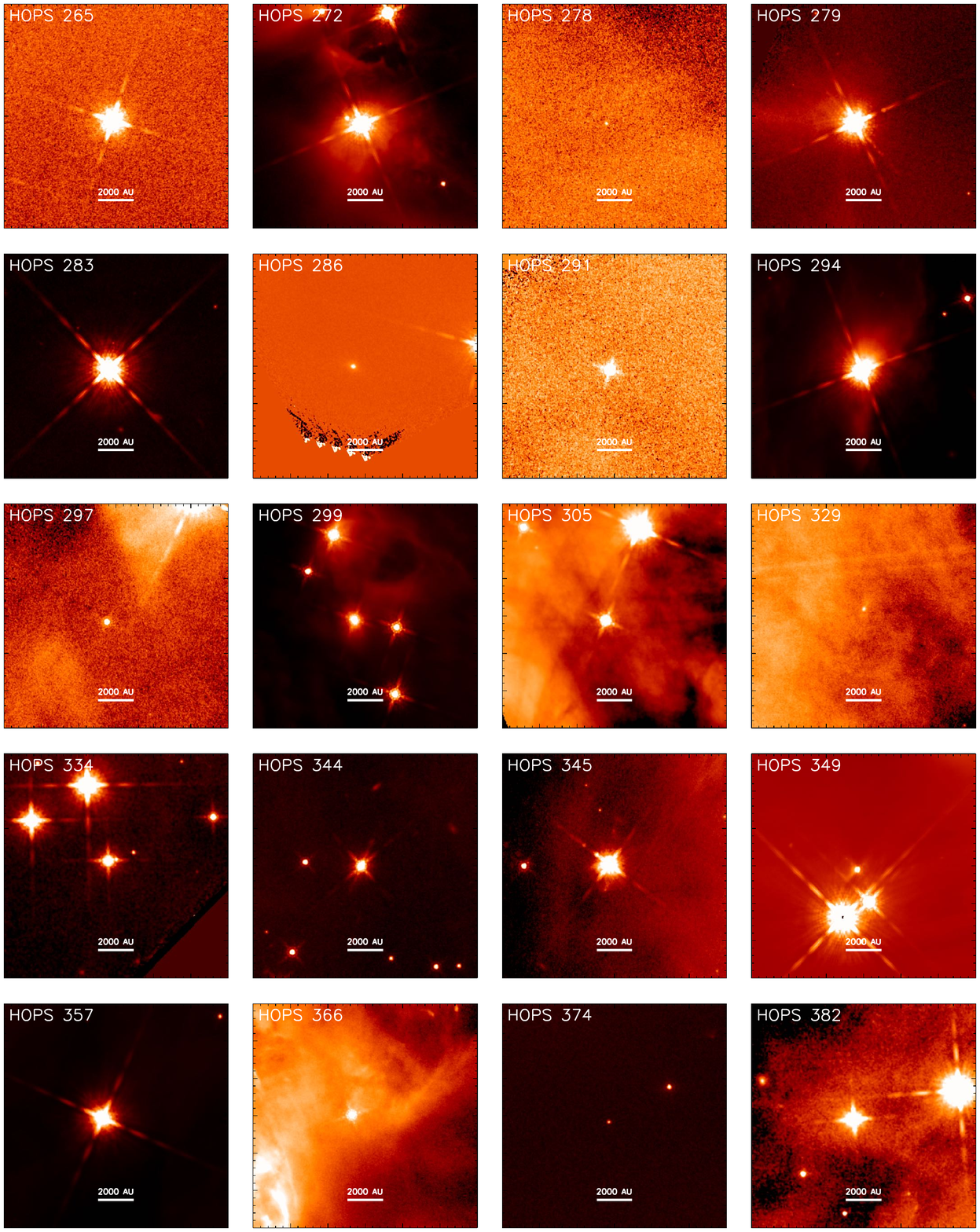}
	\caption{WFC3 F160W images of point sources without associated nebulosity, continued.}
\end{figure*}

\begin{figure*}[ht!]\centering
	\includegraphics[scale = 0.8]{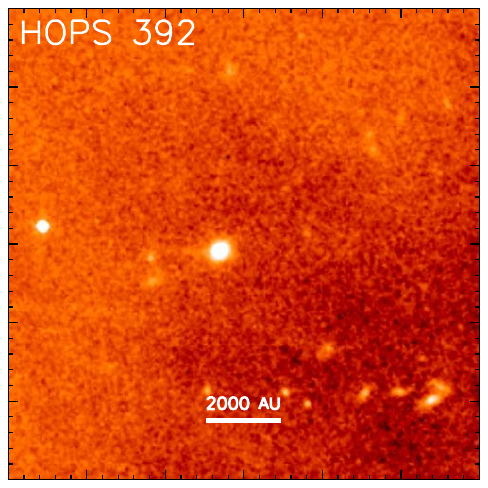}
	\caption{WFC3 F160W images of point sources without associated nebulosity, continued.}
\end{figure*}

%% file: table_HOPS_Cleaner_Format.tex
\begin{deluxetable*}{cccccccccccc}
\tablecaption{\label{tab:bigtable}}
\tablehead{\colhead{HOPS} & \colhead{RA} & \colhead{DEC} & \colhead{SED} & \colhead{$T_\text{bol}$\tablenotemark{1}} & \colhead{Instrument\tablenotemark{2}} & \colhead{Point} & \colhead{F160W} & \colhead{Volume} & \colhead{Half-Opening} & \colhead{Power-} \\
            \colhead{ID} & \multicolumn{2}{c}{ICRS} & \colhead{ Class\tablenotemark{1}} & \colhead{} & \colhead{} & \colhead{Source\tablenotemark{3}} & \colhead{Morphology} & \colhead{Cleared} & \colhead{Angle} & \colhead{Law Fit}\\
            \colhead{} & \colhead{ } & \colhead{} & \colhead{} & \colhead{(\si{\kelvin})} & \colhead{} & \colhead{} & \colhead{} & \colhead{(\%)} & \colhead{($^{\circ}$)} & \colhead{n}
          }
\startdata
1 & $05:54:12.3$ & $+01:42:35$ & I & 72.6 & NIC WFC3 & n & Unipolar & - & - & - \\ %
2 & $05:54:09.1$ & $+01:42:52$ & I & 356.5 & NIC & y & Point Source & - & - & - \\ %
3 & $05:54:56.9$ & $+01:42:56$ & flat & 467.5 & WFC3 & y & Point Source & - & - & - \\ %
4 & $05:54:53.7$ & $+01:47:09$ & I & 203.3 & WFC3 & y & Unipolar & - & - & - \\ %
5 & $05:54:32.1$ & $+01:48:07$ & I & 187.1 & WFC3 & y & Unipolar & - & - & - \\ %
6 & $05:54:18.4$ & $+01:49:03$ & I & 112.5 & NIC & y & Unipolar & - & - & - \\ %
7 & $05:54:20.0$ & $+01:50:42$ & 0 & 58.0 & NIC & n & Bipolar & - & - & - \\ %
10 & $05:35:09.0$ & $-05:58:27$ & 0 & 46.2 & NIC & n & Non Detection & - & - & - \\ %
11 & $05:35:13.4$ & $-05:57:58$ & 0 & 48.8 & NIC & n & Unipolar & - & - & - \\ %
12 & $05:35:08.6$ & $-05:55:54$ & 0 & 42.0 & NIC & n & Unipolar & - & - & - \\ %
13 & $05:35:24.5$ & $-05:55:33$ & flat & 383.6 & NIC & y & Irregular & - & - & - \\ %
14 & $05:36:19.1$ & $-05:55:30$ & flat & 464.0 & NIC & y & Irregular & - & - & - \\ %
15 & $05:36:19.0$ & $-05:55:25$ & flat & 342.0 & NIC & y & Point Source & - & - & - \\ %
16 & $05:35:00.8$ & $-05:55:25$ & flat & 361.0 & NIC & y & Point Source & - & - & - \\ %
17 & $05:35:07.1$ & $-05:52:05$ & I & 341.3 & WFC3 & y & Irregular & - & - & - \\ %
18 & $05:35:05.4$ & $-05:51:54$ & I & 71.8 & WFC3 & n & Unipolar & - & - & - \\ %
20 & $05:33:30.7$ & $-05:50:41$ & I & 94.8 & NIC & y & Unipolar & - & - & - \\ %
21 & $05:36:10.1$ & $-05:50:08$ & ex & 584.5 & WFC3 & y & Point Source & - & - & - \\ %
24 & $05:34:46.9$ & $-05:44:50$ & I & 288.9 & WFC3 & y & Point Source & - & - & - \\ %
26 & $05:35:17.3$ & $-05:42:14$ & II & 1124.9 & WFC3 & y & Point Source & - & - & - \\ %
28 & $05:34:47.2$ & $-05:41:55$ & 0 & 46.3 & WFC3 & n & Bipolar & - & - & - \\ %
29 & $05:34:49.0$ & $-05:41:42$ & I & 148.2 & WFC3 & y & Unipolar & 18.4 & 26.9 & 2.3 \\ %
30 & $05:34:44.0$ & $-05:41:25$ & I & 81.2 & NIC WFC3 & n & Non Detection & - & - & - \\ %
31 & $05:35:17.2$ & $-05:40:26$ & flat & 634.7 & WFC3 & y & Point Source & - & - & - \\ %
32 & $05:34:35.4$ & $-05:39:59$ & 0 & 58.9 & NIC & n & Unipolar & 14.4 & 28.1 & 1.3 \\ %
35 & $05:35:19.9$ & $-05:39:01$ & I & 305.2 & WFC3 & y & Point Source & - & - & - \\ %
36 & $05:34:26.4$ & $-05:37:40$ & flat & 374.6 & NIC & y & Point Source & - & - & - \\ %
38 & $05:35:04.7$ & $-05:37:12$ & 0 & 58.5 & WFC3 & n & Non Detection & - & - & - \\ %
40 & $05:35:08.5$ & $-05:35:59$ & 0 & 38.1 & WFC3 & n & Non Detection & - & - & - \\ %
41 & $05:34:29.4$ & $-05:35:42$ & I & 82.3 & NIC & n & Non Detection & - & - & - \\ %
42 & $05:35:05.0$ & $-05:35:40$ & I & 200.9 & NIC WFC3 & n & Bipolar & - & - & - \\ %
43 & $05:35:04.5$ & $-05:35:14$ & I & 75.0 & NIC WFC3 & n & Non Detection & - & - & - \\ %
44 & $05:35:10.5$ & $-05:35:06$ & 0 & 43.8 & NIC WFC3 & n & Unipolar & - & - & - \\ %
45 & $05:35:06.4$ & $-05:33:35$ & flat & 517.8 & NIC WFC3 & y & Irregular & - & - & - \\ %
48\tablenotemark{4} & $05:35:06.5$ & $-05:32:51$ & flat & 611.0 & WFC3 & y & Unipolar & - & - & - \\ %
50 & $05:34:40.9$ & $-05:31:44$ & 0 & 51.4 & WFC3 & y & Unipolar & 15.4 & 28.0 & 1.5 \\ %
53 & $05:33:57.3$ & $-05:23:30$ & 0 & 45.9 & NIC & n & Non Detection & - & - & - \\ %
56 & $05:35:19.4$ & $-05:15:32$ & 0 & 48.1 & NIC & n & Non Detection & - & - & - \\ %
57 & $05:35:19.8$ & $-05:15:08$ & flat & 421.2 & NIC & y & Irregular & - & - & - \\ %
58 & $05:35:18.5$ & $-05:13:38$ & flat & 620.0 & WFC3 & y & Irregular & - & - & - \\ %
59 & $05:35:20.1$ & $-05:13:15$ & flat & 528.4 & WFC3 & y & Irregular & - & - & - \\ %
60 & $05:35:23.3$ & $-05:12:03$ & 0 & 54.1 & WFC3 & n & Unipolar & - & - & - \\ %
62 & $05:35:24.5$ & $-05:11:29$ & flat & 1154.1 & NIC WFC3 & y & Point Source & - & - & - \\ %
63 & $05:35:24.9$ & $-05:10:01$ & flat & 544.5 & WFC3 & n & Non Detection & - & - & - \\ %
64 & $05:35:26.9$ & $-05:09:54$ & I & 29.7 & WFC3 & y & Unipolar & 0.49 & 4.1 & 3.8 \\ %
65 & $05:35:21.5$ & $-05:09:38$ & I & 545.7 & NIC & y & Unipolar & - & - & - \\ %
66 & $05:35:26.8$ & $-05:09:24$ & flat & 264.9 & WFC3 & y & Unipolar & 5.0 & 17.5 & 1.1 \\ %
67 & $05:35:22.6$ & $-05:08:34$ & I & 278.7 & WFC3 & y & Bipolar & - & - & - \\ %
68 & $05:35:24.3$ & $-05:08:30$ & I & 100.6 & WFC3 & n & Non Detection & - & - & - \\ %
69 & $05:35:25.2$ & $-05:08:23$ & flat & 31.3 & WFC3 & n & Non Detection & - & - & - \\ %
70 & $05:35:22.4$ & $-05:08:04$ & flat & 619.3 & WFC3 & y & Point Source & - & - & - \\ %
71 & $05:35:25.6$ & $-05:07:57$ & I & 277.5 & NIC WFC3 & y & Point Source & - & - & - \\ %
72 & $05:35:25.7$ & $-05:07:46$ & ex & 693.0 & NIC WFC3 & y & Point Source & - & - & - \\ %
73 & $05:35:27.7$ & $-05:07:03$ & 0 & 43.0 & WFC3 & n & Unipolar & 0.53 & 5.3 & 1.4 \\ %
74 & $05:35:24.8$ & $-05:06:21$ & flat & 516.5 & WFC3 & y & Point Source & - & - & - \\ %
75 & $05:35:26.6$ & $-05:06:10$ & 0 & 67.9 & WFC3 & n & Non Detection & - & - & - \\ %
76 & $05:35:25.7$ & $-05:05:57$ & I & 135.5 & NIC WFC3 & y & Unipolar & - & - & - \\ %
77 & $05:35:31.5$ & $-05:05:47$ & flat & 550.3 & WFC3 & y & Unipolar & - & - & - \\ %
78 & $05:35:25.8$ & $-05:05:43$ & 0 & 38.1 & WFC3 & n & Non Detection & - & - & - \\ %
79 & $05:35:27.8$ & $-05:05:36$ & flat & 666.2 & WFC3 & y & Point Source & - & - & - \\ %
80 & $05:35:25.1$ & $-05:05:09$ & flat & 275.3 & WFC3 & y & Point Source & - & - & - \\ %
81 & $05:35:27.9$ & $-05:04:58$ & 0 & 40.1 & WFC3 & n & Unipolar & 1.7 & 6.9 & 6.7 \\ %
82 & $05:35:19.7$ & $-05:04:54$ & flat & 116.4 & WFC3 & y & Point Source & - & - & - \\ %
84 & $05:35:26.5$ & $-05:03:55$ & I & 90.8 & WFC3 & y & Unipolar & 28.3 & 35.9 & 1.8 \\ %
85 & $05:35:28.1$ & $-05:03:40$ & flat & 174.2 & WFC3 & y & Point Source & - & - & - \\ %
86 & $05:35:23.6$ & $-05:01:40$ & I & 112.7 & WFC3 & y & Unipolar & - & - & - \\ %
87 & $05:35:23.4$ & $-05:01:28$ & 0 & 38.1 & WFC3 & n & Non Detection & - & - & - \\ %
88 & $05:35:22.4$ & $-05:01:14$ & 0 & 42.4 & WFC3 & n & Non Detection & - & - & - \\ %
89 & $05:35:19.9$ & $-05:01:02$ & flat & 158.3 & WFC3 & y & Point Source & - & - & - \\ %
91 & $05:35:18.9$ & $-05:00:50$ & 0 & 41.7 & WFC3 & n & Non Detection & - & - & - \\ %
92 & $05:35:18.3$ & $-05:00:32$ & flat & 186.3 & WFC3 & y & Point Source & - & - & - \\ %
93 & $05:35:15.0$ & $-05:00:08$ & I & 107.3 & NIC WFC3 & n & Bipolar & 5.1 & 17.3 & 1.2 \\ %
94 & $05:35:16.1$ & $-05:00:02$ & I & 123.0 & NIC WFC3 & y & Irregular & - & - & - \\ %
95 & $05:35:34.1$ & $-04:59:52$ & 0 & 41.8 & NIC & n & Non Detection & - & - & - \\ %
96 & $05:35:29.7$ & $-04:58:48$ & 0 & 35.6 & WFC3 & n & Non Detection & - & - & - \\ %
97 & $05:35:28.8$ & $-04:57:38$ & ex & 403.8 & WFC3 & y & Point Source & - & - & - \\ %
98 & $05:35:19.3$ & $-04:55:44$ & II & 587.5 & WFC3 & y & Irregular & - & - & - \\ %
99 & $05:34:29.4$ & $-04:55:30$ & 0 & 48.9 & NIC & n & Non Detection & - & - & - \\ %
101 & $05:35:08.2$ & $-04:54:09$ & ex & 481.2 & WFC3 & y & Point Source & - & - & - \\ %
105 & $05:35:32.2$ & $-04:46:48$ & flat & 520.3 & WFC3 & y & Point Source & - & - & - \\ %
107 & $05:35:23.3$ & $-04:40:10$ & flat & 472.0 & WFC3 & y & Unipolar & - & - & - \\ %
108 & $05:35:27.0$ & $-05:10:00$ & 0 & 38.5 & WFC3 & n & Non Detection & - & - & - \\ %
113 & $05:39:58.1$ & $-07:26:41$ & II & 583.8 & WFC3 & y & Point Source & - & - & - \\ %
114 & $05:40:01.3$ & $-07:25:38$ & I & 117.3 & WFC3 & y & Point Source & - & - & - \\ %
115 & $05:39:56.5$ & $-07:25:51$ & flat & 461.3 & WFC3 & y & Point Source & - & - & - \\ %
116 & $05:39:57.8$ & $-07:25:13$ & flat & 411.1 & WFC3 & y & Point Source & - & - & - \\ %
117 & $05:39:55.4$ & $-07:24:19$ & flat & 277.0 & WFC3 & y & Point Source & - & - & - \\ %
118 & $05:39:54.5$ & $-07:24:14$ & flat & 552.8 & WFC3 & y & Irregular & - & - & - \\ %
119 & $05:39:50.6$ & $-07:23:30$ & flat & 573.8 & WFC3 & y & Point Source & - & - & - \\ %
120 & $05:39:34.3$ & $-07:26:11$ & flat & 455.3 & WFC3 & y & Irregular & - & - & - \\ %
121 & $05:39:33.7$ & $-07:23:01$ & 0 & 34.8 & WFC3 & y & Point Source & - & - & - \\ %
123 & $05:39:33.2$ & $-07:22:57$ & 0 & 50.1 & WFC3 & n & Non Detection & - & - & - \\ %
124 & $05:39:19.9$ & $-07:26:11$ & 0 & 44.8 & NIC & n & Irregular & - & - & - \\ %
125 & $05:39:19.6$ & $-07:26:18$ & flat & 110.5 & NIC & y & Irregular & - & - & - \\ %
127 & $05:39:00.9$ & $-07:20:22$ & I & 133.3 & WFC3 & y & Unipolar & - & - & - \\ %
128 & $05:38:52.0$ & $-07:21:06$ & flat & 469.2 & WFC3 & y & Point Source & - & - & - \\ %
129 & $05:39:11.8$ & $-07:10:34$ & flat & 191.3 & WFC3 & y & Unipolar & - & - & - \\ %
130 & $05:39:02.9$ & $-07:12:52$ & I & 156.7 & WFC3 & y & Point Source & - & - & - \\ %
131 & $05:39:07.5$ & $-07:10:52$ & I & 82.3 & WFC3 & y & Point Source & - & - & - \\ %
132 & $05:39:05.3$ & $-07:11:05$ & flat & 616.3 & WFC3 & y & Unipolar & - & - & - \\ %
133 & $05:39:05.8$ & $-07:10:39$ & I & 74.6 & WFC3 & n & Bipolar & - & - & - \\ %
135 & $05:38:45.3$ & $-07:10:55$ & I & 130.3 & NIC & n & Bipolar & 22.1 & 38.4 & 1.0 \\ %
136\tablenotemark{5} & $05:38:46.5$ & $-07:05:37$ & I & 161.7 & NIC & n & Bipolar & 1.8 & 8.7 & 2.5 \\ %
138 & $05:38:48.3$ & $-07:02:43$ & 0 & 42.8 & WFC3 & y & Point Source & - & - & - \\ %
139 & $05:38:49.6$ & $-07:01:17$ & I & 84.3 & WFC3 & n & Irregular & - & - & - \\ %
140 & $05:38:46.2$ & $-07:01:53$ & I & 137.2 & WFC3 & y & Point Source & - & - & - \\ %
141 & $05:38:48.0$ & $-07:00:49$ & flat & 741.6 & NIC WFC3 & y & Point Source & - & - & - \\ %
143 & $05:38:46.1$ & $-07:00:48$ & I & 242.1 & NIC WFC3 & n & Non Detection & - & - & - \\ %
144 & $05:38:45.0$ & $-07:01:01$ & I & 99.2 & NIC WFC3 & n & Non Detection & - & - & - \\ %
145 & $05:38:43.8$ & $-07:01:13$ & I & 133.7 & NIC WFC3 & y & Irregular & - & - & - \\ %
146 & $05:38:44.1$ & $-07:00:40$ & ex & 519.7 & WFC3 & y & Point Source & - & - & - \\ %
148 & $05:38:39.5$ & $-06:59:30$ & I & 262.9 & NIC & y & Point Source & - & - & - \\ %
150 & $05:38:07.5$ & $-07:08:29$ & flat & 245.2 & WFC3 & y & Bipolar & 11.6 & 26.2 & 1.2 \\ %
152 & $05:37:58.7$ & $-07:07:25$ & 0 & 53.8 & WFC3 & n & Non Detection & - & - & - \\ %
153 & $05:37:57.0$ & $-07:06:56$ & 0 & 39.4 & WFC3 & n & Non Detection & - & - & - \\ %
154 & $05:38:20.0$ & $-06:59:04$ & I & 166.7 & WFC3 & y & Point Source & - & - & - \\ %
156 & $05:38:03.4$ & $-06:58:15$ & I & 90.1 & NIC & y & Point Source & - & - & - \\ %
157 & $05:37:56.5$ & $-06:56:39$ & I & 77.6 & NIC & n & Irregular & - & - & - \\ %
158 & $05:37:24.4$ & $-06:58:32$ & flat & 591.6 & NIC & y & Point Source & - & - & - \\ %
159 & $05:37:53.7$ & $-06:47:16$ & flat & 498.4 & NIC & y & Point Source & - & - & - \\ %
160 & $05:37:51.0$ & $-06:47:20$ & I & 80.4 & NIC & y & Unipolar & - & - & - \\ %
163 & $05:37:17.2$ & $-06:36:18$ & I & 432.3 & WFC3 & y & Point Source & - & - & - \\ %
164 & $05:37:00.4$ & $-06:37:10$ & 0 & 50.0 & WFC3 & n & Unipolar & 1.2 & 7.5 & 1.8 \\ %
165 & $05:36:23.5$ & $-06:46:14$ & I & 96.1 & NIC & n & Non Detection & - & - & - \\ %
166 & $05:36:25.1$ & $-06:44:41$ & flat & 457.1 & NIC & y & Point Source & - & - & - \\ %
167 & $05:36:19.7$ & $-06:46:00$ & flat & 568.6 & NIC & y & Point Source & - & - & - \\ %
168 & $05:36:18.9$ & $-06:45:22$ & 0 & 54.0 & NIC & n & Irregular & - & - & - \\ %
169 & $05:36:36.1$ & $-06:38:51$ & 0 & 32.5 & NIC & n & Non Detection & - & - & - \\ %
170 & $05:36:41.3$ & $-06:34:00$ & flat & 832.5 & NIC & y & Point Source & - & - & - \\ %
171 & $05:36:17.1$ & $-06:38:01$ & 0 & 61.8 & WFC3 & n & Bipolar & 8.2 & 21.3 & 1.3 \\ %
172 & $05:36:19.4$ & $-06:29:06$ & I & 149.8 & NIC & y & Point Source & - & - & - \\ %
173 & $05:36:26.0$ & $-06:25:05$ & 0 & 60.2 & WFC3 & n & Non Detection & - & - & - \\ %
174 & $05:36:25.8$ & $-06:24:58$ & flat & 350.3 & WFC3 & y & Irregular & - & - & - \\ %
175 & $05:36:24.0$ & $-06:24:54$ & I & 104.3 & NIC WFC3 & y & Point Source & - & - & - \\ %
176 & $05:36:23.5$ & $-06:24:51$ & flat & 312.2 & NIC WFC3 & y & Irregular & - & - & - \\ %
177 & $05:35:50.0$ & $-06:34:53$ & I & 84.7 & NIC & n & Irregular & - & - & - \\ %
178 & $05:36:24.6$ & $-06:22:41$ & I & 155.1 & WFC3 & y & Point Source & - & - & - \\ %
179 & $05:36:21.8$ & $-06:23:29$ & flat & 467.5 & WFC3 & y & Irregular & - & - & - \\ %
181 & $05:36:19.5$ & $-06:22:12$ & I & 131.3 & WFC3 & n & Non Detection & - & - & - \\ %
182 & $05:36:18.8$ & $-06:22:10$ & 0 & 51.9 & WFC3 & n & Non Detection & - & - & - \\ %
183 & $05:36:17.8$ & $-06:22:28$ & flat & 224.5 & WFC3 & y & Point Source & - & - & - \\ %
184 & $05:36:12.9$ & $-06:23:30$ & II & 201.3 & WFC3 & y & Point Source & - & - & - \\ %
185 & $05:36:36.9$ & $-06:14:57$ & I & 96.9 & WFC3 & y & Unipolar & 9.1 & 18.2 & 2.7 \\ %
186 & $05:35:47.2$ & $-06:26:14$ & I & 72.3 & WFC3 & n & Bipolar & - & - & - \\ %
187 & $05:35:50.9$ & $-06:22:43$ & flat & 1210.9 & WFC3 & y & Point Source & - & - & - \\ %
188 & $05:35:29.8$ & $-06:26:58$ & I & 103.3 & WFC3 & y & Irregular & - & - & - \\ %
189 & $05:35:30.8$ & $-06:26:32$ & I & 133.1 & NIC WFC3 & y & Point Source & - & - & - \\ %
190 & $05:35:28.4$ & $-06:27:01$ & I & 385.3 & WFC3 & y & Bipolar & 13.3 & 26.2 & 1.5 \\ %
191 & $05:36:17.2$ & $-06:11:10$ & I & 196.7 & NIC & y & Bipolar & - & - & - \\ %
192 & $05:36:32.4$ & $-06:01:16$ & flat & 202.5 & WFC3 & y & Unipolar & - & - & - \\ %
193 & $05:36:30.2$ & $-06:01:17$ & I & 226.7 & NIC WFC3 & y & Point Source & - & - & - \\ %
194 & $05:35:51.9$ & $-06:10:01$ & flat & 645.0 & NIC & y & Point Source & - & - & - \\ %
197 & $05:34:15.8$ & $-06:34:32$ & flat & 506.6 & NIC & y & Point Source & - & - & - \\ %
198 & $05:35:22.1$ & $-06:13:06$ & 0 & 61.4 & NIC & n & Irregular & - & - & - \\ %
199 & $05:34:39.8$ & $-06:25:14$ & flat & 576.7 & NIC & y & Point Source & - & - & - \\ %
200 & $05:35:33.2$ & $-06:06:09$ & flat & 244.4 & NIC & n & Bipolar & - & - & - \\ %
203 & $05:36:22.8$ & $-06:46:06$ & 0 & 43.7 & NIC & n & Non Detection & - & - & - \\ %
204 & $05:43:10.1$ & $-08:46:07$ & I & 85.4 & NIC & y & Unipolar & - & - & - \\ %
205 & $05:43:02.8$ & $-08:47:49$ & ex & 427.8 & WFC3 & y & Point Source & - & - & - \\ %
206 & $05:43:07.2$ & $-08:44:31$ & 0 & 65.1 & WFC3 & n & Non Detection & - & - & - \\ %
207 & $05:42:38.5$ & $-08:50:18$ & flat & 446.2 & WFC3 & y & Irregular & - & - & - \\ %
209 & $05:42:52.8$ & $-08:41:41$ & I & 554.1 & WFC3 & y & Point Source & - & - & - \\ %
210 & $05:42:58.2$ & $-08:38:05$ & flat & 204.9 & WFC3 & y & Point Source & - & - & - \\ %
211 & $05:42:58.3$ & $-08:37:43$ & flat & 87.9 & WFC3 & n & Non Detection & - & - & - \\ %
213 & $05:42:48.0$ & $-08:40:08$ & flat & 534.9 & WFC3 & y & Unipolar & - & - & - \\ %
214 & $05:42:47.2$ & $-08:36:36$ & flat & 360.8 & WFC3 & y & Point Source & - & - & - \\ %
215 & $05:43:09.5$ & $-08:29:27$ & I & 195.5 & NIC & y & Point Source & - & - & - \\ %
216 & $05:42:55.5$ & $-08:32:48$ & I & 117.7 & WFC3 & n & Bipolar & - & - & - \\ %
219 & $05:41:29.2$ & $-08:43:04$ & I & 90.0 & WFC3 & n & Bipolar & 37.6 & 49.7 & 1.1 \\ %
220 & $05:41:29.7$ & $-08:42:45$ & I & 193.6 & WFC3 & y & Bipolar & - & - & - \\ %
221 & $05:42:47.0$ & $-08:17:06$ & I & 172.3 & WFC3 & y & Unipolar & - & - & - \\ %
222 & $05:41:26.6$ & $-08:42:24$ & II & 738.2 & WFC3 & y & Point Source & - & - & - \\ %
223 & $05:42:48.4$ & $-08:16:34$ & I & 247.5 & WFC3 & y & Irregular & - & - & - \\ %
224 & $05:41:32.0$ & $-08:40:09$ & 0 & 48.6 & NIC & n & Non Detection & - & - & - \\ %
225 & $05:41:30.3$ & $-08:40:17$ & flat & 432.5 & NIC & y & Point Source & - & - & - \\ %
226 & $05:41:30.0$ & $-08:40:09$ & flat & 350.2 & NIC & y & Point Source & - & - & - \\ %
228 & $05:41:34.1$ & $-08:35:27$ & I & 293.0 & NIC & n & Unipolar & - & - & - \\ %
229 & $05:42:47.3$ & $-08:10:08$ & flat & 471.6 & WFC3 & y & Point Source & - & - & - \\ %
232 & $05:41:35.4$ & $-08:08:22$ & I & 187.9 & WFC3 & y & Bipolar & - & - & - \\ %
233 & $05:41:52.3$ & $-08:01:21$ & I & 106.2 & WFC3 & y & Bipolar & 0.71 & 4.9 & 3.9 \\ %
234 & $05:41:49.9$ & $-08:01:26$ & I & 79.8 & WFC3 & y & Irregular & - & - & - \\ %
235 & $05:41:25.3$ & $-08:05:54$ & flat & 680.1 & WFC3 & y & Point Source & - & - & - \\ %
236 & $05:41:30.2$ & $-08:03:41$ & flat & 332.8 & WFC3 & y & Unipolar & - & - & - \\ %
237 & $05:41:28.9$ & $-08:03:25$ & I & 177.7 & NIC WFC3 & y & Point Source & - & - & - \\ %
238 & $05:41:26.6$ & $-08:03:12$ & I & 269.1 & WFC3 & y & Point Source & - & - & - \\ %
239 & $05:41:27.0$ & $-08:00:54$ & I & 116.2 & WFC3 & y & Point Source & - & - & - \\ %
240 & $05:41:25.9$ & $-08:01:15$ & I & 191.0 & WFC3 & y & Point Source & - & - & - \\ %
241 & $05:41:26.3$ & $-08:01:02$ & I & 100.3 & WFC3 & n & Non Detection & - & - & - \\ %
242 & $05:40:48.5$ & $-08:11:08$ & flat & 836.7 & WFC3 & y & Point Source & - & - & - \\ %
243 & $05:41:01.6$ & $-08:06:44$ & 0 & 50.8 & WFC3 & n & Non Detection & - & - & - \\ %
244 & $05:41:01.9$ & $-08:06:01$ & I & 127.3 & WFC3 & n & Unipolar & 10.5 & 23.4 & 1.4 \\ %
245 & $05:41:22.8$ & $-07:58:55$ & flat & 302.1 & WFC3 & y & Point Source & - & - & - \\ %
246 & $05:40:47.1$ & $-08:09:47$ & I & 95.6 & WFC3 & y & Irregular & - & - & - \\ %
247 & $05:41:26.2$ & $-07:56:51$ & 0 & 42.8 & WFC3 & n & Irregular & - & - & - \\ %
248 & $05:41:22.0$ & $-07:58:02$ & flat & 484.3 & WFC3 & y & Point Source & - & - & - \\ %
249 & $05:40:52.8$ & $-08:05:48$ & flat & 268.5 & WFC3 & y & Point Source & - & - & - \\ %
250 & $05:40:48.8$ & $-08:06:57$ & 0 & 69.4 & WFC3 & y & Unipolar & 29.8 & 38.4 & 1.5 \\ %
251 & $05:40:54.0$ & $-08:05:13$ & flat & 345.7 & WFC3 & y & Unipolar & - & - & - \\ %
252 & $05:40:49.9$ & $-08:06:08$ & flat & 329.2 & WFC3 & y & Irregular & - & - & - \\ %
253 & $05:41:28.7$ & $-07:53:50$ & flat & 321.1 & WFC3 & y & Unipolar & - & - & - \\ %
254 & $05:41:24.5$ & $-07:55:07$ & I & 114.7 & WFC3 & n & Unipolar & - & - & - \\ %
255 & $05:40:50.5$ & $-08:05:48$ & flat & 572.0 & WFC3 & y & Point Source & - & - & - \\ %
256 & $05:40:45.2$ & $-08:06:42$ & 0 & 72.4 & WFC3 & y & Irregular & - & - & - \\ %
257 & $05:41:19.8$ & $-07:55:46$ & flat & 292.6 & WFC3 & y & Irregular & - & - & - \\ %
258 & $05:41:24.7$ & $-07:54:08$ & flat & 385.7 & WFC3 & y & Irregular & - & - & - \\ %
259 & $05:40:20.8$ & $-08:13:55$ & flat & 410.3 & WFC3 & y & Unipolar & - & - & - \\ %
260 & $05:40:19.3$ & $-08:14:16$ & flat & 600.1 & WFC3 & y & Irregular & - & - & - \\ %
261 & $05:41:18.8$ & $-07:55:29$ & I & 149.5 & WFC3 & y & Irregular & - & - & - \\ %
262 & $05:41:23.9$ & $-07:53:41$ & flat & 202.4 & WFC3 & y & Point Source & - & - & - \\ %
263 & $05:41:23.6$ & $-07:53:46$ & I & 145.1 & WFC3 & n & Non Detection & - & - & - \\ %
265 & $05:41:20.3$ & $-07:53:10$ & flat & 635.1 & WFC3 & y & Point Source & - & - & - \\ %
267 & $05:41:19.6$ & $-07:50:41$ & I & 186.2 & NIC & y & Point Source & - & - & - \\ %
268 & $05:40:38.3$ & $-08:00:35$ & I & 113.9 & NIC & n & Irregular & - & - & - \\ %
270 & $05:40:40.5$ & $-07:54:39$ & I & 96.6 & WFC3 & n & Bipolar & - & - & - \\ %
272 & $05:40:20.5$ & $-07:56:39$ & II & 559.2 & WFC3 & y & Point Source & - & - & - \\ %
273 & $05:40:20.8$ & $-07:56:24$ & I & 243.3 & WFC3 & y & Unipolar & 22.8 & 34.4 & 1.4 \\ %
274 & $05:40:20.7$ & $-07:54:59$ & flat & 546.5 & WFC3 & y & Irregular & - & - & - \\ %
275 & $05:40:36.3$ & $-07:49:06$ & I & 146.4 & WFC3 & y & Unipolar & 22.5 & 35.7 & 1.3 \\ %
278 & $05:40:20.3$ & $-07:51:14$ & I & 96.3 & WFC3 & y & Point Source & - & - & - \\ %
279 & $05:40:17.7$ & $-07:48:25$ & flat & 382.0 & NIC WFC3 & y & Point Source & - & - & - \\ %
280\tablenotemark{5} & $05:40:14.9$ & $-07:48:48$ & I & 121.2 & WFC3 & y & Bipolar & 16.0 & 28.4 & 1.5 \\ %
281 & $05:40:24.6$ & $-07:43:08$ & flat & 189.3 & WFC3 & y & Irregular & - & - & - \\ %
282 & $05:40:26.0$ & $-07:37:31$ & I & 95.1 & WFC3 & n & Irregular & - & - & - \\ %
283 & $05:40:44.6$ & $-07:29:54$ & II & 807.9 & WFC3 & y & Point Source & - & - & - \\ %
284 & $05:38:51.4$ & $-08:01:27$ & flat & 913.9 & WFC3 & y & Unipolar & - & - & - \\ %
286 & $05:39:58.6$ & $-07:31:12$ & I & 123.7 & NIC WFC3 & y & Point Source & - & - & - \\ %
287 & $05:40:08.7$ & $-07:27:27$ & I & 117.8 & WFC3 & n & Bipolar & 6.3 & 15.4 & 2.7 \\ %
288 & $05:39:55.9$ & $-07:30:27$ & 0 & 48.6 & WFC3 & n & Unipolar & - & - & - \\ %
289 & $05:39:56.7$ & $-07:30:06$ & I & 331.1 & WFC3 & y & Bipolar & - & - & - \\ %
290 & $05:39:57.4$ & $-07:29:33$ & 0 & 47.3 & WFC3 & n & Non Detection & - & - & - \\ %
291 & $05:39:57.9$ & $-07:28:57$ & flat & 340.1 & WFC3 & y & Point Source & - & - & - \\ %
294 & $05:40:51.7$ & $-02:26:48$ & flat & 606.8 & WFC3 & y & Point Source & - & - & - \\ %
295 & $05:41:28.9$ & $-02:23:19$ & I & 86.6 & WFC3 & y & Bipolar & - & - & - \\ %
297 & $05:41:23.2$ & $-02:17:35$ & I & 274.9 & WFC3 & y & Point Source & - & - & - \\ %
298 & $05:41:37.1$ & $-02:17:16$ & I & 169.3 & WFC3 & y & Irregular & - & - & - \\ %
299 & $05:41:44.5$ & $-02:16:06$ & I & 277.0 & NIC WFC3 & y & Point Source & - & - & - \\ %
300 & $05:41:24.2$ & $-02:16:06$ & I & 93.7 & WFC3 & y & Unipolar & 1.0 & 6.8 & 2.0 \\ %
301 & $05:41:44.7$ & $-02:15:55$ & flat & 518.8 & WFC3 & y & Unipolar & - & - & - \\ %
305 & $05:41:45.3$ & $-01:51:56$ & flat & 300.7 & WFC3 & y & Point Source & - & - & - \\ %
310 & $05:42:27.6$ & $-01:20:00$ & 0 & 51.8 & WFC3 & n & Unipolar & - & - & - \\ %
311 & $05:43:03.0$ & $-01:16:28$ & flat & 383.0 & NIC WFC3 & y & Irregular & - & - & - \\ %
312 & $05:43:05.7$ & $-01:15:54$ & 0 & 46.7 & WFC3 & n & Non Detection & - & - & - \\ %
316 & $05:46:07.2$ & $-00:13:22$ & 0 & 55.2 & NIC & y & Irregular & - & - & - \\ %
317 & $05:46:08.5$ & $-00:10:38$ & 0 & 47.5 & WFC3 & n & Unipolar & - & - & - \\ %
321 & $05:46:33.1$ & $+00:00:02$ & I & 78.6 & WFC3 & n & Unipolar & - & - & - \\ %
322 & $05:46:46.4$ & $+00:00:16$ & I & 71.3 & WFC3 & n & Unipolar & 13.5 & 27.4 & 1.3 \\ %
323 & $05:46:47.6$ & $+00:00:25$ & I & 82.9 & WFC3 & n & Non Detection & - & - & - \\ %
324 & $05:46:37.5$ & $+00:00:34$ & I & 89.9 & WFC3 & n & Unipolar & - & - & - \\ %
325 & $05:46:39.2$ & $+00:01:14$ & 0 & 49.2 & WFC3 & y & Irregular & - & - & - \\ %
326 & $05:46:39.5$ & $+00:04:16$ & 0 & 58.8 & WFC3 & n & Non Detection & - & - & - \\ %
329 & $05:47:01.6$ & $+00:17:58$ & I & 89.2 & WFC3 & y & Point Source & - & - & - \\ %
331 & $05:46:28.3$ & $+00:19:49$ & flat & 82.5 & WFC3 & n & Non Detection & - & - & - \\ %
333\tablenotemark{5} & $05:47:22.8$ & $+00:20:58$ & flat & 240.9 & WFC3 & y & Bipolar & 14.9 & 29.1 & 1.5 \\ %
334 & $05:46:48.5$ & $+00:21:28$ & flat & 506.7 & WFC3 & y & Point Source & - & - & - \\ %
335 & $05:47:05.8$ & $+00:22:38$ & I & 81.1 & WFC3 & n & Irregular & - & - & - \\ %
337 & $05:46:55.0$ & $+00:23:34$ & I & 128.8 & WFC3 & n & Unipolar & - &-  & - \\ %
338 & $05:46:57.3$ & $+00:23:50$ & 0 & 53.7 & WFC3 & n & Non Detection & - & - & - \\ %
340 & $05:47:01.2$ & $+00:26:21$ & 0 & 40.6 & NIC & n & Unipolar & 2.4 & 10.4 & 1.8 \\ %
341 & $05:47:00.9$ & $+00:26:22$ & 0 & 39.4 & NIC & n & Non Detection & - & - & - \\ %
342 & $05:47:57.0$ & $+00:35:27$ & I & 312.6 & WFC3 & y & Bipolar & - & - & - \\ %
343 & $05:47:59.0$ & $+00:35:32$ & I & 82.1 & WFC3 & n & Irregular & - & - & - \\ %
344 & $05:47:24.7$ & $+00:37:35$ & I & 408.2 & WFC3 & y & Point Source & - & - & - \\ %
345 & $05:47:38.9$ & $+00:38:36$ & I & 219.4 & WFC3 & y & Point Source & - & - & - \\ %
346 & $05:47:42.9$ & $+00:40:57$ & flat & 649.5 & WFC3 & y & Unipolar & - & - & - \\ %
347 & $05:47:15.8$ & $+00:21:23$ & 0 & 33.5 & WFC3 & n & Non Detection & - & - & - \\ %
349 & $05:35:26.1$ & $-05:08:33$ & - & - & WFC3 & y & Point Source & - & - & - \\ %
351 & $05:35:31.4$ & $-05:04:47$ & ex & 217.1 & WFC3 & n & Non Detection & - & - & - \\ %
352 & $05:35:26.8$ & $-05:04:02$ & - & - & WFC3 & n & Non Detection & - & - & - \\ %
353 & $05:54:13.3$ & $+01:43:03$ & - & - & WFC3 & n & Non Detection & - & - & - \\ %
354 & $05:54:24.2$ & $+01:44:19$ & 0 & 34.8 & WFC3 & n & Non Detection & - & - & - \\ %
355 & $05:37:17.0$ & $-06:49:49$ & 0 & 44.9 & WFC3 & n & Non Detection & - & - & - \\ %
357 & $05:41:39.0$ & $-01:52:07$ & flat & 628.2 & WFC3 & y & Point Source & - & - & - \\ %
358 & $05:46:07.2$ & $-00:13:29$ & 0 & 41.7 & NIC & n & Non Detection & - & - & - \\ %
359 & $05:47:24.8$ & $+00:20:59$ & 0 & 36.7 & WFC3 & n & Non Detection & - & - & - \\ %
360 & $05:47:27.0$ & $+00:20:33$ & I & 43.2 & WFC3 & n & Non Detection & - & - & - \\ %
361 & $05:47:04.7$ & $+00:21:42$ & 0 & 69.0 & WFC3 & y & Irregular & - & - & - \\ %
363 & $05:46:43.1$ & $+00:00:52$ & flat & 367.6 & WFC3 & y & Irregular & - & - & - \\ %
364 & $05:47:36.5$ & $+00:20:06$ & I & 96.7 & WFC3 & n & Unipolar & 5.2 & 16.9 & 1.3 \\ %
365 & $05:47:10.6$ & $+00:21:14$ & I & 160.3 & WFC3 & n & Unipolar & - & - & - \\ %
366 & $05:47:03.9$ & $+00:22:10$ & I & 292.2 & WFC3 & y & Point Source & - & - & - \\ %
367 & $05:54:36.2$ & $+01:53:54$ & I & 249.4 & WFC3 & y & Bipolar & 0.91 & 6.4 & 1.9 \\ %
368 & $05:35:24.7$ & $-05:10:30$ & I & 137.5 & WFC3 & n & Bipolar & - & - & - \\ %
369 & $05:35:26.9$ & $-05:10:17$ & flat & 379.2 & WFC3 & y & Irregular & - & - & - \\ %
370 & $05:35:27.6$ & $-05:09:33$ & I & 71.5 & WFC3 & n & Irregular & - & - & - \\ %
374 & $05:41:25.4$ & $-07:55:18$ & 0 & 56.9 & WFC3 & y & Point Source & - & - & - \\ %
376 & $05:38:18.1$ & $-07:02:26$ & flat & 492.0 & WFC3 & y & Irregular & - & - & - \\ %
377 & $05:38:45.5$ & $-07:01:02$ & 0 & 53.7 & NIC WFC3 & n & Non Detection & - & - & - \\ %
380 & $05:36:25.2$ & $-06:25:02$ & 0 & 36.6 & WFC3 & n & Non Detection & - & - & - \\ %
382 & $05:35:21.6$ & $-05:37:57$ & I & 204.4 & WFC3 & y & Point Source & - & - & - \\ %
383 & $05:35:29.8$ & $-04:59:51$ & 0 & 45.8 & WFC3 & n & Non Detection & - & - & - \\ %
386 & $05:46:08.4$ & $-00:10:02$ & I & 147.4 & WFC3 & y & Irregular & - & - & - \\ %
387 & $05:46:07.8$ & $-00:10:00$ & I & 118.3 & WFC3 & y & Bipolar & - & - & - \\ %
389 & $05:46:47.0$ & $+00:00:27$ & 0 & 42.8 & WFC3 & y & Irregular & - & - & - \\ %
391 & $05:47:17.0$ & $+00:20:53$ & 0 & 58.1 & WFC3 & n & Non Detection & - & - & - \\ %
392 & $05:46:16.4$ & $+00:21:36$ & 0 & 62.4 & WFC3 & y & Point Source & - & - & - \\ %
393 & $05:46:42.4$ & $+00:23:01$ & I & 250.5 & WFC3 & y & Bipolar & - & - & - \\ %
394 & $05:35:23.9$ & $-05:07:53$ & 0 & 45.5 & WFC3 & y & Unipolar & - & - & - \\ %
397 & $05:42:48.8$ & $-08:16:10$ & 0 & 46.1 & WFC3 & n & Non Detection & - & - & - \\ %
399 & $05:41:24.9$ & $-02:18:08$ & 0 & 31.1 & WFC3 & n & Non Detection & - & - & - \\ %
405 & $05:40:58.4$ & $-08:05:36$ & 0 & 35.0 & WFC3 & n & Non Detection & - & - & - \\ %
406 & $05:47:43.3$ & $+00:38:22$ & 0 & 24.6 & WFC3 & n & Non Detection & - & - & - \\ %
407 & $05:46:28.2$ & $+00:19:27$ & 0 & 26.8 & WFC3 & n & Non Detection & - & - & - \\ %
408 & $05:39:30.7$ & $-07:23:59$ & 0 & 37.9 & WFC3 & n & Non Detection & - & - & - \\ %
410\tablenotemark{6} & $05:46:53.2$ & $+00:22:10$ & 0 & 39.6 & WFC3 & y & Bipolar & - & - & - \\ %
\enddata
\tablenotetext{1}{~\citet{furlan_herschel_2016}}
\tablenotetext{2}{~Objects observed with the instrument NICMOS are noted with the abbreviation "NIC."}
\tablenotetext{3}{~Denotes, (yes or no), if the object is observed with an apparent central point source.}
\tablenotetext{4}{~This source is coincident with the tip of a large pillar in the surrounding cloud. We judge features surrounding it to be consistent with a unipolar source but cannot rule out a point source incidentally appearing with a "false" cavity.}
\tablenotetext{5}{~ Averaged parameters are reported for these bipolar sources where both cavities were measured.}
\tablenotetext{6}{~Coordinates, class and $T_\text{bol}$ are from \citet{stutz_herschel_2013}.}

\end{deluxetable*}